\newif\ifsubmode

\submodefalse

\newif\ifcmtmode
\cmtmodetrue

\ifsubmode
  \documentclass[12pt,preprint]{aastex}
  \usepackage{natbib,amsmath,verbatim}
  \received{}
  \revised{}
  \accepted{}
\else
  \documentclass[12pt,preprint,iop]{emulateapj}
  \usepackage{natbib,amsmath,verbatim}

\fi

\usepackage[usenames]{color}

\newcommand{\lta}{\lesssim}                                               
\newcommand{\gta}{\gtrsim}                                                
                                       
\newcommand{\microns}{\,\mu \rm m}                                       

\def\Ha{\ensuremath{{\rm H}\alpha}}

\def\Hb{\ensuremath{{\rm H}\beta}}

\def\NII{\ensuremath{[{\rm N~II}]}}

\def\ergcmsqpersec{\ensuremath{{\rm erg\,cm^{-2}\,s^{-1}}}}

\def\Om{\ensuremath{\Omega_{\rm m}}}
\def\Ot{\ensuremath{\Omega_{\rm tot}}}

\def\OL{\ensuremath{\Omega_{\Lambda}}}

\def\Ho{\ensuremath{H_0}}

\def\arcsec{\ensuremath{^{\prime\prime}}} 

\def\Hubble{{\it Hubble}}

\def\Spitzer{{\it Spitzer}}
\def\Chandra{{\it Chandra}}
\def\Herschel{{\it Herschel}}

\newenvironment{inlinefigure}{
\def\@captype{figure}
\noindent\begin{minipage}{0.999\linewidth}\begin{center}}
{\end{center}\end{minipage}\smallskip}

\newcommand{\insertfigurewide}[2] {
\ifsubmode
\else
\begin{figure*}
\begin{center}
\resizebox{\textwidth}{!}{\includegraphics{{#1}}}
\caption{{#2}}
\end{center}
\end{figure*}
\fi
}

\newcommand{\insertfigure}[2] {
\ifsubmode

\else
\begin{inlinefigure}
\begin{center}
\resizebox{\textwidth}{!}{\includegraphics{{#1}}}
\figcaption{{#2}}
\end{center}
\end{inlinefigure}
\fi
}

\newcommand{\insertfigend}[2] {
\newpage
\begin{inlinefigure}
\begin{center}
\resizebox{\textwidth}{!}{\includegraphics{{#1}}}
\figcaption{{#2}}
\end{center}
\end{inlinefigure}
}

\newcommand{\note}[1]{}

\ifsubmode
   \newcommand{\cmt}[1]{}

\else
   \newcommand{\cmt}[1]{\textcolor{red}{#1}}
\fi

\newcommand{\figmorph}{ \label{fig:morphology} 
Comparison of galaxy structural parameters derived from
Wide-quality 4/3-orbit images (synthesized from HUDF F160W) 
vs.~values obtained from the full 28-orbit data.
In order, the quantities
plotted are effective radius, Sersic index, axial ratio, Gini, and 
M$_{20}$.  The adopted useful limit for Wide
data is the vertical dotted line at $H = 24.0$\,mag;
Deep data will go 0.7\,mag fainter, which 
is $M_* \approx 2 \times 10^{10}$ M$_{\odot}$ at $z \approx 2.5$ (see Table 
\ref{tab:numbers_of_galaxies}).
}

\newcommand{\figmerger}{ \label{fig:merger} 
Top panels show simulated rest-frame $urz$ images of a
  gas-rich major merger.  The image was created using a hydrodynamic
  simulation of a binary galaxy merger including star formation, black
  hole accretion, and AGN feedback
  \protect\citep{2005Natur.433..604D}. The simulation results were
  post-processed through the dust and radiative transfer code Sunrise
  \protect\citep{2010MNRAS.403...17J}.  Successive merger stages are labeled
  (1--6). The middle and bottom panels show the predicted IR luminosity
  and rest-frame optical asymmetry vs.~merger stage for the same
  simulation. This picture predicts that the maximal asymmetry should
  be measured during the relatively early stages of the merger, while
  the object would be identified as an obscured or optical QSO in the
  late stages, when the two nuclei may have largely coalesced and
  faint tidal features may be the only remaining signs of
  morphological disturbance (see also \protect\citealt{2008ApJS..175..356H}).
}

\newcommand{\figmontage}{ \label{fig:smallmontage} 
Four-orbit images of HUDF galaxies
from ACS vs.~2-orbit 
images from WFC3/IR illustrate the 
importance of WFC3/IR for studying distant galaxy structure. WFC3/IR
unveils the true stellar mass distributions of these 
galaxies unbiased by young stars and obscuring dust.  The new structures
that emerge in many cases
inspire revised interpretations of these objects, as indicated.
}

\newcommand{\figlimits}{ \label{fig:limits} 
Limiting magnitudes of CANDELS photometry compared to
  existing photometry and to the SEDs of
  model blue and red galaxies.  \textit{HST} observations are shown in
  two-toned triangles.  Each triangle encodes three brightness limits.
  The bottoms of the light-colored inverted triangles are $5\sigma$
  point-source limits for Deep data.  The bottoms of the darker
  triangles show the approximate $5\sigma$ limits for aperture
  photometry of a fiducial Lyman-break galaxy with a half-light radius
  of $0.25''$, also in Deep data. The tops of the triangles show the
  approximate $5\sigma$ limits for LBGs in the CANDELS/Wide
  survey using the exposure times from Table \ref{tab:areadepth} (except for F814W,
  see below).  The green triangles at left show the GOODS/ACS data ($BViz$),
  the red triangles are for WFC3/IR ($YJH$).  The yellow triangles denote
  CANDELS F814W, with the triangle bottoms corresponding to the
  fiducial 28\,ks exposure in CANDELS/Deep and the top corresponding to a
  1-orbit exposure.  The points of the solid green triangles in the IR
  show point-source depths (valid for distant galaxies) for the GOODS
  VLT ISAAC $JHK$ observations \protect\citep{2010A&A...511A..50R} and the GOODS
  {\it Spitzer} IRAC data at 3.6$\micron$ and 4.5$\micron$
  \protect\citep{2003mglh.conf..324D}.  
      The blue SEDs are for $L^*$ Lyman-break galaxies at redshifts
      $z=3,5,$ and 7; the red SED shows a maximally old galaxy at
      $z=3$ with rest-frame $M_V = -21$\,mag.  For comparison,
      we also plot the $5\sigma$ sensitivities of the deepest existing
      \textit{HST} images (gray triangles), covering the much smaller 
      areas of the HUDF ($BViz$) and the HUDF09 ($YJH$) as reported by 
      \protect\citet{2011arXiv1105.2297O}.
}

\newcommand{\figlbglf}{ \label{fig:lbglf} 
  Collected data on luminosity functions (LFs) of distant LBGs
  at $z \approx 7$ (left) and $z \approx 8$ (right).  The colored
  regions show where data from each level of the three-tiered survey
  strategy is strongest.  The vertical right-hand edges of the shaded
  areas indicate the $5\sigma$ point-source limits from Table
  \ref{tab:numbers_of_galaxies}.  The horizontal edges indicate the
  level in the LF (black curves, from 
  \citealt{2010MNRAS.403..960M}) where the number of galaxies
  detected per magnitude bin in the survey is $\sim 10$.  In a 
  well-designed survey, the colored areas should overlap.  This requirement
  helps to define the areas of the Deep and Wide surveys, respectively.
}

\newcommand{\figebl}{ \label{fig:ebl} Extragalactic background fluctuations.
The green curve is a model \citep{2004ApJ...606..611C, 2007ApJ...671....1T}\note{(Cooray et al.
2004)} that provides the star formation necessary to reionize the
universe by $z \approx 7$--8.  The purple points are the \citet{2007ApJ...657..669T} measurement of
fluctuations in the NICMOS HUDF.  The red points with error bars illustrate the 
proposed CANDELS measurements, which are set to match the NICMOS measurements on 
small scales and the predicted linear fluctuations due to reionization sources
on scales of $10'$. Under these assumptions, the measurements
would yield a $7\sigma$ detection of the expected spatial clustering at 
scales larger than $1'$. The dashed line shows the ``shot noise''
Poisson fluctuations due to the finite number of faint galaxies. The solid
curve shows the linear clustering of halos at $z \approx 10$ projected on the sky.
}

\newcommand{\figlargemontage}{ \label{fig:largemontage} 
Clumpy $z \approx 2$ galaxies
from WFC3 Early Release Science (ERS) images taken in F160W (2-orbit depth)
versus F775W (3-orbit depth).  
Object IDs are from the FIREWORKS \citep{2008ApJ...682..985W} and
MUSYC \citep{2010ApJS..189..270C} catalogs.  Galaxies
are arranged in two groups.  On the left are
objects for which the two images
look quite similar, while objects on the right look substantially
different.  Cases like W4934,
W4973, and W5609 on the right appear to have underlying disks composed
of older stars and may be candidates for inherently
regular galaxies in which clump formation has occurred via
{\it in situ} disk instabilities.  Cases on the left like W4032,
W5201, and W5353 lack such underlying disks
and are more likely to be separate galaxies now
merging.  F160W also
highlights the presence and location of 
central potential wells, as in W4934, W5160, and W5229.
}

\newcommand{\figzphot}{ \label{fig:zphot}
Plotted is the difference between two sets of photometric redshifts
for galaxies 
in the ERS region.  One set uses the best available space-based and ground-based
photometry at optical and near-IR wavelengths.  The second set substitutes
new $YJH$ data from WFC3/IR. For brighter galaxies ($H<25$\.mag), the substitution 
makes little difference, but for fainter galaxies ($H>25$\,mag), photometric redshifts change
significantly with the newer (and presumably more accurate) WFC3/IR values.
}

\newcommand{\figuv}{
\label{fig:uv}
The composite LBG spectrum of 
\citet{2003ApJ...588...65S}\note{Shapley et al. 2003} is
shifted to $z=2.4$ (the spectrum at
$\lambda_{\rm rest}<912$\,\AA\ is an estimate).  Overplotted 
(with arbitrary scaling) are the
transmission curves of the UVIS filters
being used in the GOODS-N CVZ day-side observations.  At $z>2.38$,
the Lyman limit shifts redward of any significant transmission
in the F275W filter.  Therefore, this filter will probe the escaping
Lyman continuum radiation for galaxies at $2.38<z<2.55$ (where the
upper limit is dictated by IGM opacity).  Also demonstrated 
is the ability to efficiently select star-forming galaxies via a
decrement in the F275W flux due to Lyman continuum opacity from
H{\sc i} in both the IGM and the interstellar medium of the galaxy (see text for
details).
}

\newcommand{\figsndndz}{
\label{fig:sn_dndz}
Predicted redshift distributions of detected SNe~Ia assuming a 52
day separation between observational epochs and assuming that the
search was done entirely in each filter shown. The magnitude limit
assumed for detecting SNe corresponds to the $S/N=5$ limit reached
in one orbit for each filter. The curves are normalized so that the
integral under the distributions recovers the total number of SNe~Ia
expected in the combined CANDELS and CLASH programs for the 
mixed model of Figure \ref{fig:sn_rates}.
}

\newcommand{\figsnrates}{
\label{fig:sn_rates}
Evolution of the SN~Ia rate. The points represent observed
rates, where black dots are from 
\citet{2008A&A...479...49B}\note{Botticella et al. 2008, A&A, 479, 49} and
\citet{2010ApJ...713.1026D}\note{Dilday et al. 2010, ApJ, 713,
1026}, while red dots are from 
\citet{2008ApJ...681..462D}\note{Dahlen et al. 2008, ApJ, 681, 462}. 
The solid red curve shows a fit to the
data assuming a $\sim 3$\,Gyr delay between the formation of the
progenitor star and the explosion of the SN 
\citep{2010ApJ...713...32S}.\note{Strolger et al. 2010}
The blue line shows the prediction from a model where the rate is
dominated by a prompt component 
\citep{2005ApJ...629L..85S}.\note{Scannapieco \& Bildsten 2005}
The green line is based on a ``mixed'' model that gives similar weight to the
prompt and delayed components 
\citep{2007AIPC..924..373P}.\note{Panagia et al. 2007}
Finally, the
purple line and dots are also based on the Dahlen et al.~measured
data but assume that extinction due to dust in host galaxies
of SNe~Ia increases significantly at higher redshift, requiring
large corrections to the derived rates at $z>1.5$. Asterisks
show the predicted rates for the combined CANDELS and CLASH programs
assuming the four different rate scenarios.
}

\newcommand{\figgds}{
\label{fig:gds}
A simplified footprint of the CANDELS observations in the GOODS-S 
field with WFC3/IR.  The ``Wide'' portion of the
CANDELS observations (green) and the pre-existing observations
of the WFC3 ERS (brown) are reproduced
faithfully.  However the ``Deep'' portion of the
field, represented here as a $3\times5$ raster (blue), will
in practice be observed across several epochs with 
slightly varying geometries.  The individual epochs, as
well as the footprints of the ACS parallel observations, 
are illustrated in Figures \ref{fig:gdsepa} and \ref{fig:gdsepb}.
}

\newcommand{\figgdn}{
\label{fig:gdn}
A simplified footprint of the CANDELS observations in the GOODS-N 
field with WFC3/IR.  The Wide portion of the
CANDELS observations (green) is reproduced
faithfully.  However the Deep portion of the
field, represented here as a $3\times5$ raster (blue), will
in reality will be observed across several epochs with 
slightly varying geometries.  At the time of writing, the
precise layout of each Deep epoch had not yet been 
determined.  The ensemble will qualitatively resemble
the progression in Figures  \ref{fig:gdsepa} and 
\ref{fig:gdsepb}.
}

\newcommand{\figudslayout}{
\label{fig:uds_layout}
Footprint of the CANDELS observations in the UDS field.  The WFC3/IR
prime exposures are shown in blue and the ACS/WFC parallel exposures shown in
magenta.}

\newcommand{\figcosmoslayout}{
\label{fig:cosmos_layout}
Footprint of the CANDELS observations in the COSMOS field
with WFC3/IR prime exposures shown in blue and ACS/WFC parallel exposures
shown in magenta.}

\newcommand{\figegslayout}{
\label{fig:egs_layout}
Footprint of the CANDELS observations in the EGS field
with WFC3/IR prime exposures shown in blue and ACS/WFC parallel exposures
shown in magenta. Just over half of the area was observed in \textit{HST} Cycle 
18, while the remainder will be observed in \textit{HST} Cycle 20.  The two halves 
are shown independently in Figure \ref{fig:egs_combo}.}

\newcommand{\figegscombo}{
\label{fig:egs_combo}
Footprints of the two CANDELS campaigns in the EGS 
 field.  Slightly over half the total area was observed in \textit{HST} Cycle 
 18 (\textit{left}), while the remainder will be observed in \textit{HST} Cycle 20 
(\textit{right}).  The WFC3/IR prime exposures are shown in blue, 
and the ACS/WFC parallel exposures are shown in magenta. 
}

\newcommand{\figgdsepa}{
\label{fig:gdsepa}
Footprints of the first eight epochs of 
CANDELS observations in the GOODS-S field with WFC3/IR and ACS/WFC.
The blue and orange outlines are individual WFC3/IR pointings in the $J/H$ and
$Y$ filters, respectively.  The corresponding ACS/WFC parallels shown in
magenta, red, or green according to the ACS filter selection.  The black box
is the fiducial CANDELS/Deep region, to guide the eye.  
At the top of each epoch is shown the pre-existing  
exposure map from the WFC3 ERS 
observations in this field (purple).  These ERS observations 
are intermediate in depth between the CANDELS Deep and Wide 
components. 
}

\newcommand{\figgdsepb}{
\label{fig:gdsepb}
Footprints of the last seven epochs of 
CANDELS observations in the GOODS-S field with WFC3/IR and ACS/WFC.
See Figure \ref{fig:gdsepa} for a description.
}

\newcommand{\figudsexpmaps}{
Exposure maps for the CANDELS UDS region.  
The ACS exposure maps (F606W and F814W) are
on a linear stretch from 0 to 12\,ks.  The WFC3 exposure
maps (F350LP, F125W, and F160W) are on a linear stretch 
from 0 to 4\,ks.  The dotted red line indicates the bounds of
the F160W coverage.\label{fig:udsexpmaps}
}

\newcommand{\figcosmosexpmaps}{
Exposure maps for the CANDELS COSMOS region, including legacy
ACS/WFC exposures.  The ACS exposure maps (F606W and F814W) are
on a linear stretch from 0 to 12\,ks.  The WFC3 exposure
maps (F350LP, F125W, and F160W) are on a linear stretch 
from 0 to 4\,ks.  The dotted red line indicates the bounds of
the F160W coverage.\label{fig:cosexpmaps}
}

\newcommand{\figegsexpmaps}{
Exposure maps for the CANDELS EGS region, including legacy
ACS/WFC exposures.  The ACS exposure maps (F606W and F814W) are
on a linear stretch from 0 to 12\,ks.  The WFC3 exposure
maps (F350LP, F125W, and F160W) are on a linear stretch 
from 0 to 4\,ks.  The dotted red line indicates the bounds of
the F160W coverage.\label{fig:egsexpmaps}
}

\newcommand{\figwidead}{
Cumulative area vs.~depth profiles of the CANDELS/Wide fields, 
including legacy ACS/WFC exposures.  The areas are limited to the 
footprints of the CANDELS F160W images in each field.
\label{fig:widead}
}

\newcommand{\figgdsloexpmaps}{
Exposure maps for the bluer filters of the CANDELS 
GOODS-S region, including legacy ACS and WFC3 exposures.  
All exposure maps are plotted with 
a logarithmic stretch from 0.4 to 200\,ks.  The small extreme-depth 
portions of the exposure maps are the HUDF and its parallel fields.
The dotted red line indicates the fiducial GOODS ACS boundary.
\label{fig:gdsloexpmaps}
}

\newcommand{\figgdshiexpmaps}{
Exposure maps for the redder filters of the CANDELS 
GOODS-S region, including legacy ACS and WFC3 exposures.  
All exposure maps are plotted with 
a logarithmic stretch from 0.4 to 200\,ks.  The small extreme-depth 
portions of the exposure maps are the HUDF and its parallel fields.
The dotted red line indicates the fiducial GOODS ACS boundary.
\label{fig:gdshiexpmaps}
}

\newcommand{\figgdsad}{
Cumulative area vs.~depth profiles of the CANDELS 
GOODS-S field, including legacy ACS and WFC3 exposures.  
The areas are limited to the footprint covered by the union of
ERS and CANDELS images.
\label{fig:gdsad}
}

\newcommand{\figgdnloexpmaps}{
Exposure maps for the bluer filters of the CANDELS 
GOODS-N region, including legacy ACS exposures.  
All exposure maps are plotted with 
a logarithmic stretch from 0.4 to 80\,ks.  The F606W will receive
additional exposures that are currently shown as F814W (Fig.~\ref{fig:gdnhiexpmaps}).
The dotted red line indicates the fiducial GOODS ACS boundary.
\label{fig:gdnloexpmaps}
}

\newcommand{\figgdnhiexpmaps}{
Exposure maps for the redder filters of the CANDELS 
GOODS-N region, including legacy ACS exposures.  
All exposure maps are plotted with 
a logarithmic stretch from 0.4 to 80\,ks.  Virtually all of the 
F814W exposures shown here are from the CANDELS program. A small 
portion of these will instead be taken with the
F606W and F850LP filters, when our GOODS-N filter assignments
are finalized.
The dotted red line indicates the fiducial GOODS ACS boundary.
\label{fig:gdnhiexpmaps}
}

\newcommand{\figgdnad}{
Cumulative area vs.~depth profiles of the CANDELS 
GOODS-N field, including legacy ACS exposures.  
The areas are limited to the footprint of the CANDELS images.
The curves for F606W and F850LP will be slightly augmented at the 
expense of F814W, once our GOODS-N filter assignments have 
been finalized.
\label{fig:gdnad}
}

\newcommand{\figepochs}{
Timetable of CANDELS epochs.  The individual fields are
color coded as follows: GOODS-S (cyan), UDS (orange), EGS (blue),
COSMOS (yellow), and GOODS-N (magenta). 
See text for 
a full description of the columns.\label{fig:epochs}
}

\shorttitle{CANDELS Multi-Cycle Treasury Program}
\shortauthors{Grogin et al.}

\begin{document}

\title{CANDELS: The Cosmic Assembly Near-infrared Deep Extragalactic Legacy Survey}

\author{
Norman A.~Grogin\altaffilmark{1},
Dale D.~Kocevski\altaffilmark{2},
S.~M.~Faber\altaffilmark{2},
Henry C.~Ferguson\altaffilmark{1},
Anton M.~Koekemoer\altaffilmark{1},
Adam G.~Riess\altaffilmark{3},
Viviana Acquaviva\altaffilmark{4},
David M.~Alexander\altaffilmark{5},
Omar Almaini\altaffilmark{6},
Matthew L.~N.~Ashby\altaffilmark{7}
% ...
}

\altaffiltext{1}{Space Telescope Science Institute}

\altaffiltext{2}{UCO/Lick Observatory, University of California, Santa Cruz}

\altaffiltext{3}{The Johns Hopkins University}

\altaffiltext{4}{Rutgers, The State University of New Jersey}

\altaffiltext{5}{Durham University}

\altaffiltext{6}{University of Nottingham}

\altaffiltext{7}{Harvard-Smithsonian Center for Astrophysics}

% ...

\begin{abstract}
The Cosmic Assembly Near-infrared Deep Extragalactic Legacy Survey (CANDELS) is
designed to document the first third of galactic evolution, over the
approximate redshift ($z$) range 8--1.5.
It will image $>$250,000 distant galaxies using three separate cameras
on the \Hubble\ \textit{Space Telescope}, from the mid-ultraviolet to 
the near-infrared, and will find and measure Type Ia supernovae at $z>1.5$ to
test their accuracy as
standardizable candles for cosmology. Five premier multi-wavelength sky regions are
selected, each with extensive ancillary data.
The use of five widely
separated fields mitigates cosmic variance and yields statistically robust and
complete samples of galaxies down to 
a stellar mass of $10^9 {\rm M}_\odot$ to $z \approx 2$, reaching
the knee of the ultraviolet luminosity function (UVLF) of galaxies to $z \approx 8$.
The survey covers approximately 800 arcmin$^2$ and is
divided into two parts. The CANDELS/Deep survey 
($5\sigma$ point-source limit $H=27.7\,$mag)
covers $\sim 125$ arcmin$^2$ within GOODS-N and GOODS-S\@. 
The CANDELS/Wide survey includes GOODS and three additional 
fields (EGS, COSMOS, and UDS) and covers the full area to 
a $5\sigma$ point-source limit of $H \gtrsim 27.0$\,mag.
Together with the \Hubble\ Ultra Deep Fields, the 
strategy creates a three-tiered ``wedding cake'' approach 
that has proven efficient for extragalactic surveys.
Data from the survey are nonproprietary and are
useful for a wide variety of science investigations. In this
paper, we describe the basic motivations for the survey, the
CANDELS team science goals and the resulting observational requirements,
the field selection
and geometry, and the observing design.
The \Hubble\ data processing and products are described in a
companion paper.
\end{abstract}

\keywords{ 
Cosmology: observations ---
Galaxies: high-redshifts --- 
}

%
%
%

%

\section {Introduction}\label{sec:intro}

During the past decade, the {\it Hubble Space Telescope} ({\it HST}) and other telescopes have fueled a series
of remarkable discoveries in cosmology that would have seemed impossible
only a few short years ago.  Galaxies are now routinely found when the
Universe was only 5\% of its current age and before 99\% of present-day
stars had formed.  Distant Type Ia supernovae (SNe~Ia) showed
that the expansion of the Universe was decelerating for the first 
$\sim 9$\,Gyr, followed by acceleration due to a mysterious ``dark
energy'' whose nature remains completely unknown.  The troika of \Hubble,
\Spitzer, and \Chandra\ revealed a complex interplay between galaxy
mergers, star formation, and black holes over cosmic time that spawned
the new concept of galaxy/black hole ``co-evolution.''

This rapid progress can be attributed in part to an unprecedented degree
of coordination between observatories across the electromagnetic spectrum
in which a few small regions of sky were observed as deeply as possible
at all accessible wavelengths.  Such regions have become ``magnet''
regions whose total scientific value now far exceeds
that of their individual surveys.

The Great Observatories Origins Deep Survey
\citep[GOODS;][]{2004ApJ...600L..93G} possesses the deepest data on
the sky from virtually every telescope:  \Hubble, \Spitzer, \Chandra,
\Herschel, the VLA, and many other observatories both in space and on
the ground. However, the GOODS-North and GOODS-South fields together
subtend only about 300 arcmin$^2$, which makes them too small for science
goals involving rare and/or massive objects. Samples are limited,
and count fluctuations tend to be large owing to the high intrinsic
bias and cosmic variance of massive halos.  GOODS \textit{HST} images
furthermore probe only optical wavelengths, which are strongly biased
toward ongoing star-forming regions and miss the older stars beyond
redshift $z \approx 1.3$.

With a survey speed gain of a factor of $\sim 30$ over NICMOS for galaxy
imaging, the new WFC3/IR (infrared) camera enables much more ambitious near-IR
surveys than were previously possible with \Hubble.  For example,
the single-filter GOODS NICMOS Survey \citep[GNS;][]{2011MNRAS.413...80C}
required 180 \textit{HST} orbits to survey 45 arcmin$^2$ to a limiting
magnitude of $H_{\rm AB}\approx 26.5$.  By comparison, WFC3/IR
enables a two-filter survey of five times the GNS area to limiting
magnitudes of $H_{\rm AB} \gtrsim 27.1$ and $J_{\rm AB} \gtrsim 27.0$
--- using \textit{half} the orbits required for the GNS\@.  The
capacity for longer-wavelength and larger-area surveys now allows \Hubble\
to: follow galaxies well into the reionization era; measure spectra and
light curves for SNe~Ia in the deceleration era; and measure rest-frame
optical shapes and sizes of galaxies at a time ($z \approx 2$) when 
cosmic luminosity peaked for both star formation and active galactic nuclei (AGNs), 
and when the Hubble sequence was starting to take shape.
The power of WFC3/IR imaging for distant galaxies is demonstrated in
Figure \ref{fig:smallmontage}, which compares GOODS-depth 4-orbit F775W
images from the \Hubble\ Advanced Camera for Surveys (ACS) with 2-orbit
images from WFC3/IR\@.  Regions with red colors due to heavy dust or old
stars leap out with the WFC3/IR, in many cases leading to a new interpretation
of the object.

\ifsubmode
\placefigure{fig:smallmontage}
\else
  \begin{figure*}[!ht]
  \plotone{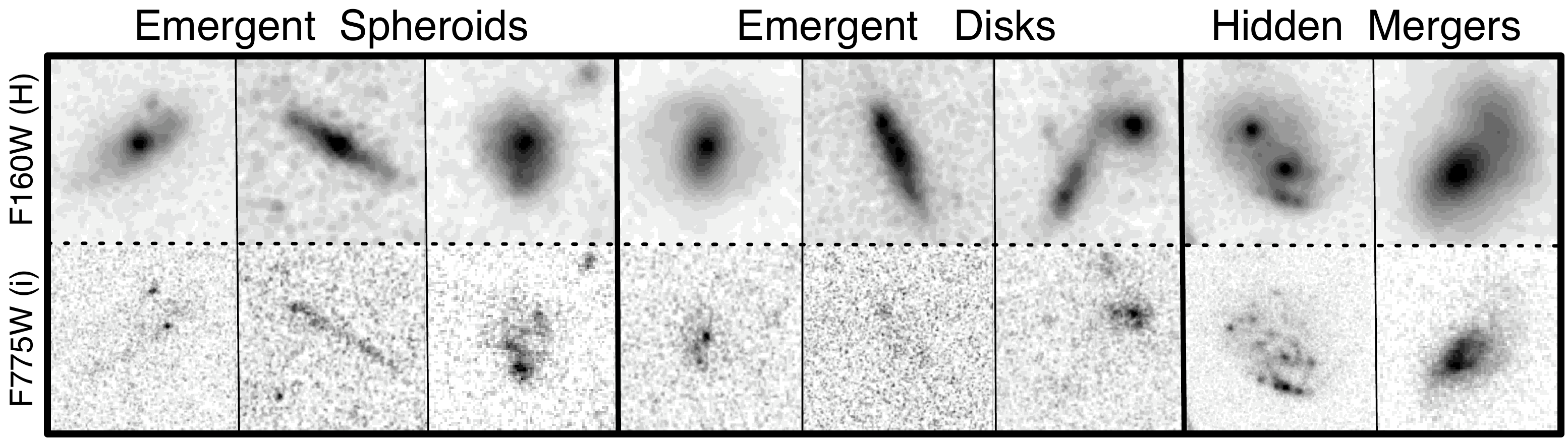}\caption{\figmontage}
  \end{figure*}
\fi

Given the power of this gain, there is strong motivation to extend deep-field
WFC3/IR imaging beyond the GOODS regions to larger areas.  
Three well-studied regions in the sky are natural candidates for this
extension: COSMOS \citep{2007ApJS..172....1S}, the Extended Groth
Strip \citep[EGS;][]{2007ApJ...660L...1D}, and the UKIDSS Ultra-deep
Survey field \citep[UDS;][]{2007MNRAS.379.1599L,2007MNRAS.380..585C}.
These fields are larger than GOODS and already have high-quality (though
generally shallower) multi-wavelength data.  Establishing multiple,
statistically independent WFC3/IR regions over the sky would also minimize
cosmic variance and facilitate follow-up observations by ground-based telescopes.

Several teams responded to the \Hubble\ Multi-Cycle Treasury Program
Call for Proposals with programs targeting high-latitude fields using WFC3/IR
to study galaxy evolution.  Two teams targeted the above five fields and
also proposed to find high-redshift SNe~Ia to improve constraints on
cosmic deceleration and acceleration.  The program led by Ferguson proposed to observe
the full 300 arcmin$^2$ of the GOODS fields to uniform depth in
$YJH$. This proposal contained time for both spectroscopic and photometric
SN~Ia follow-up observations and took advantage of GOODS-N in the \Hubble\ Continuous
Viewing Zone (CVZ) to obtain far-ultraviolet (UV) images on the day side
of the orbit when the sky is too bright for broadband IR imaging. The
second program, led by Faber, proposed imaging half the area of the
two GOODS fields to about twice the depth of the Ferguson program and
also added shallower imaging over $\sim 1000$ arcmin$^2$ in EGS, COSMOS,
UDS, and the Extended \Chandra\ Deep Field South.  ACS parallels were included to broaden total wavelength
coverage, deepen existing ACS mosaics, and add a new ACS mosaic in UDS,
where none existed.  SN~Ia searches were also included, but the proposal
did not contain time for SN follow-up observations, nor did it feature 
UV imaging.

The \Hubble\ time-allocation committee (TAC) saw merit in both proposals
and charged the two teams to craft a joint program to retain the
best features of both programs yet fit within 902 orbits.  This was
challenging, owing to four requirements mandated by the TAC: (1) visit
all WFC3/IR tiles at least twice with the proper cadence for finding SNe~Ia
($\sim 60$ days; this severely restricts the range of allowable dates
and ORIENTs in each field), (2) discriminate SNe~Ia candidates from other
interlopers (requires very specific multi-wavelength data at each visit),
(3) put as many ACS parallels as possible on top of each WFC3/IR tile
(further restricting \Hubble\ observation dates and ORIENTs), and (4)
maximize the overlap of \Hubble\ data on top of existing ancillary
data (compatible dates and ORIENTs become vanishingly small).
Further complicating matters, the \Hubble\ TAC also approved the 
CLASH program on clusters of galaxies by Postman et al.~(GO 12065), 
including SN discovery and follow-up observations, with the 
mandate that it be coordinated with the SN~Ia program here.  
The SN portions of both proposals were
consolidated under a separate program by Riess et al.~(GO 12099), and 
the SN~Ia follow-up orbits from both programs were pooled.  Our program 
takes prime responsibility for the highest-redshift SNe ($z > 1.3$),
while CLASH addresses SNe at lower redshifts.

The resulting observing program, now entitled the
Cosmic Assembly Near-infrared Extragalactic Legacy Survey (CANDELS), 
targets five distinct fields (GOODS-N, GOODS-S, EGS, UDS, and COSMOS) 
at two distinct depths. Henceforth, we will refer to the deep portion
of the survey as ``CANDELS/Deep'' and the shallow portion as ``CANDELS/Wide.''
Adding in the \Hubble\ Ultra Deep Fields (HUDF) makes a three-tiered 
``wedding cake'' approach, which has proven to be
very effective with extragalactic surveys.
CANDELS/Wide has exposures in all five CANDELS fields, while 
CANDELS/Deep is only in GOODS-S and GOODS-N.  
 
The outline of this paper is as follows.  We first provide
a brief synopsis of the survey in \S\ref{sec:synopsis}.
We follow in \S\ref{sec:science_goals} with a detailed
description of the major science goals along with their
corresponding observational requirements that CANDELS addresses.
We synthesize the combined observing requirements 
in \S\ref{sec:requirements} with regard to facets of
our survey.  A description of the particular survey fields 
and an overview of existing ancillary data are 
provided in \S\ref{sec:fields}.  Section \ref{sec:observing_strategy}
describes the detailed observing plan, including the
schedule of observations. 
Section \ref{sec:discussion} summarizes the paper,
along with a brief description of the CANDELS 
data reduction and data products; a much more complete
description is given by Koekemoer et al.~(2011), which is 
intended to be read as a companion paper to this one.

Where needed, we adopt the following cosmological parameters: 
$\Ho = 70$\,km\,s$^{-1}$\,Mpc$^{-1}$;
$\Ot,\OL,\Om = 1, 0.3, 0.7$ (respectively), though numbers used in individual
calculations may differ slightly from these values.  All
magnitudes are expressed in the AB system (Oke \& Gunn 1983).

\section{CANDELS Synopsis}\label{sec:synopsis}

Table \ref{tab:at_a_glance} provides a convenient summary
of the survey, listing 
the various filters and corresponding total exposure within each field,
along with each field's coordinates and dimensions.
The \Hubble\ data are of several different types, 
including images from WFC3/IR and WFC3/UVIS 
(both UV and optical) plus extensive ACS parallel exposures.
Extra grism and direct images will also be included for SN~Ia follow-up
observations (see \S\ref{sec:sne}), but
their exposure lengths and locations are not pre-planned.  They are 
not included in Table \ref{tab:at_a_glance}.
In perusing the table, it may be useful to look ahead
at Figures \ref{fig:gds}--\ref{fig:egs_layout}, which
illustrate the layout of exposures on the sky.

%
\ifsubmode 
  \begin{deluxetable}{lllllll}
  \tabletypesize{\scriptsize}
\else
  \begin{deluxetable*}{lllllll}
\fi
\tablecaption{CANDELS at a glance \label{tab:at_a_glance}}
\tablehead{%
Field	& Coordinates			& Tier	& WFC3/IR Tiling	& HST Orbits/Tile	& 
IR Filters\tablenotemark{a}	& UV/Optical Filters\tablenotemark{b}}
\startdata
GOODS-N	& 189.228621, $+$62.238572	& Deep	& $\sim 3 \times 5$		& $\sim$13		& YJH		& UV,UI(WVz)		\\
GOODS-N	& 189.228621, $+$62.238572	& Wide  & 2 @ $\sim 2 \times 4$		& $\sim$3		& YJH		& Iz(W)		\\
GOODS-S	& 53.122751, $-$27.805089	& Deep  & $\sim 3 \times 5$		& $\sim$13		& YJH		& I(WVz)		\\
GOODS-S	& 53.122751, $-$27.805089	& Wide  & $\sim 2 \times 4$		& $\sim$3		& YJH		& Iz(W)		\\
COSMOS	& 150.116321, $+$2.2009731	& Wide  &  $4 \times 11$			& $\sim$2		& JH		& VI(W)		\\
EGS	& 214.825000, $+$52.825000	& Wide  &  $3 \times 15$			& $\sim$2		& JH		& VI(W)		\\
UDS	& 34.406250,  $-$5.2000000	& Wide  &  $4 \times 11$			& $\sim$2		& JH		& VI(W)
\enddata
\tablenotetext{a}{WFC3/IR filters Y $\equiv$ F105W, J $\equiv$ F125W, H $\equiv$ F160W\@.}
\tablenotetext{b}{WFC3/UVIS filters UV $\equiv$ F275W, W $\equiv$ F350LP; ACS filters V $\equiv$ F606W, I $\equiv$ F814W,
z $\equiv$ F850LP\@.  Parenthesized filters indicate incomplete and/or relatively shallow coverage of the indicated field.}
\ifsubmode 
  \end{deluxetable}
\else
  \end{deluxetable*}
\fi
\clearpage

The three
purely Wide fields (UDS, COSMOS, and EGS, 
Figs.~\ref{fig:uds_layout}--\ref{fig:egs_layout}) consist of a 
contiguous
mosaic of overlapping WFC3/IR tiles (shown in blue) along with
a contemporaneous mosaic of ACS parallel exposures (shown in magenta).   
The two cameras are offset by $\sim 6'$, but overlap
between them
is maximized by choosing the appropriate telescope roll angle. 
The Wide exposures are taken over the course of two \textit{HST}
orbits, with exposure time allocated roughly 2:1 between
F160W and F125W\@.  The observations are scheduled in two visits separated  
by $\sim 52$ days in order to find SNe~Ia.  The stacked exposure
time is effectively twice as long in ACS (i.e., 4 orbits)
on account of its larger field of
view, and its time is divided roughly 2:1 between F814W
and F606W\@.  In the small region where the WFC3/IR lacks ACS parallel 
overlap, we sacrifice WFC3/IR depth to obtain a short exposure in the 
WFC3/UVIS ``white-light'' filter F350LP for SN type discrimination.

The GOODS fields contain both Deep and Wide exposure regions.
In addition, GOODS-S also contains the ERS data  
\citep{2011ApJS..193...27W}, which we have taken into account 
in our planning and include in the CANDELS quoted areas
because its filters and exposure times match well to
ours (see \S\ref{sec:deepstrategy}).  The two GOODS layouts are shown
in Figures \ref{fig:gds} and \ref{fig:gdn}.  The Deep
portion of each one is a central region of approximately $3 \times 5$
WFC3/IR tiles, which is observed to an effective depth of
3 orbits in F105W and 4 orbits in F125W and F160W\@.  To the
north and south lie roughly rectangular ``flanking fields''
each covered by 8--9 WFC3/IR tiles, which are observed using the
Wide strategy of 2 orbits in $J$+$H$.  The flanking fields additionally
receive an orbit of F105W\@.  The net result is coverage over most of
the GOODS fields to at least $\sim 1$-orbit depth in $YJH$, plus 
deeper coverage in all three filters within the Deep areas. 

Executing the GOODS exposures requires $\sim 15$ visits to each
field, and virtually all $J$+$H$ orbits are employed in SN~Ia
searching.   The filter layouts
at each visit are shown for GOODS-S in Figures \ref{fig:gdsepa}
and \ref{fig:gdsepb} (detailed visits for GOODS-N have
not yet been finalized).  Because the telescope roll
angle cannot be held constant across so many visits, the matching ACS parallel
exposures (which are always taken) are distributed around the
region in a complicated way.  These parallels are taken in F606W, F814W,
and F850LP according to a complex scheme explained in
\S\ref{sec:deepstrategy}.  For now, it is sufficient to note
that most of the ACS parallels use F814W, for the purpose of
identifying high-$z$ Lyman-break ``drop-out'' galaxies.  Because
there is poor overlap between WFC3/IR exposures and their ACS parallels
during any given epoch, \textit{all} GOODS $J$+$H$ orbits include a short
WFC3/UVIS F350LP exposure as noted above for CANDELS/Wide.
Finally, Table \ref{tab:at_a_glance} also lists
special WFC3/UVIS exposures taken during GOODS-N CVZ 
opportunities, of total duration $\sim 13$\,ks in F275W and
$\sim 7$\,ks in F336W\@.  
Exposure maps of the expected final data
in GOODS-S are shown in Figures \ref{fig:gdsloexpmaps}
and \ref{fig:gdshiexpmaps},
including all previous legacy exposures in 
\textit{HST} broadband filters (the GOODS-N maps 
in Figs.~\ref{fig:gdnloexpmaps}
and \ref{fig:gdnhiexpmaps} are preliminary).  

Realizing the full science potential of this extensive but complex
dataset will require closely interfacing with many other 
ground-based and space-based surveys.  Among these we particularly mention 
SEDS\footnote{http://www.cfa.harvard.edu/SEDS}, the \textit{Spitzer}
Warm Mission Extended Deep Survey,
whose deep 3.6\,$\micron$ and 4.8\,$\micron$ data points 
provide vital stellar masses 
(the CANDELS fields are completely embedded within the
SEDS regions).
The total database is rich, far richer than our team can
exploit.  Because of this, and the Treasury nature of the
CANDELS program, we are moving speedily to process and make the
\Hubble\ data public (for details, see Koekemoer et al.~2011).  
The first CANDELS data release occurred 
on 12 January 2011, 60 days after the first epoch was acquired 
in GOODS-S\@.  As a further service to the community, we are 
constructing separate websites for each CANDELS field to collect
and serve the ancillary data.  For further information,
please visit the CANDELS website\footnote{http://candels.ucolick.org}.

\section {Science Goals}\label{sec:science_goals}

The Multi-Cycle Treasury (MCT) Program was established to 
address high-impact science questions that
require \Hubble\ observations on a scale that cannot be accommodated
within the standard time-allocation process. MCT programs 
are also intended to seed a wide variety of compelling scientific 
investigations. Deep WFC3/IR observations of well-studied fields 
at high Galactic latitudes naturally meet these 
two criteria.

In this section we outline the CANDELS science goals, prefixed by
a brief discussion in \S\ref{sec:theory_support}
of the theoretical tools that are being developed for
the CANDELS program.  Most of our investigations of galaxies and AGNs
divide naturally into two epochs.  
In \S\ref{sec:cosmic_dawn} (``cosmic dawn''), we
discuss studies of very early galaxies during the reionization era. In
\S\ref{sec:high_noon} (``cosmic high noon''), we discuss the growth
and transformation of galaxies during the era of peak star formation
and AGN activity. Section~\ref{sec:uv} describes science goals enabled by UV
observations that exploit the GOODS-N CVZ\@. 
Section~\ref{sec:sne} describes the use of high-$z$ SNe to
constrain the dynamics of dark energy, measure the evolution of
SN rates, and test whether SNe~Ia remain viable as standardizable
candles at early epochs. Finally, \S\ref{sec:grism} describes science
goals enabled by the grism portion of the program.  The complete list
of goals is collected for reference in Table~\ref{tab:science_goals}.
Work is proceeding within the team on all of these topics.





\ifsubmode
  \begin{deluxetable}{l p{6in}}
  \tabletypesize{\scriptsize}
\else
  \begin{deluxetable*}{l p{6in}}
\fi






\tablecaption{CANDELS Primary Scientific Goals\label{tab:science_goals}}


\tablehead{\colhead{No.} &  \colhead{Goal} } 

\startdata

\multicolumn{2}{c}{{\bf Cosmic Dawn} (CD): Formation and early evolution of galaxies and AGNs}\\

CD1 & Improve constraints on the bright end of the galaxy LF at $z \approx 7$ and 8 and make $z \approx 6$ measurements more
robust. Combine with WFC3/IR data on fainter magnitudes to constrain the UV luminosity
density of the Universe at the end of the reionization era. \\

CD2 & Constrain star-formation rates, ages, metallicities, 
stellar masses, and dust contents of galaxies at the end of the 
reionization era, $z \approx 6$--10. 
Tighten estimates of the evolution of stellar mass, 
dust and metallicity at $z = 4$--8 by combining WFC3 data with 
very deep \textit{Spitzer} IRAC photometry. \\

CD3 & Measure fluctuations in the near-IR background light, at sensitivities sufficiently faint and angular scales sufficiently large to constrain reionization models. \\

CD4 & Use clustering statistics to estimate the dark-halo masses of high-redshift galaxies with double the area of prior \Hubble\ surveys. \\

CD5 & Search deep WFC3/IR images for AGN dropout candidates at $z > 6$--7 and constrain the AGN LF\@. \\

\noalign{\vspace{5pt}}

\multicolumn{2}{c}{{\bf Cosmic High Noon} (CN): The peak of star formation and AGN activity}\\

CN1 & Conduct a mass-limited census of galaxies down to $M_* = 2\times10^9$ M$_{\odot}$ at $z \approx 2$ and determine redshifts, star-formation rates, and stellar masses from broadband spectral-energy distributions (SEDs).  Quantify patterns of star formation versus stellar mass and other variables and measure the cosmic-integrated stellar mass and star-formation rates to high accuracy. \\

CN2 &Obtain rest-frame optical morphologies and structural parameters of $z \approx 2$ galaxies, 
including morphological types, radii, stellar mass surface densities, and
quantitative disk, spheroid, and interaction measures.  
Use these to address the relationship between galactic structure, 
star-formation history, and mass assembly. \\ 

CN3 & Detect galaxy sub-structures and measure their stellar masses. Use these data to assess disk instabilities, quantify internal patterns of star formation, and test bulge formation by clump migration to the centers of galaxies. \\

CN4 & Conduct the deepest and most unbiased census yet of active galaxies at $z \gta 2$ selected by X-ray, IR, optical spectra, and optical/NIR variability.  Test models for the co-evolution of black holes and galaxies and triggering mechanisms using demographic data on host properties, including morphology and interaction fraction. \\


\noalign{\vspace{5pt}}

\multicolumn{2}{c}{{\bf Ultraviolet Observations} (UV): Hot stars at $ 1 < z < 3.5 $}\\

UV1 & Constrain the Lyman-continuum escape fraction for galaxies at $z \approx 2.5$.  \\

UV2 & Identify Lyman-break galaxies at $z \approx 2.5$ and compare their properties to higher-$z$ Lyman-break galaxy samples. \\

UV3 & Estimate the star-formation rate in dwarf galaxies to $z > 1$ to test whether dwarf galaxies are ``turning on'' as the UV background declines at low redshift. \\

\noalign{\vspace{5pt}}

\multicolumn{2}{c}{{\bf Supernovae} (SN): Standardizable candles beyond $ z \approx 1 $}\\

SN1 & Test for the evolution of SNe~Ia as distance indicators by observing them at $z > 1.5$, where the effects of dark energy are expected to be insignificant but the effects of the evolution of the SN~Ia white-dwarf progenitor masses ought to be significant. \\

SN2 & Refine constraints on the time variation of the cosmic equation-of-state parameter, on a path to more than doubling the strength of this crucial test of a cosmological constant by the end of {\it HST}'s life. \\

SN3 & Measure the SN~Ia rate at $z \approx 2$ to constrain progenitor models by detecting the offset between the peak of the cosmic star-formation rate and the peak of the cosmic rate of SNe~Ia. \\

\enddata




\ifsubmode
\end{deluxetable}
\else
\end{deluxetable*}
\fi
\clearpage

\subsection{Theory Support}\label{sec:theory_support}

Theoretical predictions have been an integral
part of the project's development since its inception. We have
extracted merger trees from the new Bolshoi N-body simulation
\citep{2010arXiv1002.3660K}, which also track the evolution of sub-halos. We then
extract lightcones that mimic the geometry of the CANDELS fields, with
many realizations of each field in order to study cosmic
variance. These lightcones are populated with galaxies using several
methods: (1) Sub-Halo Abundance Matching (SHAM), a variant of Halo
Occupation Distribution modeling, in which stellar masses or
luminosities are assigned to (sub-)halos such that the observed galaxy
abundance is reproduced \citep{2010ApJ...717..379B}; (2) semi-analytic models
(SAMs), which use simplified recipes to track the main physical
processes of galaxy formation. We are using three different,
independently developed SAM codes, based on updated versions of the
models developed by \citet{2008MNRAS.391..481S}, 
\citet{2006MNRAS.365...11C}, and \citet{2010arXiv1004.2518L}, 
that are being run in the same Bolshoi-based merger
trees. This will allow us to explore the impact of different model
assumptions on galaxy observables. 

All three SAM models include treatments
of radiative cooling of gas, star formation, stellar feedback, and
stellar population synthesis. The Somerville and Croton SAMs also
include modeling of black hole formation and growth, and so can track
AGN activity. We are developing more detailed and accurate modeling of
the radial sizes of disks and spheroids in the SAMs, using an approach
based on the work of \citet{2009MNRAS.396..141D} in the case of the former, and
using the recipe based on mass ratio, orbit parameters, and gas
content taken from merger simulations by \citet{2011MNRAS.415.3135C} for the
latter. Using simple analytic prescriptions for dust extinction, we
will use the SAMs to create synthetic images based on these mock
catalogs, assuming smooth parameterized light profiles for the
galaxies. A set of mock catalogs, containing physical properties such
as stellar mass and star-formation rate (SFR), as well as observables such
as luminosities in all CANDELS bands, will be released to the public
through a queryable database\footnote{See, for example, Darren Croton's
website at http://web.me.com/darrencroton/Homepage/Downloads.html.}. The
synthetic images will also be made publicly available.

In addition, several team members are pursuing N-body and hydrodynamic
simulations using a variety of approaches.  To complement Bolshoi,
Piero Madau is computing Silver River, a higher-resolution version of
his previous N-body {\it Via Lactea} Milky Way simulation.  Romeel
Dav\'e is using a proprietary version of Gadget-2 to track gas infall,
star formation, and stellar feedback in very 
high-redshift galaxies \citep{2011MNRAS.410.1703F}.  
Working with multiple teams, Avishai Dekel is
guiding the computation of early disky galaxies at $z \approx 2$ with
particular reference to clump formation; early results were presented
in \citet{2010MNRAS.404.2151C}.  The simulation efforts make use of a wide
range of codes and numerical techniques, including ART, ENZO, RAMSES,
GADGET, and GASOLINE\@.  Finally, the theory effort includes 
postprocessing of hydrodynamic simulations with the {\it SUNRISE}
radiative transfer code to produce realistic images and spectra,
including the effects of absorption and scattering by dust
\citep{2010MNRAS.403...17J}.  Our goal is to create a library of $z
\ga 2$ galaxy images that can be used to help interpret the CANDELS
observations. These images will also be released to the public.

\subsection{Galaxies and AGNs: Cosmic Dawn}\label{sec:cosmic_dawn}

The science aims of CANDELS for redshifts $z \gta 3$ are collected together
under the rubric ``cosmic dawn.'' 
Within a few months of installation, WFC3 proved its power for
galaxy studies in this era by uncovering a wealth of new objects at $z > 6$
(e.g., \citealt{2010ApJ...709L.133B, 2010MNRAS.403..960M, 2010RAA....10..867Y}). 
WFC3/IR can potentially detect objects as distant as $z \gtrsim 10$, for which all 
bands shortward  
of F160W drop out (\citealt{2011Natur.469..504B}).  It also enables proper 3-band color selection of 
Lyman-break galaxies (LBGs) at $5<z<10$,
which ACS would detect in only two, one, or zero filters, and it offers 
multiple bands for better 
constraining stellar populations and reddening of galaxies at $3<z<7$.  

We describe five science drivers for the cosmic dawn epoch, and
summarize them in Table~\ref{tab:science_goals}.

{\bf(CD1) Improve constraints on the bright end of the galaxy
  luminosity function at {\boldmath$z \approx 7$}--8 and make {\boldmath$z \approx 6$} measurements
  more robust.  Combine with WFC3/IR data on fainter magnitudes to constrain the UV
  luminosity density of the Universe at the end of the reionization
  era.}

Quasi-stellar object (QSO) spectra \citep{2006ARA&A..44..415F}\note{Fan et
  al.\ 2006} and WMAP polarization
(\citealt{2007ApJS..170..335P, 2007ApJS..170..377S})\note{Page et
  al.\ 2007; Spergel et al.\ 2007} both indicate that the intergalactic
medium (IGM) was reionized between 0.5 and 1\,Gyr after the Big Bang.
Moreover, the IGM was seeded with metals to $Z \approx 4 \times 10^{-4}$
within the first billion years, and the energy released by the stars
that produced these metals appears sufficient to reionize the IGM
(\citealt{2001ApJ...561L.153S, 2006MNRAS.371L..78R}).\note{Songaila
  2001; Ryan-Weber et al.\ 2006} However, the exact stars emitting
this radiation have not yet been identified.  The bright end of the
UVLF of star-forming galaxies is evolving
rapidly at $4 < z < 7$ (e.g., \citealt{2004ApJ...600L..99D},
\citealt{2007ApJ...670..928B, 2008ApJ...686..230B, 2011ApJ...737...90B}), but
the UV flux from these bright galaxies thus far appears insufficient to
explain reionization. Estimates of stellar masses and ages at $z = 5$--6
also hint that some galaxies may have experienced a rapid burst of
star formation in the first Gyr (\citealt{2005ApJ...634..109Y,
  2006ApJ...651...24Y, 2005MNRAS.364..443E, 2007MNRAS.374..910E,
  2005ApJ...635..832M, 2008ApJ...676..781W}),\note{Yan et al.\ 2005,
  2006; Eyles et al.\ 2005; Mobasher et al.\ 2005; Wiklind et
  al.\ 2007} but the connection with reionization is unclear.
Finally, the bright end of the UVLF is
itself worthy of study, as these bright high-$z$ galaxies are
statistically the most likely forerunners of massive galaxies
today (Papovich et al.~2011).

The rapid evolution  of the UVLF reflects an interplay between the growth of
dark matter halos, fueling of star formation, regulation by feedback, and dust
obscuration.  Empirical constraints on these basic 
elements of early galaxy growth are essential.  
CANDELS will provide several fundamental tools needed
for this, including measurements of
the rest-frame UVLF, the stellar-mass function, the
color-luminosity and size-luminosity relations, and the angular
correlation function. Many CANDELS galaxies will be bright enough
for detailed morphological study and will be detected individually
beyond $3\,\microns$ by the {\it Spitzer}/SEDS IRAC survey. For others, CANDELS
will provide large statistical samples for IRAC stacking.





\ifsubmode
  \begin{deluxetable}{lclcc}
  \tabletypesize{\scriptsize}
\else
  \begin{deluxetable*}{lclcc}
\fi





\tablecaption{Numbers of Representative Objects \label{tab:numbers_of_galaxies}}


\tablehead{\colhead{Object class} &  \colhead{Redshift range}  & \colhead{$H\tablenotemark{a}$} 
& \colhead{Deep+Wide\tablenotemark{b}} & \colhead{Deep only\tablenotemark{c}}} 

\startdata
$M_{\rm UV} < -19.0$  &  6.5--7.5  & 27.9  &  \nodata      & 280--480 \\
$M_{\rm UV} < -20.0$  &  6.5--7.5  & 26.9  &  160--240  & 25--35 \\
$M_{\rm UV} < -21.3$  &  6.5--7.5  & 25.6  &  $\sim 10$ & $\sim 1$ \\
 & & & & \\
$M_{\rm UV} < -19.0$  &  7.5--8.5  & 28.1  &  \nodata      & 120--280 \\
$M_{\rm UV} < -20.0$  &  7.5--8.5  & 27.1  &  70--150   & 10--20 \\
$M_{\rm UV} < -21.2$  &  7.5--8.5  & 25.9  &  $\sim 10$ & $\sim 1$ \\
 & & & & \\
$M_*>10^9$M$_{\odot}$      & 1.5-2.5  & 28.0\tablenotemark{d} &  \nodata   &   1000  \\
$M_*>10^{10}$M$_{\odot}$    & 1.5-2.5  & 25.5\tablenotemark{d}  &  3000   &   450  \\
$M_*>10^{11}$M$_{\odot}$    & 1.5-2.5  & 23.0\tablenotemark{d}  &  300   &   40  \\
$M_*>10^{11.6}$M$_{\odot}$  & 1.5-2.5  & 21.5\tablenotemark{d}  &  $\sim$10  & $\sim$1 \\
 & & & & \\
Detailed morphologies, Wide\tablenotemark{e} & 1.5--2.5  & 24.0  &  1200  &   \nodata  \\
Detailed morphologies, Deep\tablenotemark{e} & 1.5--2.5  & 24.7  &  \nodata  &  250   \\
X-ray sources\tablenotemark{f} & 1.5--2.5  & \nodata &  200  &   30  \\
Mergers\tablenotemark{g} & 1.5--2.5  & \nodata &  300  &  45  \\
\enddata


\tablecomments{$H$-band magnitude limits for various
classes of galaxies, together with the estimated number of objects brighter 
than this limit within CANDELS\@.}
\tablenotetext{a}{$H$-band magnitude at the indicated stellar mass limit or the
$M_{\rm UV}$ limit at the far edge of the indicated redshift
bin.}
\tablenotetext{b}{Number of galaxies expected in total survey, Deep + Wide
(0.22 deg$^2$).  The higher numbers come from 
LFs by \citet{2011ApJ...737...90B}, 
while the smaller numbers come from \citet{2010MNRAS.403..960M}
(see Fig.~\ref{fig:lbglf}).}
\tablenotetext{c}{As in previous column but for Deep
area only (0.033 deg$^2$).}
\tablenotetext{d}{Magnitude of a red galaxy at the indicated
stellar mass limit at $z=2.5$ (blue galaxies of the same mass are
brighter), based on stellar mass estimates from
FIREWORKS \citep{2008ApJ...682..985W}.}
\tablenotetext{e}{Number of 
galaxies for which detailed morphologies are possible to measure, 
i.e., $H \leq 24.0$\,mag
in Wide and $\leq 24.7$\,mag in Deep (see Fig.~\ref{fig:morphology}).}
\tablenotetext{f}{Total taken from Table \ref{tab:numbers_of_xraysources} 
assuming $ \gtrsim 800$\,ks depth over the whole survey.}  
\tablenotetext{g}{Assumes that 10\% of all
galaxies with $M_*>10^{10}$M$_{\odot}$ are detectable mergers.}


\ifsubmode
\end{deluxetable}
\else
\end{deluxetable*}
\fi
\clearpage

Nailing down the number of bright galaxies impacts
the shape of the LF as a whole, and hence
the estimate of total UV density.
The distribution of galaxy luminosities is 
traditionally characterized by the 
\citet{1976ApJ...203..297S} function: 
\begin{equation}\label{eq1}
N(M)dL = \phi^* (L/L^*)^\alpha \exp(-L/L^*) d(L/L^*),
\end{equation}
which thus far fits distant galaxies fairly well.  
The goal is to robustly identify samples of high-redshift galaxies
and measure the characteristic luminosity $L^*$, the space density $\phi^*$, and
faint-end slope $\alpha$.  However,
the LF cannot be reliably constrained with galaxies 
in only a narrow range of sub-$L^\ast$ luminosities alone --- accurate
measurements at the bright end and near the knee are also needed,
which CANDELS will provide.

At $z \approx 8$, current LFs are based on small fields 
with only a handful of objects, most with $L_{\rm UV} \lta L^\ast$ (there 
are only $\sim 1.0$ 
galaxies with $L > L^\ast$ per $\Delta z = 1$ per WFC3/IR field at $z \geq 6$;
see Table \ref{tab:numbers_of_galaxies}).  
The uncertainty in the luminosity density at $L>L^*$ 
prior to the WFC3 era was $\sim 100$\%\@.  
The population of $z \gtrsim 8$ galaxies now has grown to
the low dozens from early \Hubble\ WFC3/IR data 
\citep{ 2011ApJ...727L..39T, 2011A&A...532A..33G,
2011ApJ...728L..22Y, 2011ApJ...737...90B, 2011arXiv1102.4881M, 2011arXiv1105.2297O}
and some heroic efforts from the ground 
\citep{2009ApJ...706.1136O, 2010ApJ...723..869O, 2010ApJ...722..803O, 
2010A&A...524A..28C, 2011ApJ...730L..35V}; recent
results are summarized in Figure \ref{fig:lbglf}.  
Nonetheless, it is worth noting that there is no 
particularly good physical reason to adopt the exponential
cutoff at the bright end of the LF given by Equation \ref{eq1}.
If one drops this assumption, then estimates of the overall UV luminosity 
density are as sensitive to the shape of the LF at the bright end as they are
to the slope at the faint end.

\ifsubmode
  \placefigure{fig:lbglf}
\else
  \insertfigure{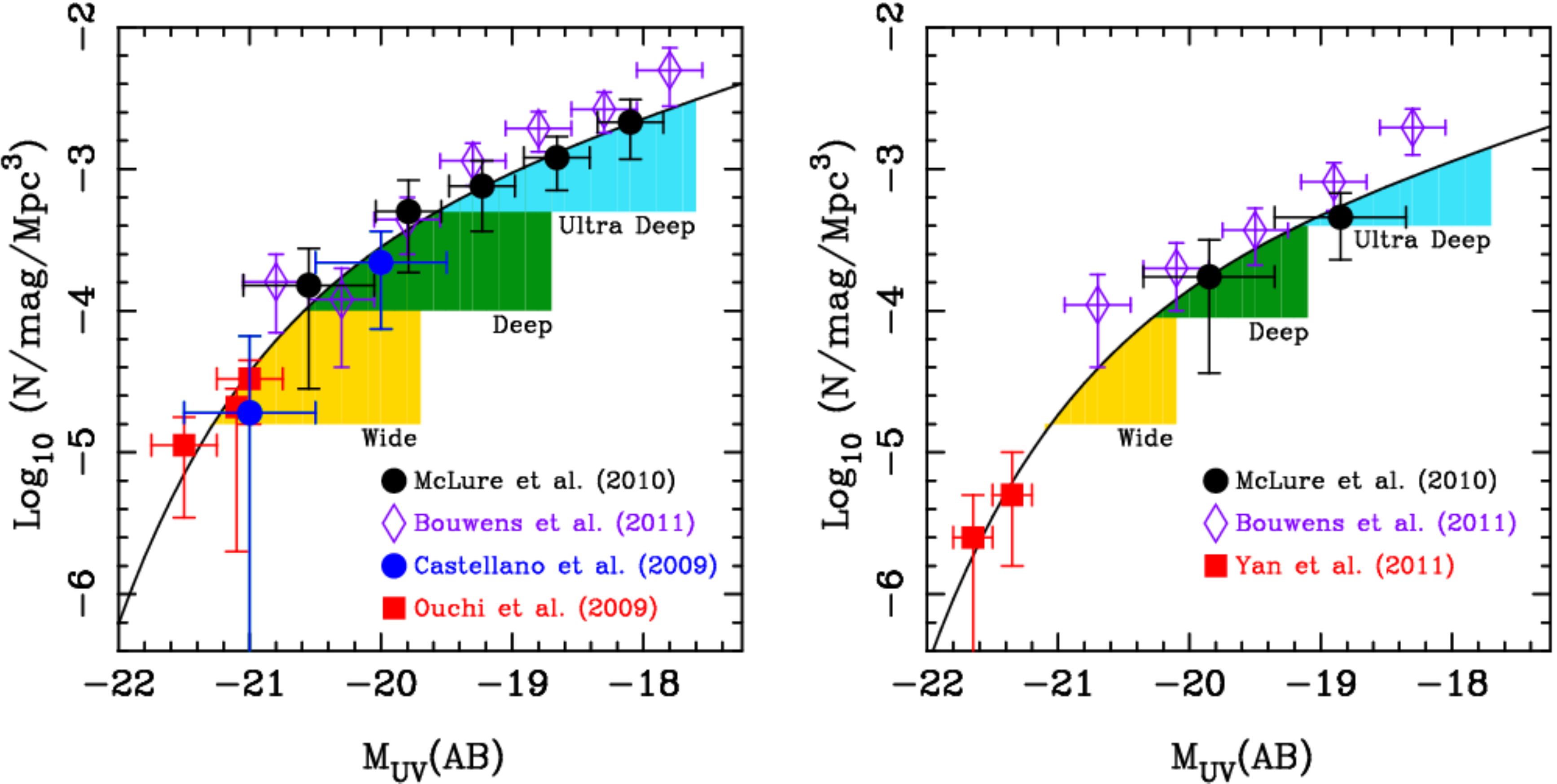}{\figlbglf}
\fi

The expected sensitivity of CANDELS data is compared to other existing
datasets in Figure 
\ref{fig:limits}.  Here we have used the
deepest photometry samples now available over sizeable regions of the 
sky for studying distant
galaxy evolution.  The tips of the triangles
are $5\sigma$ point-source limits, as appropriate for 
$z>6$ galaxies, most of which are nearly point-like in all detectors.  
Green triangles show the data pre-CANDELS, and red and
yellow triangles show what CANDELS will contribute.
Sample SEDs of distant galaxies are superposed to 
illustrate how the combination of deep \textit{HST} and
{\Spitzer}/SEDS data can be used to distinguish old red
foreground interloper galaxies from true distant blue galaxies
with similar brightness at $\sim 1.3\,\mu$m.  Note that existing
ground-based $JHK$ photometry falls well short of
matching the AB sensitivity of associated \textit{HST}/ACS and 
\textit{Spitzer}/IRAC\@.  This shortfall spans crucial wavelengths
where Balmer/4000~\AA\ breaks occur for
$z \approx 2$--5 galaxies and Lyman breaks occur for $z \gtrsim 8$ galaxies.
      
The new CANDELS data fill most of this photometry
gap.   Noteworthy in Figure 
\ref{fig:limits} is the yellow triangle, which
is designed to verify the Lyman breaks of very distant
galaxies and is achieved by stacking 
28\,ks of parallel F814W  
exposure time in the CANDELS/Deep regions
(we are indebted to R.~Bouwens for suggesting
this maximally efficient filter).  With the combination of \Hubble\
$V_{606}$, $i_{775}$, $I_{814}$, $z_{850}$, $Y_{105}$, $J_{125}$, and $H_{160}$, the
Deep filter set will enable robust LBG color 
selection at $\langle z \rangle = 5.8$, 6.6, and 8.0, 
a necessary requirement since most objects
will be too faint for spectroscopic 
confirmation in the near future.  

\ifsubmode
  \placefigure{fig:limits}
\else
  \insertfigurewide{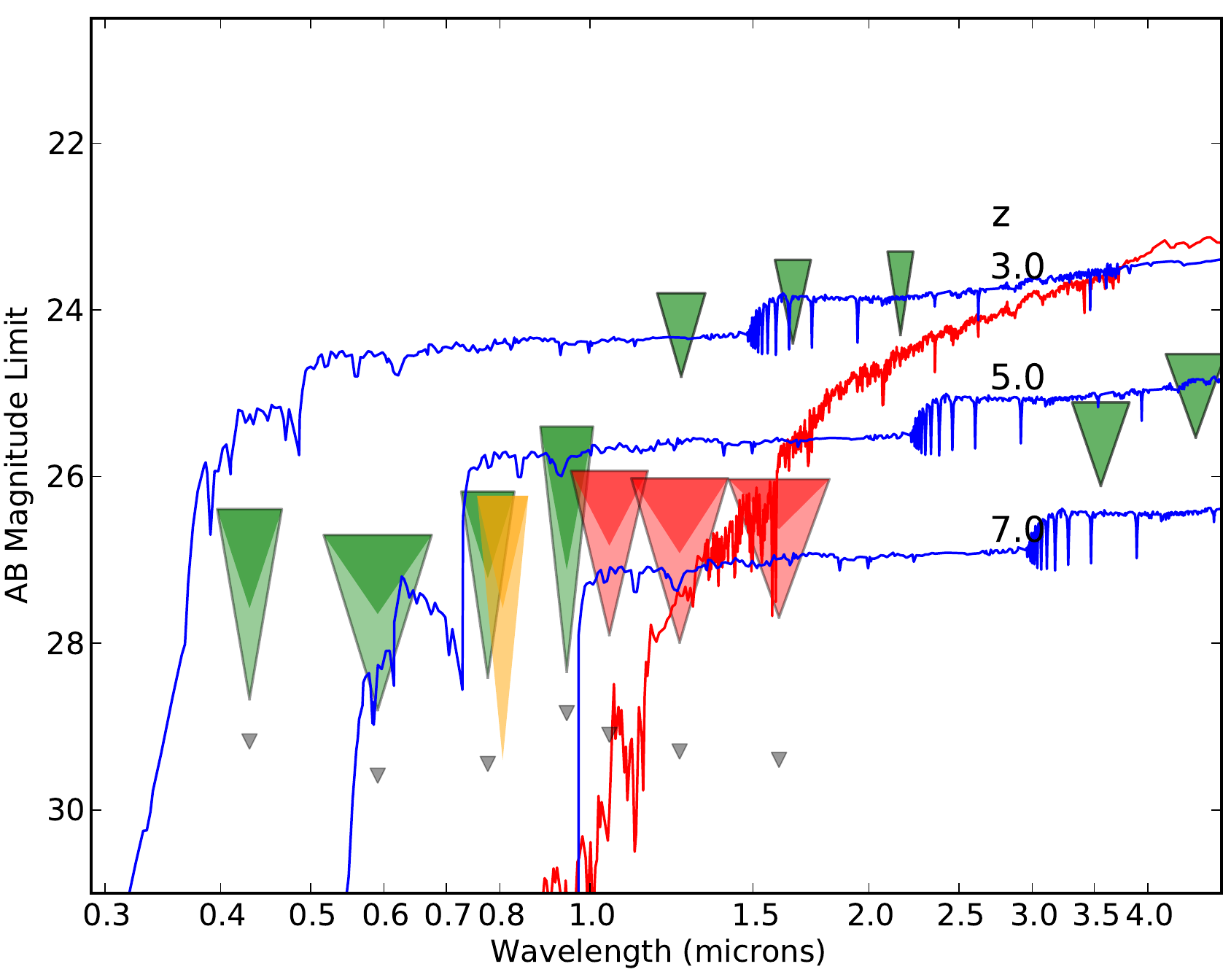}{\figlimits}
\fi

Figure \ref{fig:lbglf} illustrates the
three-tiered observing strategy for distant
objects, and Table \ref{tab:numbers_of_galaxies}
sets forth estimated numbers of sources
in various redshift and magnitude ranges.
Based on the LFs of \citet{2011ApJ...737...90B} and 
\citet{2011arXiv1102.4881M}, we anticipate finding 
hundreds of galaxies at $z \approx 7$ and $z \approx 8$ 
down to $H = 27.9$\,mag in CANDELS/Deep (green areas
in Fig.~\ref{fig:lbglf}).
These objects also have the deepest available observations
from \Spitzer, \Chandra, \Herschel, the VLA, and other major facilities, which
is important if we are to have any near-term hope of constraining their stellar
populations, dust content, and  AGN contribution.

CANDELS/Wide is designed to firm up measurements of brighter high-$z$ 
galaxies over a larger area.  The Wide survey consists of two orbits of
$J$+$H$, accompanied by four {\it effective} parallel orbits in 
$V$+$I$ over nearly three-quarters of the area
(see a sample Wide layout in Fig.~\ref{fig:uds_layout}).  
The Wide filter depths are well
matched to each other and to {\it Spitzer}/SEDS for detecting LBG
galaxies in the range $z \approx 6.5$--8.5 down to $M_{\rm UV} \approx
-20.0$\,mag ($H \approx 27.0$\,mag; yellow regions in Fig.~\ref{fig:lbglf}).
Several hundred such objects are expected.  More precise redshifts will
eventually require either spectra or adding $z$+$Y$ photometry, but
the number density is well matched to current multi-object spectrographs.  The
Wide portions within GOODS, including ERS in GOODS-S, already have $z$-band
imaging and already or soon will have $Y$-band imaging.

Eventual spectroscopic confirmation of galaxy redshifts at $z \gta 7$
will be challenging but not impossible.  For $J = 27$\,mag at $z =
8$, a Ly$\alpha$ line with rest-frame equivalent width
30\,\AA\ yields a line flux of $4 \times
10^{-18}$\,erg\,s$^{-1}$\,cm$^{-2}$, which approaches the present
capabilities of near-IR spectroscopy. Current attempts have already yielded plausible
single-line detections \citep{2010Natur.467..940L,
  2011ApJ...730L..35V} and useful upper limits
\citep{2010ApJ...725L.205F}.  As part of CANDELS itself, 
there are opportunities for spectroscopic confirmation of
high-$z$ candidates with grism observations that will be taken as part
of the SN follow-up observations (see \S\ref{sec:grism}). Ultimately, objects at
the {\Spitzer}/SEDS sensitivity limit of $\sim 26$\,mag should be
spectroscopically accessible with {\it JWST}.

{\it Observational requirements: }
The prime goal for Deep data
is to reach 1\,mag
below $L^{\ast}$ at $z = 8$ with $5\sigma$ accuracy,
corresponding to $M_{\rm UV} \approx -19.0$\,mag  and hence $H = 27.9$\,mag 
(see green regions in Fig.~\ref{fig:lbglf}).
All three $YJH$ filters are required in the Deep program 
for secure detection and redshifts of 
the faintest galaxies.  
This strategy complements the HUDF09 (GO-11563; P.I.~Illingworth), 
which used the same three filters, 
going $\sim 1.1$\,mag deeper, but over $1/10$ the area.
Our SED-fitting simulations suggest that achieving secure redshifts 
further requires $I_{814}$ data to be 
1.5\,mag deeper than $YJH$, and thus a minimum total exposure time in F814W of
28\,ks in the Deep regions.   Placing that much $I$-band exposure time 
in ACS parallels uniformly across both Deep
regions is a major driver for the CANDELS observing
strategy (\S\ref{sec:deepstrategy}).

The required area of the Deep fields is set by various counting
statistics.  First, the horizontal lines in Figure \ref{fig:lbglf} 
correspond to counting 10 objects per mag, which is roughly
the number needed to reduce Poisson noise below cosmic variance
(\S\ref{sec:cosmicvariance}).  Increasing the area of any one tier
of the survey extends its ``sweet spot'' downward and to the left.  A basic
requirement is that all three
tiers --- HUDF, Deep, and Wide --- need to be large enough to
make their sweet spots overlap.  Figure \ref{fig:lbglf}
shows that this has been achieved with the adopted survey sizes.  Second,   
observing only half of each GOODS field in the Deep program 
yields cosmic-variance uncertainties
that are only 20\% worse than two whole GOODS fields, at a savings of
half the observing time.  As shown in \S\ref{sec:cosmicvariance},
the resulting cosmic variance for $6.5<z<7.5$
galaxies in the Deep survey is $\sim 20$\%, sufficient to detect number
density changes from bin to bin of a factor of two at $3\sigma$.

In the Wide program (yellow regions in Fig.~\ref{fig:lbglf}), 
two orbits total in $J$+$H$ reach to
27.0--27.1\,mag in each filter individually, which corresponds 
to $L^{\ast}$ for galaxies at $z \gta 7$ .
The prime goal of Wide is to return a rich sample of luminous high-$z$
candidate galaxies for future follow-up observations.   
A total field size in Deep+Wide of 0.22 deg$^2$ is required to obtain a
total of $\sim 30$ {\it bona fide} $z = 6.5$--8.5
galaxies to $H = 26$\,mag and 200--400 galaxies to $H = 27$\,mag.  If 
divided into several separate fields, the
resulting cosmic variance is again $\sim 20$\% per $\delta z = 1$. 

\textbf{(CD2) Constrain star-formation rates, ages, metallicities, 
stellar masses, and dust contents of galaxies at the end of the 
reionization era, {\boldmath$z \approx 6$}--10. 
Tighten estimates of the evolution of stellar mass, 
dust, and metallicity at {\boldmath$z = 4$}--8 by combining WFC3/IR data with 
very deep \emph{Spitzer}/IRAC photometry.}

Existing data have revealed tantalizing trends in the 
stellar populations of high-redshift galaxies, which are providing important
clues to the progress of star formation and its dependence on galaxy properties.
\citet{2010ApJ...708L..69B} claim that the UV continuum slopes of 
galaxies at $z \approx 7$ are very steep, implying that these galaxies are
younger, less dusty, and/or much lower in metallicity 
($< 10^{-3}\, Z_\odot$) than LBGs at lower redshift. 
Furthermore, the
data, though noisy, 
show a {\it constant ratio} of SFR to stellar mass 
at all redshifts $z>3$ \citep{2010ApJ...713..115G}.  This constant specific
SFR conflicts strongly with the traditional assumption that SFRs
of galaxies are either constant or exponentially declining 
and indeed implies that SFR and stellar mass are both {\it exponentially
increasing} \citep{2011MNRAS.412.1123P}. 
This implies that the long-sought era of galaxy turn-on has 
finally been detected, at redshifts that are accessible to \Hubble. 

CANDELS data are crucial for following up these results. 
At $z>6$, current trends are based on fewer than 100 galaxies
spanning a small dynamic range in luminosity. CANDELS observations will
increase both the number of available galaxies and the dynamic range. 
Estimates of stellar masses at $z \approx 6$--8 connect these galaxies to reionization 
by constraining the total number of 
ionizing photons emitted by previous stellar generations and
to their progenitors and descendants at other redshifts. 
At $z<6$, the combination of WFC3/IR and SEDS/IRAC
bridges the Balmer break and removes much of the degeneracy
between dust and age.  Color criteria 
also reveal whether there are  non-star-forming galaxy candidates lurking 
at these redshifts.  Such searches have been 
severely hampered by the lack of adequate NIR data, 
and the few reports of massive aging galaxies at high $z$ are 
highly controversial 
\citep{2005ApJ...635..832M, 2008ApJ...676..781W, 2007ApJ...665..257C, 
2007MNRAS.376.1054D}.
\note{(Mobasher et al.\ 2005; Wiklind et al.\ 2008; Chary et al.\ 2007, 
Dunlop et al.\ 2007)}
Finally, better photometric redshifts from more accurate SEDs will
inform all of this work.

{\it Observational requirements:}
Most of the requirements for modeling high-$z$ stellar populations
are already met by goal CD1.  The main new requirement 
is that fields have both deep optical images and deep \Spitzer/IRAC.
The latter reaches rest-frame $B$ to $V$ and is crucial for measuring stellar
masses (see Fig.~\ref{fig:limits}).
Choosing fields within the \Spitzer/SEDS survey satisfies
this need, increasing the total sample of galaxies at $z>6$ with 
suitably deep IRAC+WFC3/IR data by tenfold. 
Other high-$z$ WFC3/IR surveys (e.g., those with WFC3/IR in parallel)
lack \Spitzer\ imaging as well as all the other multi-wavelength 
data needed for excellent photometric redshifts and SED modeling.
A few dozen bright LBGs at $z > 6$ will be visible
individually in the {\it Spitzer}/SEDS Deep regions, and 
stacking can extend IRAC constraints to fainter samples \citep{2010ApJ...716L.103L, 
2011ApJ...735....5F}.
\note{ (Labb\'e et al.\ 2009; Finkelstein et al.\ 2011)} 

{\bf (CD3) Measure fluctuations in the near-IR background light,
at sensitivities sufficiently faint and angular scales sufficiently
large to constrain reionization models. }

Most galaxies prior to reionization --- including those dominated by 
Population III
stars --- are too faint to be detected individually even with WFC3, but they 
contribute to the 
spatial fluctuations of the extragalactic background light (EBL).  
First-light galaxies should be highly biased and trace the 
linear regime of clustering at scales of tens of arcminutes when projected
on the sky  
\citep{2004ApJ...606..611C, 2010ApJ...710.1089F}.\note{(Cooray et al.\ 2004; 
Fernandez et al.\ 2009)}
Recent attempts to detect the EBL have been controversial. 
\citet{2005Natur.438...45K}\note{Kashlinsky et al.\ (2005)} 
detected fluctuations in deep \Spitzer/IRAC
observations and interpreted these as evidence for a
large surface density of reionization sources.
Subsequent work, which stacked NICMOS data 
\citep{2007ApJ...657..669T, 2007ApJ...666..658T}\note{Thompson et al} at
at the positions of
$z_{850} \approx 27$ sources 
\citep{2008ApJ...681...53C}\note{Chary et al.\ 2008}, 
suggested that about half the power in the \Spitzer\
fluctuations can be attributed to $z \approx 1.5$ galaxies.

The deepest current observations appear to show a declining
SFR density at $z>6$ among bright LBGs.  This could
imply that galaxies below the WFC3/IR detection limit are producing
the bulk of the ionizing photons.  Alternately, the very small volume surveys
that probe these redshifts (such as the HUDF) could be probing unusually
low-density parts of the Universe due to bad luck with cosmic variance. 
Observations of lensing cluster fields trace too small a volume and
suffer from uncertainties in magnification that vitiate
robust galaxy LFs at the faint end. 
Prior to {\it JWST}, EBL fluctuations in WFC3/IR 
data are arguably the most powerful
probe of very faint sources responsible for reionization.

{\it Observational requirements: }
EBL measurements are essentially impossible from the ground because of
high and variable sky background. 
EBL measurements are best done in fields with deep ACS observations, where
simultaneous optical and near-IR fluctuation measurements can be used to 
constrain
the redshift distribution of the faint sources.  Relative to existing 
\Hubble\ near-IR deep fields, 
CANDELS observations will increase the angular scales over which the 
fluctuations can be measured by a factor of 2--3, which
will enable the contribution of the first-light sources to
be measured at $\sim 0.8$\,nW\,m$^{-2}$\,sr$^{-1}$,
at wavelengths where the contribution of these sources to the EBL is 
thought to peak. This is illustrated in Figure~\ref{fig:ebl}, which shows
the power spectrum of fluctuations on arcminute scales expected if the
reionization sources trace linear clustering on large scales at $z \approx 10$
(green curve), along with the predicted uncertainties from CANDELS
and the measured values from deep NICMOS observations \citep{2007ApJ...657..669T}.
The observing strategy and data reduction will require careful 
attention to flatfielding, scattered
light, bad pixels, and latent images from previous observations. 

Fast-reionization scenarios, involving rapid ($\Delta z \approx 2$--4)
transition from a neutral to an ionized universe ending at
$z \approx 6$--7, are consistent with the WMAP measured optical depth and
the Sloan observations of the Gunn-Peterson trough in $z \approx 6$ QSOs
(Chary \& Cooray, in prep.). Because the redshift range is narrow
and the sources are brighter than in scenarios extending to higher
redshift, the EBL fluctuations are expected to be stronger in such
fast scenarios.  A nondetection of the fluctuations in the proposed
data would point toward a more extended epoch of reionization with
significant contributions from fainter sources.

\textbf{(CD4) Use clustering statistics to estimate the dark-halo masses
of high-redshift galaxies with double the area of previous \emph{Hubble}
surveys.}

The biggest challenge in galaxy formation studies is relating the
visible parts of galaxies to the properties of their invisible dark
matter halos. Galaxy clustering statistics such as two-point
correlation functions can provide strong constraints on the masses of
the dark matter halos that harbor a given observed
population 
\citep{2006ApJ...647..201C,2006ApJ...642...63L,2009ApJ...695..368L}. 
This is done by taking advantage of our excellent understanding of the
clustering properties of dark matter halos, and adopting a
prescription to populate halos with galaxies (often called ``Halo
Occupation Distribution'' or ``Conditional Luminosity Function''
formalism (e.g., \citealt{2007MNRAS.376..841V}). The CANDELS fields
will likely be too small for clustering to yield strong constraints at
the highest redshifts, but it should be possible to apply these kinds
of techniques at the lower range of the ``cosmic dawn'' redshift
window ($z \approx 6$--7), which is still very interesting.

{\it Observational requirements:} Cosmic variance is a dominant source
of error in clustering measurements for small-area surveys, and it can
be mitigated in two ways. The most effective is to break up the
imaging area into several separate regions that are widely distributed
on the sky.  Secondly, elongated field geometries sample
larger scales that are more weakly correlated, and thus suffer less
cosmic variance than compact geometries.  However, the narrowest
field dimension should not drop too close to the correlation length
$r_0$, which at moderate redshifts is a few Mpc.  The adopted sizes and
layouts of the CANDELS fields are well optimized in these respects.

\ifsubmode
\placefigure{fig:ebl}
\else
  \insertfigure{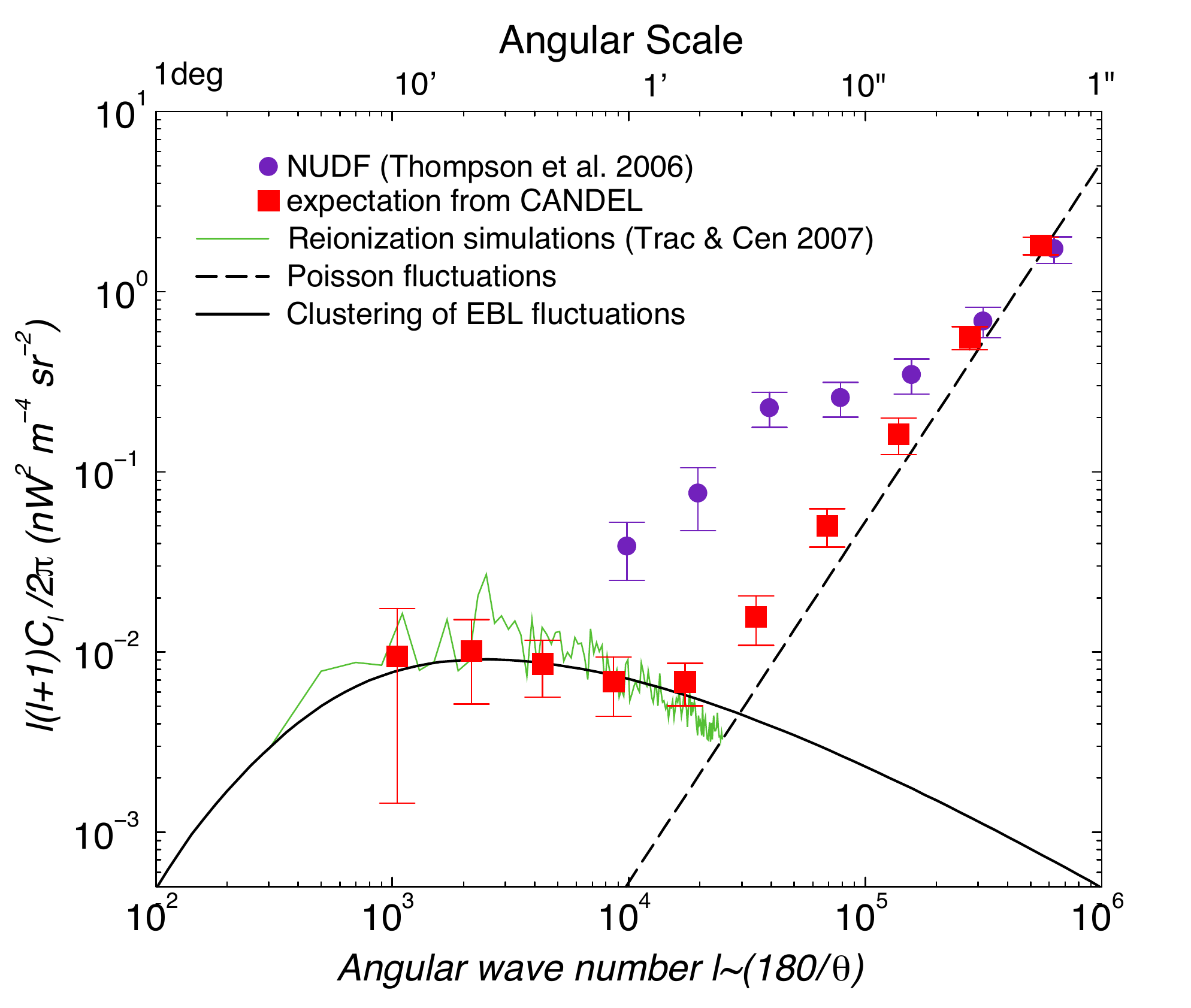}{\figebl}
\fi

\textbf{(CD5) Search the deep WFC3/IR images for AGN drop-out 
candidates at {\boldmath$z > 6$}--7 and constrain the AGN luminosity function.}

In the very early universe, AGNs are expected to play a significant role
both in the evolution of the first galaxies and the reionization of the
universe. Wide-area optical/NIR surveys have found QSOs out to $z \approx 6$
\citep[e.g.,][]{2003AJ....125.1649F} but are limited to the most luminous
and massive objects; the populations of both more typical ($L^*$) and
obscured AGNs are almost completely unknown. Deep X-ray surveys
can readily detect $L^*$ (and fainter) AGNs at $z > 6$, but extremely
deep near-IR data are required to verify them. 
Candidate high-redshift AGNs have already been postulated among the ``EXOs'': 
X-ray sources without optical counterparts even in deep {\it HST}/ACS images
\citep{2004ApJ...600L.123K}.  Drop-out techniques
should be effective for identifying high-$z$ AGNs, just as for galaxies; ultra-deep
near-IR data are critical for distinguishing true high-$z$ candidates from
interlopers via SED fitting.

Semi-analytic models make significantly varying predictions
of the low-luminosity AGN population at high $z$, differing by
over an order of magnitude \citep[e.g.,][]{2008MNRAS.389..270R,
2008MNRAS.385.1846M}. CANDELS will distinguish decisively
between these models: we estimate $\sim 10$ AGNs at $z > 6$ in the
Wide survey accounting for the X-ray detection limits in all fields,
but with a factor $\sim 2$--3 uncertainty in each direction depending
on which evolutionary model is employed \citep{2008MNRAS.387..883A,
2009ApJ...693....8B,2009A&A...493...55E}. Once
appropriate candidates have been found, it will be possible to probe the physical
processes associated with the formation of the first AGNs. For example,
if early black hole growth is mainly merger-driven \citep{2007ApJ...665..187L},
one would expect both coeval star formation and a substantial stellar
population associated with the AGN\@. CANDELS multi-wavelength data will
provide the first clues to this.

{\it Observational requirements:} Many of the requirements for detecting
the host galaxies of these high-$z$ AGNs are already met by the requirements
for goal CD1, namely the combination of area and depth in $J$+$H$ as well
as the bluer bands, to enable robust selection of a sufficient number
of drop-out sources. Additional requirements specifically
related to high-$z$ AGNs are the existence of sufficiently deep X-ray 
imaging, to enable selection of $L^*$ AGNs up to at least $z \approx 6$--7, 
as well as deep \textit{Spitzer} mid-IR
imaging for detection of warm dust emission from the AGN torus
the can expand the sample to obscured/Compton-thick sources. These
datasets exist to varying depths across all 5 CANDELS fields, thereby
enabling a sufficient number of high-$z$ AGNs to be identified in
conjunction with the deep \textit{HST} imaging.

\subsection {Galaxies and AGNs: Cosmic High Noon}\label{sec:high_noon}

Our use of the term ``cosmic high noon'' 
covers roughly $z = 1.5$ to 3.  This epoch 
features several critical transformations in galaxy evolution, as follows.
(1) The cosmic SFR and 
number density of luminous QSOs peaked.
(2) Nearly half of all stellar mass formed in this interval.
(3) Galaxies had much higher surface densities
and gas fractions than now, and the 
nature of gravitational instabilities and 
the physics of star
formation may have been different or more extreme.
(4) Despite the overall high SFR,
SFRs in the most massive galaxies had 
begun to decline, and red central bulges had appeared in some 
star-forming galaxies.  Mature, settled disks were starting to 
form in some objects.
 
Theoretical models of galaxy assembly suggest that the 
$z \approx 2$ era was particularly important. 
The cosmic-integrated accretion rate of fresh gas likely
peaked at $z \approx 2$, and stellar and AGN-powered 
outflows were probably also at maximum strength.
Certain massive galaxies may have been transitioning from accreting gas via
dense, filamentary ``cold flows" to forming a lower density halo of hot
gas (\citealt{2005MNRAS.363....2K, 2006MNRAS.368....2D}\note{Keres et al.~2005, Dekel \& Birnboim 2006}), susceptible to 
heating by radio jets powered by the recently formed population of massive black
holes, and this transition may be a factor in the onset of ``senescence''
among massive galaxies.   Galaxy mergers may also be a driving force
in the assembly,  star-formation, and black-hole accretion of massive galaxies
at this epoch,  turning star-forming disks into quenched spheroidal systems
hosting massive black holes \citep{2006ApJS..163....1H}. 

The above complex of phenomena 
make cosmic high noon a unique era for studying 
processes that transformed the 
rapidly star-forming galaxies of cosmic dawn into the mature 
Hubble types of today. We outline below four science goals
to be addressed at this redshift.

{\bf (CN1) Conduct a mass-limited census of galaxies 
down to {\boldmath$M_* = 2\times10^9 {\rm M}_{\odot}$} at 
{\boldmath$z \approx 2$} and determine redshifts, SFRs, and 
stellar masses from broadband spectral-energy distributions.  
Quantify patterns of star formation versus stellar mass
and other variables and measure the cosmic-integrated 
stellar mass and star formation rates to high accuracy.}

CANDELS measurements will be used to compare the universally
averaged SFR and the stellar mass density over cosmic
time.  Current estimates of integrated cosmic SFR
disagree with the observed stellar mass density by roughly a factor of
two at redshifts $z < 3$ (\citealt{2006ApJ...651..142H,
  2008MNRAS.385..687W}\note{Hopkins \& Beacom~2006; Wilkins et
  al.~2008a}), leading to suggestions that the stellar initial mass
function may be non-universal.  However, better constraints on the
properties of low-mass galaxies could also resolve this discrepancy
\citep{2008ApJS..175...48R}\note{(Reddy et al.~2009)}.  Therefore it
is critical to push measurements of the stellar mass function well
below current ground-based data stellar-mass limits of $10^{10}$
M$_{\odot}$ at $z \approx 2$ (e.g., \citealt{2009ApJ...701.1765M}\note{Marchesini et al.~2009}).

CANDELS will also be used to study how star formation proceeds and is
shut down in individual galaxies during this epoch of rapid galaxy
assembly.  The individual SFRs in star-forming
galaxies correlate closely with their stellar masses in a narrow
``star-forming main sequence'' seen at $z < 3$
(\citealt{2004MNRAS.351.1151B, 2007ApJ...660L..43N,
  2007ApJ...670..156D, 2008ApJ...675..234P}\note{Brinchmann et
  al.~2004; Daddi et al.~2007; Elbaz et al.~2007; Per\'ez-Gonz\'alez
  et al.~2008}).  The zero-point of this sequence declines steadly
from $z \approx 2.5$ to $z \approx 0$ (\citealt{2007ApJ...660L..43N,
  2007ApJ...670..156D}\note{Noeske et al.~2007, Daddi et al.~2007}),
resulting in the rapid decline of the cosmic SFR over
this epoch (\citealt{1996MNRAS.283.1388M, 1996ApJ...460L...1L}\note{Madau et al.~1996, 
Lilly et al.~1996}).  As early as $z \approx 3$,
star formation begins to shut down rapidly in massive galaxies (e.g.,
\citealt{2005ApJ...619L.135J, 2006ApJ...640...92P}\note{Juneau et
  al.~2005, Papovich et al.~2006}), leading to the formation of a
quenched ``red sequence'' that grows with time
(\citealt{2004ApJ...608..752B, 2007ApJ...665..265F,
  2007ApJ...654..858B}\note{Bell et al.~2004, Faber et al.~2007,
  Brown et al.~2007}).  Improved photometric redshifts, stellar
masses, and SFRs at cosmic high noon will place tighter constraints on
the origin and evolution of the SFR-stellar mass correlation and
galaxy quenching.

\ifsubmode
  \placefigure{fig:zphot}
\else
  \insertfigure{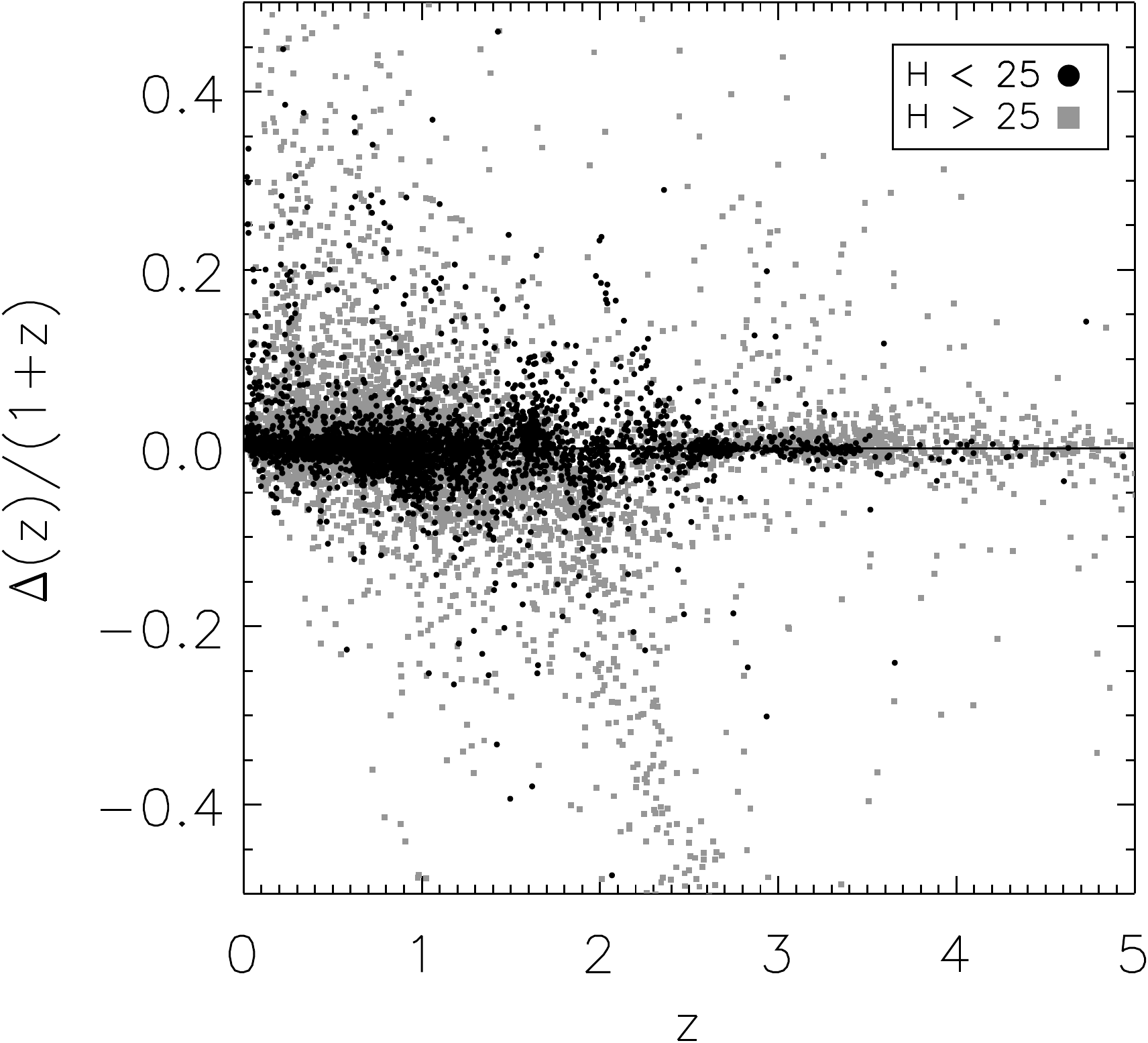}{\figzphot}
\fi

{\it Observational requirements:} The above science goals require a
mass-limited census of galaxies down to $2\times10^9$ M$_{\odot}$ out to
$z \approx 2.5$.  The reddest galaxies at this mass will have $H
\approx 27.3$\,mag and will be detected in the Deep regions only.  We estimate
finding $\sim 500$ objects in total above $2\times10^9$ M$_{\odot}$
in the redshift range $z = 1.5$--2.5 (Table
\ref{tab:numbers_of_galaxies}).  The Wide survey limit is roughly 1\,mag 
brighter and will sample $\sim 5000$ objects above
$5\times10^{9}$ M$_{\odot}$.  Mass limits for blue galaxies are
nearly ten times smaller.  Both Deep and Wide survey limits improve
several-fold over current limits.  As shown in
\S\ref{sec:cosmicvariance}, we expect that CV errors with the current
survey design will be $<$20\% for both massive galaxies ($m_{\rm star}
\approx 10^{11}$ M$_{\odot}$) in the Wide survey and also for low-mass
galaxies ($m_{\rm star} \approx 2\times10^9$ M$_{\odot}$) in the Deep
regions, each for redshifts $z = 1.5$--2.5.

High quality photometric redshifts and SEDs are needed for all
galaxies, requiring multi-band WFC3/IR data, ACS parallel imaging and
other ancillary data deep enough to match WFC3/IR for all galaxy
colors.  At CANDELS faintest levels, photometric redshifts will be the
main source of redshifts for the foreseeable future;
$YJH$ filters span the Balmer/4000\,\AA\ break at $z \approx 2$, which
makes them crucial for photometric redshifts (Figure
\ref{fig:limits}).  Figure \ref{fig:zphot} compares photometric
redshifts with and without recent high-quality \Hubble\ ERS $YJH$
photometry.  Brighter than $K = 24.3$\,mag, there is little
difference, but below $K = 24.3$\, mag, the differences can be huge,
particularly at cosmic high noon.  Better $YJH$ data will also tighten
SEDs across the Balmer break, helping to disentangle the age-dust
degeneracy and improving stellar population estimates from SED
fitting.  Existing GOODS ACS images do not reach faint red galaxies at
$z \approx 2$, and thus the $\gtrsim 28$\,ks F814W parallels in the Deep fields,
designed for high-$z$ galaxy detection (see goal CD1,
\S\ref{sec:cosmic_dawn}), are essential.

Combined with SEDS/IRAC photometry from {\it Spitzer}, CANDELS will
obtain good stellar masses for all $z \approx 2$ 
galaxies regardless of color down to the $H$-band magnitude limit
(Fig.~\ref{fig:limits}).  The resulting stellar-mass
inventory should be a considerable advance over
present data, especially when paired with improved
SFRs from \textit{HST} rest-frame UV, 
{\it Spitzer}, and {\it Herschel} mid/far-IR data
available in all CANDELS fields (see \S\ref{sec:fields}).

\textbf{(CN2) Obtain rest-frame optical morphologies 
and structural parameters of {\boldmath$z \approx 2$} galaxies, 
including morphological types, radii, stellar mass surface densities, 
and quantitative disk, spheroid and interaction measures.  
Use these to address the relationship between galactic structure, 
star-formation history, and mass assembly.} 

Previous \textit{HST} studies have characterized the rest-frame
UV structure of galaxies at $1.5 < z < 2.5$ below the Balmer
break.  The impression from these images is that a ``galactic
metamorphosis'' occurred near $z \approx 1.5$: galaxies at earlier
times displayed a much higher degree of irregularity than later
generations (e.g., \citealt{1995ApJ...453...48D, 1996MNRAS.279L..47A,
  1996ApJ...470..189G}\note{Driver et al.~1995, Abraham et al.~1996,
  Giavalisco et al.~1996}).  However, Figure \ref{fig:smallmontage}
shows that rest-frame UV images can be biased toward young stars and
may not show the true underlying structure of the galaxy, which is
better seen in the older, redder stars.  Recent results using WFC3
have demonstrated the importance of high-resolution NIR for characterizing
morphology of the massive galaxies in this redshift range 
\citep{2011arXiv1105.2522C,2011arXiv1107.3137L}.  
Covering much larger area than these early WFC3/IR studies,
CANDELS will more comprehensively measure the frequency 
of disk, spheroidal and peculiar/irregular structures in rest-frame
optical light for $1.5 < z < 3$ galaxies, as evidenced by visual
properties and quantitative measures such as Sersic index,
bulge-to-disk ratio, axial ratio distributions, and
concentration/Gini-M$_{20}$.

\ifsubmode
\placefigure{fig:largemontage}
\else
  \insertfigurewide{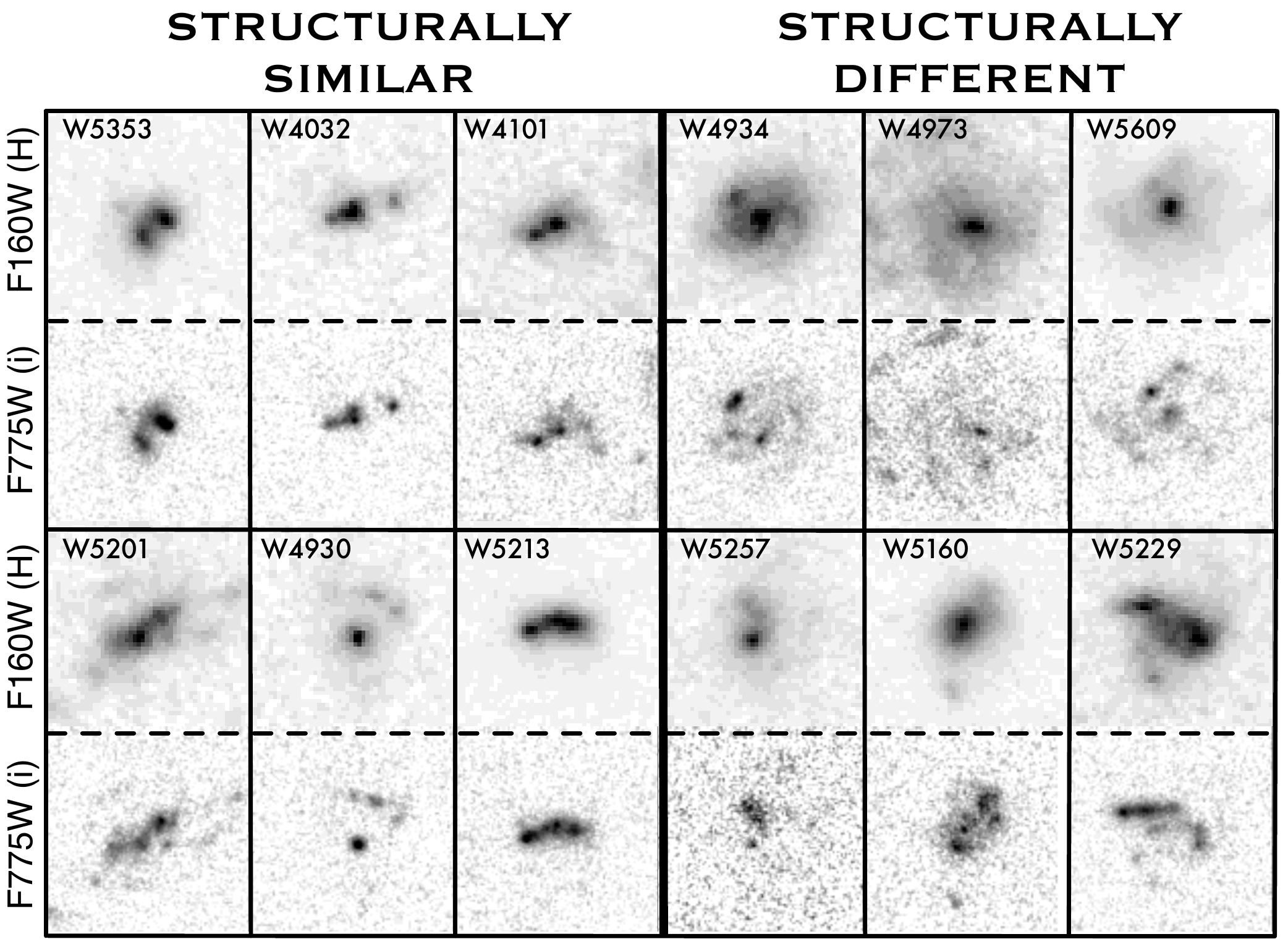}{\figlargemontage}
\fi

Theoretical models of disk formation make quantitative predictions for
disk scaling relations and their evolution with cosmic time
\citep{2008ApJ...672..776S,2011MNRAS.410.1660D}, which may be compared
with the rest-frame optical sizes that will be measured by CANDELS.
The distribution of disk axial ratios and the frequency
of bars and spiral structures will be used to probe the degree of
disk settling and the development of dynamically cold and
gravitationally unstable disk structures
(\citealt{2009ApJ...706.1364F, 2006ApJ...652..963R}\note{
Fo\"rster-Schreiber et al.~2009; Ravindranath et al.~2007}).
Radii also yield stellar mass surface densities, which have been
implicated as a threshold parameter for quenching and evolution to the
red sequence (\citealt{2003MNRAS.341...33K, 2008ApJ...688..770F}\note{
Kauffmann et al.~2004, Franx et al.~2008}).  
We will attempt to link the structural and dynamical properties of
disks with their SFRs and to confront the latest
theoretical ideas for the processes that regulate and stimulate star
formation with these observations.

Major mergers occupy a central role in hierarchical theories of galaxy
formation, and are a favorite mechanism for building spheroids,
triggering BH growth and igniting starbursts (e.g.,
\citealt{1988ApJ...325...74S, 2006ApJS..163....1H,
  2008MNRAS.391..481S}\note{Sanders et al.~1988, Hopkins et al.~2006,
  Somerville et al. 2008b}).  Quenching by mergers and associated
feedback (e.g., \citealt{2006ApJS..163....1H}\note{Hopkins et
  al.~2006}) tends to destroy disks, whereas quenching through
declining accretion (or ``strangulation'') tends to preserve them.
Measuring disk and spheroid fractions of red sequence galaxies versus
mass and time is therefore a clue to how galaxies have transited to
the red sequence at different epochs.

Passive galaxies must have evolved substantially along the red
sequence as well.  Galaxies with very low specific SFRs are present to
at least $z = 2.5$ but with sizes that are much smaller than those of
nearby ellipticals of the same mass (e.g.,
\citealt{2005ApJ...626..680D, 2007MNRAS.382..109T,
  2008A&A...482...21C, 2008ApJ...677L...5V}\note{Daddi et al.~2005;
  Trujillo et al.~2006; Cimatti et al.~2008; van Dokkum et al.~2008}).
If these sizes are correct, the implication is that massive
spheroids must have accreted considerable stellar mass after $z \approx
2$ and that their extended stellar envelopes grew by dry minor mergers
(\citealt{2009ApJ...699L.178N, 2010ApJ...709.1018V}\note{Naab,
  Johansson \& Ostriker 2009; van Dokkum et al. 2010}).  These outer
envelopes may be quite red and can be observed using stacked images
from WFC3/IR, from which evolution in both radii and concentration
indices can be measured.

Direct measurements of major and minor merger rates are also 
needed to inform theory,  especially at $z \approx 2$, where QSOs peak.
Galaxy interactions, disturbances, and mergers will be identified
directly in the CANDELS images using both visual and quantitative 
methods such as CAS \citep{2003AJ....126.1183C}\note{Conselice et al.~2003} 
and Gini/M$_{20}$ \citep{2004AJ....128..163L}\note{Lotz et
al.~2004}.  Detailed numerical simulations of merging galaxies are 
able to predict the visibility of mergers using different methods 
(e.g., \citealt{2008MNRAS.391.1137L, 
2006ApJ...638..686C}\note{Lotz et al.~2008b, Conselice et al.~2006}).  
Thus observed galaxy merger fractions may be converted 
into major and minor merger rates and compared to theoretical predictions.  

\ifsubmode
  \placefigure{fig:morphology}
\else
  \insertfigure{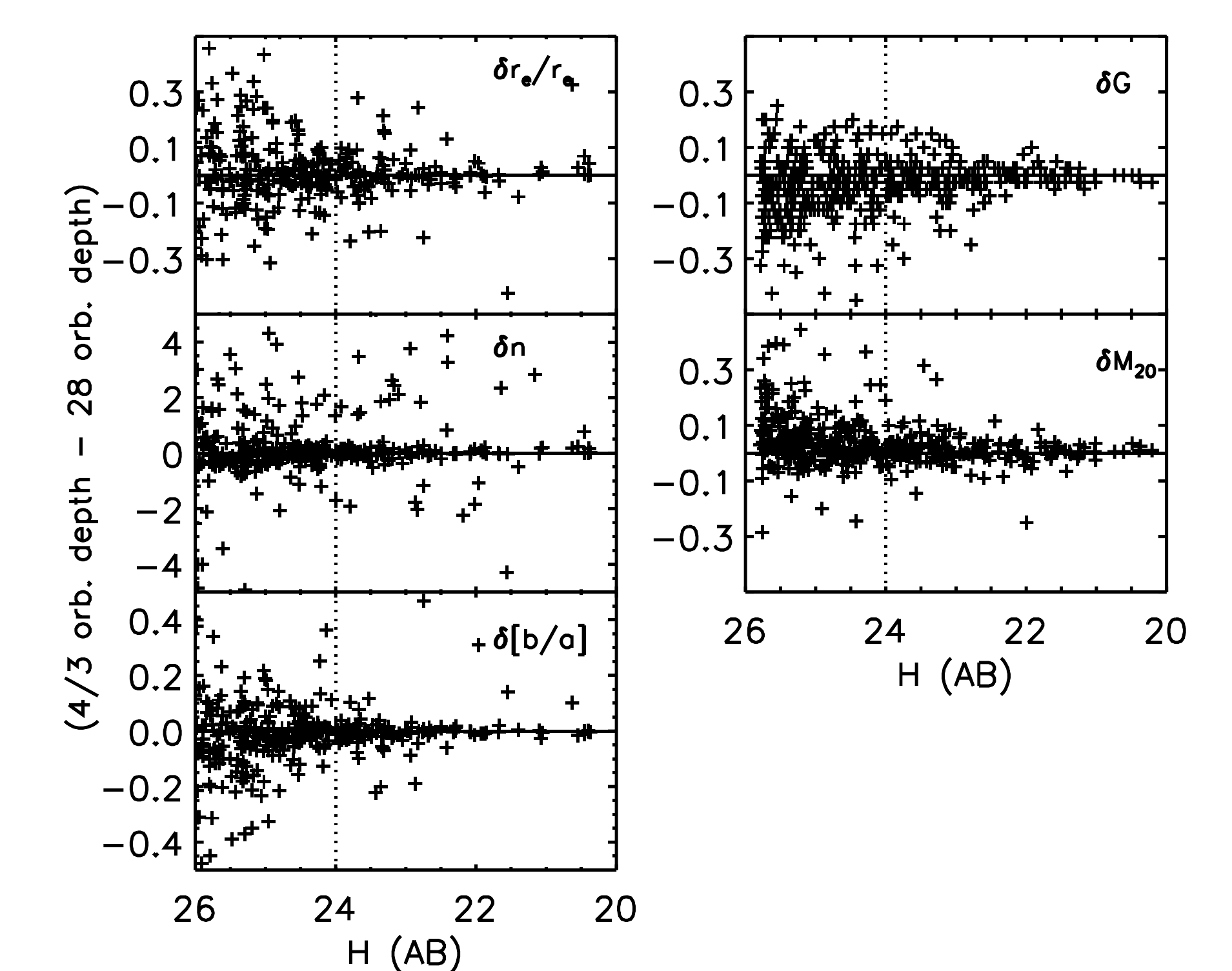}{\figmorph}
\fi

{\it Observational requirements:} 
Answering these questions requires measuring
accurate radii and axial ratios,  tracking spheroids and disks, 
searching for substructure in the form of bars and spiral arms, and identifying
merger-induced distortions. 
Given the small angular sizes of most $z \approx 2$ galaxies, 
detailed structural parameters depend on extracting the highest possible 
spatial resolution from WFC3/IR images, which requires 
multiple exposures and fine dithering.   The WFC3 Y, J, and H bandpasses
will be used to track evolution of galaxy structure at a 
uniform rest-frame optical ($\sim B$-band) wavelength from $z \approx 1.5$ to $z \approx 3$.

The quality of structural data to be expected in the Wide program is
illustrated in Figure \ref{fig:morphology}, which shows several
quantitative morphological statistics (effective radius, Sersic index,
axial ratio, Gini, and M$_{20}$) computed on 4/3-orbit F160W data
compared with the same quantities computed with 28-orbit data
from the HUDF\@.  The vertical line is the adopted ``acceptable'' limit,
which is at $H = 24.0$ for Wide, and $H = 24.7$\,mag for Deep.  
These correspond to stellar masses of $4\times10^{10}$ M$_{\odot}$ and
$2\times10^{10}$ M$_{\odot}$ respectively at $z = 2.5$ (Table
\ref{tab:numbers_of_galaxies}), which is deep enough to capture most
galaxies transiting to the red sequence if they do so at a similar
mass as the observed transition mass in the local Universe ($\sim 3 \times
10^{10}$ M$_{\odot}$, \cite{2003MNRAS.341...33K}).

Interpolating in Table \ref{tab:numbers_of_galaxies} predicts
$\sim 225$ galaxies in the Deep regions above this level in the range
$z = 1.5$--2.5.  The Wide program will mainly sample rarer objects such as
massive galaxies and mergers.  The expected limit in Wide is two times
brighter ($H = 24.0$), 
and $\sim 1200$ objects are expected above
this level in the same redshift range.  If major mergers involve
roughly $\ge$ 10\% of galaxies at $z \approx 2$ (Lotz et al.~2008),
$\geq$75 bright mergers will be available
for detailed morphological modeling in Wide+Deep.

\textbf{(CN3) Detect galaxy sub-structures and measure their stellar masses.
Use these data to assess disk instabilities, quantify internal patterns of 
star formation, and test bulge formation by clump migration to the centers of galaxies.}

Many star-forming galaxies at $z \approx 2$ have compact sizes, high
surface densities and high gas fractions 
\citep{2010Natur.463..781T}\note{(Tacconi et al.~2010)}, 
and exhibit star-forming clumps more
massive than local star clusters \citep{2009ApJ...692...12E}.  This
is qualitatively consistent with state-of-the-art cosmological simulations,
which indicate that massive galaxies can acquire a large fraction of
their baryonic mass via quasi-steady cold gas streams that penetrate
effectively through the shock-heated hot gas within massive dark
matter halos (e.g., \citealt{2005MNRAS.363....2K, 2006MNRAS.368....2D,
 2009Natur.457..451D}\note{Keres et al.~2005; Dekel \& Birnboim
  2006; Dekel et al.~2009}).  Angular momentum is largely preserved
as matter is accreted, early disks survive and are replenished, and
instabilities in fragmenting disks create massive self-gravitating
clumps that rapidly migrate towards the center and coalesce to form a
young bulge (\citealt{2009Natur.457..451D}\note{Dekel et al.~2009}; see
also \citealt{1999ApJ...514...77N, 2004ApJ...611...20I,
  2007ApJ...670..237B, 2008ApJ...688...67E}\note{ Noguchi 1999;
  Immeli et al.~2004; Bournaud et al.~2007; Elmegreen et al.~2008}).

Testing this picture requires distinguishing between clumps
generated from internal disk instabilities and substructure accreted
by external interactions and mergers.  As shown in Figure
\ref{fig:largemontage}, high-spatial resolution multi-band \textit{HST}
images may be able to do this, permitting us to classify disturbed 
galaxies into the two camps (though this needs to be checked with
simulations).
Rest-frame UV images from ACS provide a map of where unobscured stars
are forming, while rest-frame optical images from WFC3/IR give
improved dust maps and clump stellar masses.  The numbers, spatial
distribution, stellar masses and SFRs of the clumps
can then be compared to cosmological simulations of forming galaxies
to test the new paradigm of cold streams and their impact on disk star
formation.

{\it Observational requirements:} 
Deep and high-spatial resolution imaging is 
needed to extract reliable clump information, and 
may require deconvolved high signal-to-noise ratio ($S/N$) Deep WFC3/IR images
for resolved stellar population studies.   
A full inventory of star-formation 
properties for clumpy star-forming galaxies will 
require both WFC3/IR and ACS/optical imaging 
as well as extensive ancillary data on integrated
SEDs, SFRs, emission line strengths, and dust temperatures
from a wide range of ground and space telescopes. 

\textbf{(CN4) Conduct the deepest and most unbiased census yet of 
active galaxies at {\boldmath$z \gta 2$} selected by X-ray, IR, optical spectra, and optical/NIR variability.  
Test models for the co-evolution of black holes 
and galaxies and triggering mechanisms
using demographic data on host properties, including morphology and 
interaction fraction.}

The discovery of the remarkably tight relation between black hole
masses and host spheroid properties (\citealt{1998AJ....115.2285M,
  2000ApJ...539L..13G, 2000ApJ...539L...9F, 
  2004ApJ...604L..89H}\note{Magorrian et al.~1998, Gebhardt et al.~2000, Ferrarese \& Merrit
  2000, H\"aring \& Rix 2004}) has given birth to a new paradigm: the
co-evolution of galaxies and their central supermassive black holes
(BHs).  However, since only a small fraction of galaxies appear to be
building BHs at any instant, it is unclear how the galaxy-black hole
connection was either first established or subsequently maintained.  One suggested
mechanism is major mergers, which scramble disks into spheroids, feed
BHs, and quench star formation via AGN or starburst driven winds (e.g.,
\citealt{1988ApJ...325...74S, 2006ApJS..163....1H}\note{Sanders et
  al.~1988, Hopkins et al.~2006}).  A testable feature of this model
is the strong merging and disturbance signatures predicted for AGN
host galaxies, as shown in Figure \ref{fig:merger}.  The total Wide
volume in the 1.6 Gyr period $1.5 < z < 2.5$ contains roughly 3000
galaxies with stellar mass $>10^{10}$ M$_{\odot}$ (Table
\ref{tab:numbers_of_galaxies}).  Models predict that a typical galaxy
suffered one major merger during this period.  If this picture is
correct, we would therefore expect $\gta$300 galaxies in the highly
disturbed phase and 300 in the visible QSO phase.  \Hubble \ imaging
has not thus far shown much evidence for excess merger signatures in
X-ray sources at $z \approx 1$ (\citealt{2005ApJ...627L..97G,
  2007ApJ...660L..19P}\note{Grogin et al.~2005; Pierce et al.~2007}),
but triggering processes in brighter QSOs, which are more common at $z
\approx 2$, may differ.  A second prediction, though not unique to this
model, is that massive black holes should be found only in galaxies
possessing major spheroids.  Suggestion of this from the GOODS
AGN hosts at $z<1.3$ (\citealt{2005ApJ...627L..97G}) may now be 
scrutinized over a much broader redshift range, $z\lesssim3$, with CANDELS. 

\ifsubmode
  \placefigure{fig:merger}
\else
  \insertfigurewide
  {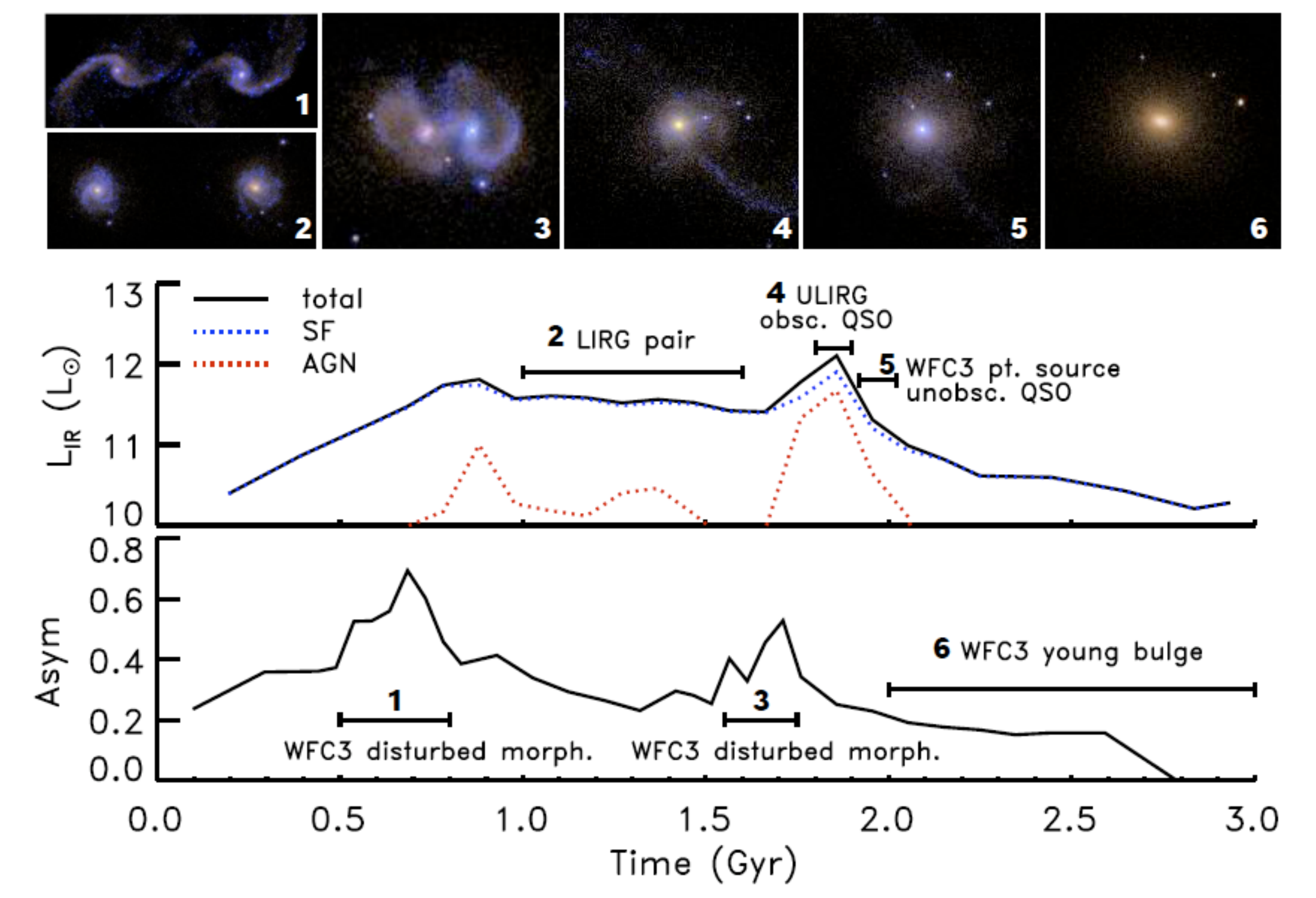}{\figmerger}
\fi

The CANDELS fields, with their superior X-ray imaging area and depth 
(now extended to an unprecedented 4\,Ms in GOODS-S), plus multi-epoch 
ACS imaging, ultra-deep {\it Spitzer}, {\it Herschel}, and radio coverage 
(see \S\ref{sec:fields}), provide by far the best data
with which to identify and study distant AGNs --- not only X-ray AGNs but also
their heavily obscured counterparts.  Between 20 and 50\% of $z \approx
2$ galaxies may host a Compton-thick AGNs undetected in X-rays
\citep{2007ApJ...670..173D}\note{Daddi et al.~2007b} but appearing 
as luminous IR power-law SEDs in {\it Spitzer} and {\it Herschel}
data (\citealt{2006ApJ...640..167A, 2008ApJ...687..111D}, Juneau \&
Dickinson, in prep.)\note{Alonso-Herrero et al.~2006; Donley et
  al.~2008; Juneau \& Dickinson 2011 in prep}.  These
highly obscured sources might hide key phases of BH growth.  They might also
be the very ones most likely to show mergers and asymmetries (Figure
\ref{fig:merger}; see also \citealt{2008ApJS..175..356H}).
The multi-wavelength ancillary data combined with CANDELS \textit{HST}
imaging will also allow us to connect and compare this active BH
accretion phase of galaxy formation to active star-formation phases
that produce ultra-luminous IR and sub-mm galaxies
(e.g., \citealt{2010ApJ...719.1393D, 2008ApJ...675.1171P,
  2008AJ....135.1968A, 2008MNRAS.389...45C}\note{Donley et al. 2010;
  Pope et al. 2008; Alexander et al. 2008; Coppin et al. 2008}).






\ifsubmode
\begin{deluxetable}{lllllll}
\tabletypesize{\scriptsize}
\else
\begin{deluxetable*}{lllllll}
\fi





\tablecaption{Number of X-ray Sources \label{tab:numbers_of_xraysources}}


\tablehead {
\colhead{Field} &  \colhead{Area}  & \colhead{Telescope} & \colhead{Exposure} & 
\colhead{Total number}  &  \colhead{Number in}   &  {$F_{lim}$} \\
\colhead{} &  \colhead{arcmin$^2$}  & \colhead{} & \colhead{} & 
\colhead{in field}  &  \colhead{$z = 1.5$--2.5}   &  {erg s$^{-1}$ cm$^{-2}$} \\
}

\startdata
COSMOS  & 176  &  Chandra   &  200 ks  &  106   &  10    &  $2\times 10^{-16}$ \\ 
EGS     & 180  &  Chandra   &  800 ks &  240   &  40    &  $3\times 10^{-17}$ \\
GOODS-N & 132  &  Chandra   &  2 Ms  &  175   &  32    &  $2\times 10^{-17}$ \\
GOODS-S & 124  &  Chandra   &  4 Ms  &  317   &  90    &  $1\times 10^{-17}$ \\
UDS     & 176  &  XMM       &  100 ks  &  70    &   5    &  $6\times 10^{-16}$ \\ 
\enddata


\tablecomments{Table shows numbers of X-ray sources, in total over
each CANDELS field and in the $z=1.5$--2.5 redshift bin.  The former is
based on actual counts, while the latter is based on extrapolations
from sources with measured redshifts and is quite uncertain.}


\ifsubmode
\end{deluxetable}
\else
\end{deluxetable*}
\fi
\clearpage

{\it Observational requirements:}
In addition to multi-wavelength data for AGN 
identification, co-evolution studies need the redshifts, stellar
masses, stellar population and dust data, bulge and disk fractions, 
and disturbance indicators to be obtained with the CANDELS \textit{HST}
data.  X-ray AGNs at $z \approx 2$ are found mainly
above $\sim 2 \times 10^{10}$ M$_{\odot}$
(Kocevski et al., in prep.),  which is fortuitously equal to the limit 
for detailed morphologies in Deep data
(goal CN2 above).  However, AGNs are rare, and 
the larger area of the Wide survey, even though it does not
go quite as deep, is clearly crucial for obtaining a valid sample of these objects. 

Predicted numbers of X-ray objects in CANDELS fields are given in
Table \ref{tab:numbers_of_xraysources} based on current observed X-ray
counts.  If all CANDELS fields were surveyed to a depth of 800\,ks
(as in EGS), roughly 200 X-ray AGNs would be found within the CANDELS
area in the range $z = 1.5$--2.5 (Table \ref{tab:numbers_of_galaxies}).
Figure \ref{fig:merger} predicts that the number of obscured AGNs 
should be similar.  Due to their extremely
red colors, some 25\% of obscured AGNs lack ACS counterparts, and 60\%
lack ground-based NIR counterparts.  WFC3/IR is the only hope for imaging
such objects in the near future.  Finally, multi-epoch ACS data in all
fields, especially GOODS, will be unparalleled for AGN variability
studies.

\subsection {Ultraviolet Observations}\label{sec:uv}

GOODS-N is in the \textit{HST} continuous viewing zone
(by design). Thus we can boost \textit{HST} efficiency
by using the bright day-side of the orbit to observe in the
UV with WFC3/UVIS (specifically the
F275W and F336W filters). We conservatively plan on 100 orbits with
UV observations, but this could be as high as $\sim 160$
orbits if we are able to make use of all the available opportunities.

Because these observations are read-noise limited, we have chosen to
bin the UVIS data to reduce read noise and gain $\sim 0.5$\,mag
of sensitivity in each filter.  To facilitate the selection of
Lyman-break drop-outs at $z \approx 2$ and to increase the sensitivity to LyC
radiation at $z \approx 2.5$, we expose twice as long in F275W 
as in F336W\@.  Due to scheduling constraints, we anticipate that
many of the orbits may not have full CVZ duration, and the UVIS exposure 
times on the day side of the 
orbit may be scaled back.  Nonetheless, we expect to get the equivalent of 
$\sim 100$ orbits of UVIS imaging in the field.  Depending upon final exposure 
times and the degradation of the UVIS detectors' charge-transfer efficiency, 
we expect depths of $\sim 27.9$ and $\sim 27.7$\,mag in F275W and
F336W, respectively ($5\sigma$ in a 0.2 arcsec$^2$ aperture).

These data enable three important scientific investigations.

\textbf{(UV1) Constrain the Lyman-continuum (LyC) escape fraction for
galaxies at {\boldmath$z \approx 2.5$}.} 

A composite spectrum of distant LBGs
is shown in Figure \ref{fig:uv}  as it would appear redshifted
to $z = 2.4$.  Overplotted are the transmission curves of the WFC3/UVIS
filters being used in the GOODS-N CVZ day-side observations.
At $z > 2.38$, the Lyman limit shifts redward of any significant
transmission in the F275W filter, and this filter therefore probes
the escaping Lyman continuum radiation for galaxies in the
range $2.38<z < 2.55$ (where the upper limit is dictated by
IGM opacity).


\ifsubmode
  \placefigure{fig:uv}
\else
  \insertfigure{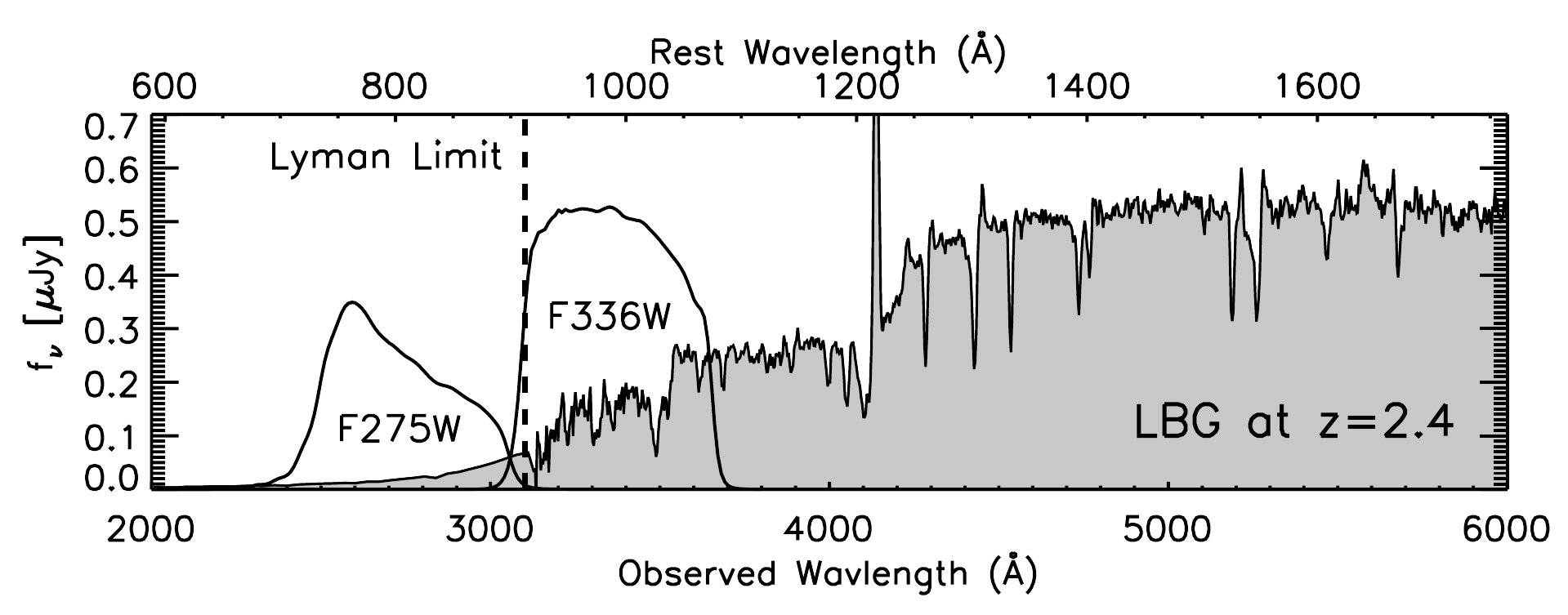}{\figuv}
\fi

Within the CANDELS/Deep portion of GOODS-N, there are about 20
LBGs (depending upon final pointings and orientations) with
spectroscopic redshifts at the optimal redshift ($2.38<z<2.55$) that are
luminous enough ($R>24.7$, $L_{\rm UV} > 0.6 L^*$) so that the Lyman
continuum escape fraction can be significantly constrained
($f_{esc,rel}<0.5$ to $3\sigma$).  Because
the spectroscopy is only $\sim 40$\% complete, we
expect to double this sample with additional spectroscopy and get strong
constraints on the LyC escape fraction in a large,
unbiased sample of more than 40 LBGs.
(cf.~\citealt{2006ApJ...651..688S,
2009ApJ...692.1287I, 2010ApJ...723..241S}).
\note{Shapley et al. 2006; Iwata et al. 2009; Siana et al. 2010}
Importantly, these galaxies are at redshifts that allow
H$\alpha$ measurements from the ground
for an independent measure of the ionizing continuum.
The resolved LyC distributions will test different mechanisms
for high $f_{esc}$\ including SN winds
\citep{2002MNRAS.337.1299C, 2002ApJ...577...11F},
and galaxy interactions
\citep{2008ApJ...672..765G}\note{(Gnedin et al. 2008)}.

\textbf{(UV2) Identify Lyman-break galaxies
at {\boldmath$z \approx 2$} and compare their properties to higher-{\boldmath$z$}
Lyman-break galaxy samples.}  

Star-forming
galaxies at $z \approx 2$ will be selected via
identification of the Lyman break in the F275W passband,
analogous to \textit{HST} Lyman break studies at $4 < z < 8$ 
\citep{2007ApJ...670..928B,2010ApJ...709L.133B}\note{
(eg. Bouwens et al. 2007, 2010)}.  
These data will help identify the large population of faint 
galaxies that are suggested by initial findings of a 
steep LF \citep{2010ApJ...725L.150O}\note{
(Oesch et al. 2010)} and provide a more accurate census of the 
SFR density at this epoch.  
Furthermore, the existing ACS GOODS data will allow an 
investigation of the dependence on the UV attenuation by 
dust as a function of UV luminosity, to compare to existing 
studies at higher redshift 
\citep{2008ApJS..175...48R,2010ApJ...708L..69B}\note{
(Reddy et al. 2008, Bouwens et al. 2010)}.

\textbf{(UV3) Estimate the star-formation rate in dwarf galaxies at {\boldmath$z
> 1$}.}  

It has been hypothesized that the UV background heats the gas in
low-mass halos enough to prevent cooling and star formation at $z >
2$, solving the missing-satellite problem \citep{1992MNRAS.255..346B,
2000ApJ...539..517B}\note{ (Babul \& Rees 1992; Bullock et al.~2000)}.
An important test of this idea is to measure the SFR
in dwarf galaxies as a function of lookback time to see if they are
beginning to form stars toward low redshift as the ionizing background
decreases.  CANDELS F275W observations can detect dwarf galaxies at $1 <
z < 1.5$ forming stars at $>0.3 M_{\odot}$ yr$^{-1}$.

\subsection {Supernova Cosmology}\label{sec:sne}

Observations of high-redshift SNe~Ia provided the
first and most direct evidence of the acceleration of the scale factor
of the Universe (\citealt{1998AJ....116.1009R, 1999ApJ...517..565P}),
indicating the presence of  ``dark energy''.  Elucidating the
nature of dark energy remains one of the most pressing priorities of
observational cosmology.  \textit{HST} and ACS have played a unique role in
the ongoing investigation of dark energy by enabling the discovery
of 23 SNe~Ia at $z>1$, beyond the reach of ground-based telescopes
(\citealt{2004ApJ...607..665R, 2007ApJ...659...98R}).  From these data we
have learned that (1) the cosmic expansion rate was decelerating before it recently
began accelerating, a critical sanity test of the model; (2)
dark energy was acting even during this prior decelerating phase; (3)
SNe~Ia from 10 Gyr ago, spectroscopically and photometrically, occupy the
small range of diversity seen locally; (4) a rapid change is {\it not}
observed in the equation-of-state parameter of dark energy, $w = P/(\rho c^2)$,
though the constraint on the time variation, ${dw \over dz}$, remains
an order of magnitude worse than on the recent value of $w$; and (5) the
rate of SNe~Ia at $z>1$ declines, suggesting a nontrivial delay between
stellar birth and SN~Ia stellar death (\citealt{2008ApJ...681..462D}).

WFC3/IR opens an earlier window into the expansion history at
$1.5<z<2.5$, beyond the reach of prior \textit{HST}/ACS $z$-band SN searches
including GOODS and the follow-on PANS (Program 10189; PI A.~Riess). CANDELS 
will exploit this added reach in order to test the foundations of
SNe~Ia as distance indicators: the nature of their progenitor systems
and their possible evolution.  Simultaneously and in parallel, \textit{HST}
will continue to find lower-redshift SNe~Ia at $z>1$, which offer additional 
constraints on the time variation of $w$.  In Figure \ref{fig:sn_dndz},
we show the predicted redshift distributions of SNe~Ia detectable at $5\sigma$ 
with one-orbit \textit{HST} observations separated by 52 days (approximately
optimal for high-$z$ SN~Ia detection) in the filters $zYJH$.  CANDELS will
be searching at one-orbit depth with a combination of $J+H$ exposures.
\ifsubmode
 \placefigure{fig:sn_dndz}
\else
  \insertfigure{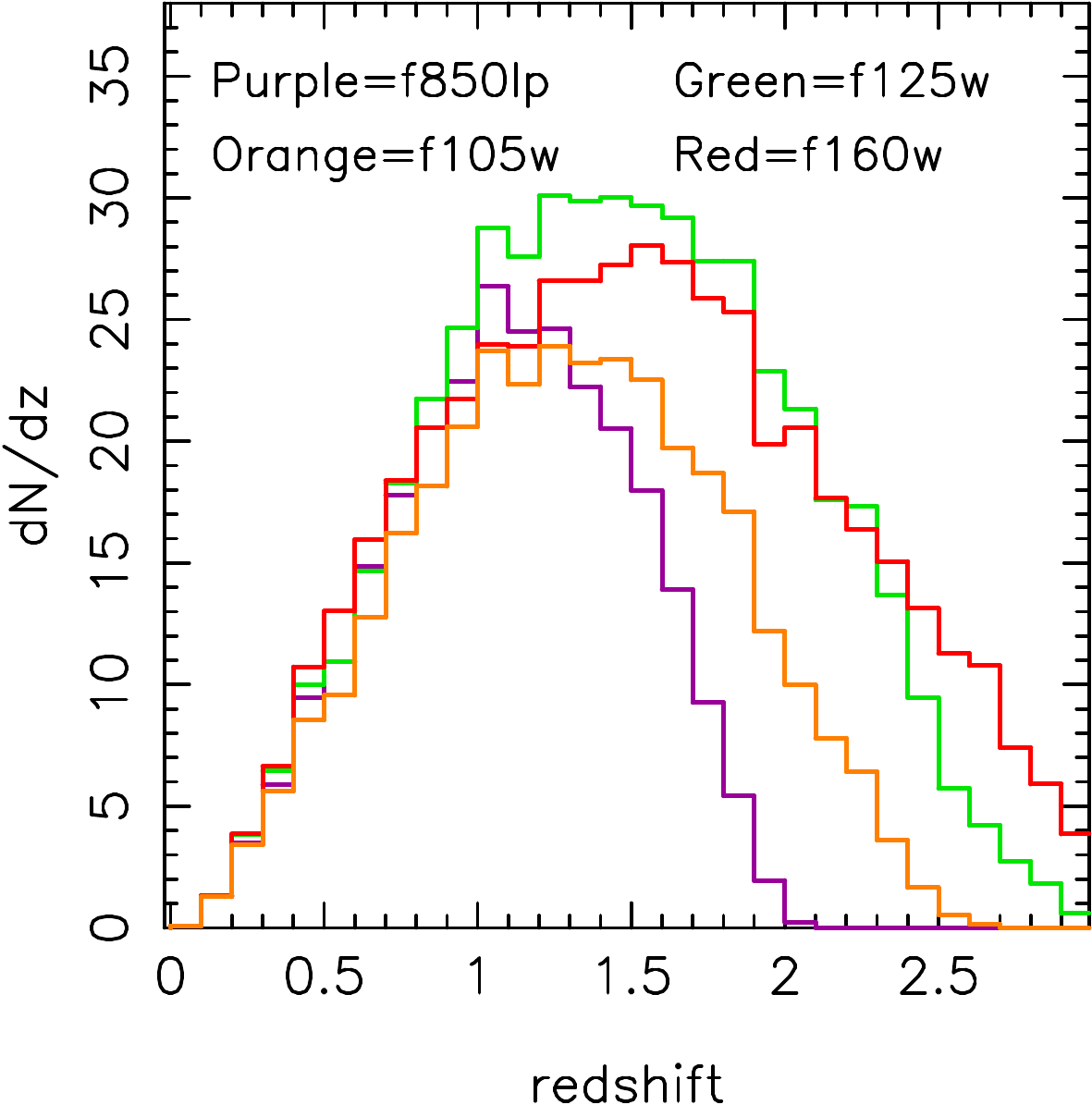}{\figsndndz}
\fi

High-redshift SNe~Ia continue to be a leading indicator on the nature of
dark energy.  The consensus goal is to look for $w_0 \neq -1$ or  ${dw
\over dz} \neq 0$, either of which would invalidate an innate vacuum energy
(i.e., cosmological constant) and would point towards a present epoch of
``weak inflation''.  Further, any discrepancy between the expansion history and the
growth history of structure expected for $w(z)$ would suggest that
general relativity suffers a scale-dependent flaw and might provide
guidance for the repair of this flaw.

The primary aim for the CANDELS SN program is assessing the evolution of 
SNe~Ia as distance indicators by observing them at redshifts $z > 1.5$,
where the effects of dark energy are expected to be insignificant
but the effects of the evolution of the SN~Ia white-dwarf progenitors
ought to be significant.  At present, the evolution of SNe~Ia as
distance indicators is the thorniest and most
uncertain contributor to the future dark-energy error budget. We will
attempt to make a {\it direct} measurement of SN evolution
(first suggested by \citealt{2006ApJ...648..884R}) that is independent
of the most uncertain aspects of the present cosmology.  By extending
the Hubble diagram of SNe~Ia to the fully dark-matter dominated epoch,
$1.5 < z < 2.5$, we can begin to distinguish the effects of dark energy
(which decays relative to dark matter as $(1+z)^{3w}$) from the 
evolution of SNe~Ia.  Since the change in distance with redshift by
$z>1.5$ depends primarily upon the matter density, a departure from the
cosmological concordance model seen among these earliest SNe~Ia --- {\it
perhaps the very first SNe~Ia} --- would directly constrain evolution
in SN~Ia distance measurements.  The near-IR capability of WFC3 
provides a timely opportunity for \textit{HST} to perform this critical test 
when results may still inform the design 
of future space-based dark energy missions.

The possibility that SNe~Ia evolve is motivated by the anticipated change
in the composition of their fuel with lookback time.   Depending on
the time elapsed between the formation of the progenitor star
and its SN (which may be 1--3 Gyr), the
formation metallicity of $z \approx 2$ SNe~Ia may be extremely low.
Although modelers do not agree on the sign of a metallicity
effect upon SN~Ia luminosity, they agree that it is one of the likeliest 
sources of evolution.

\ifsubmode
  \placefigure{fig:sn_rates}
\else
  \insertfigure{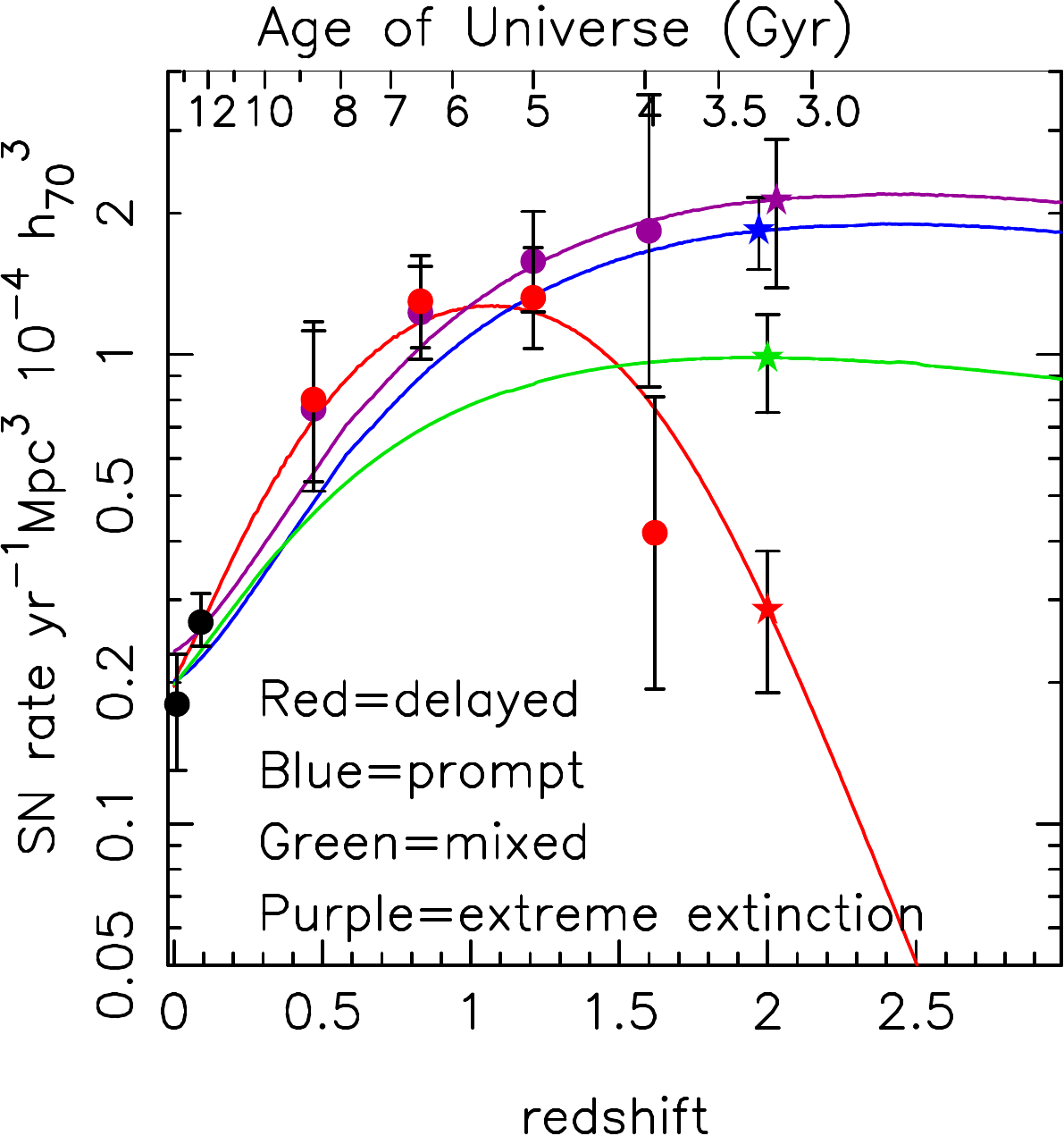}{\figsnrates}
\fi

At higher redshift, only more massive stars have time to evolve into
white dwarfs, providing another possible cause for SN~Ia evolution.
Shorter-lived, more massive stars produce white dwarfs with a
higher initial mass and a smaller ratio of carbon to oxygen.
Such white dwarfs, in a Type Ia explosion, yield less energy per 
gram and a decrease in the
$^{56}$Ni mass synthesized (e.g., \citealt{1980ApJ...237..111B}).
Figure~\ref{fig:sn_rates} (blue curve) shows the calculation from
\citet{2001ApJ...557..279D} that predicts a
change in the peak luminosity of roughly 3\% per 1~M$_\odot$ change in the
progenitor star mass.   At $z<1.5$, stars with masses 1--1.5~$M_\odot$
have time to evolve to white dwarfs and may produce the SNe~Ia we see.
At $z \approx 2$, the minimum mass rises to 2--6~$M_\odot$ for a formation
delay of a few Gyr (\citealt{2006ApJ...648..884R}).  Our program can
provide a direct, empirical constraint on the degree of such evolution.
A significant departure from the concordance model at $z>1.5$ would mean
that either SNe~Ia are evolving or that the cosmological model is even more
complex than assumed.

This first measurement of the SN~Ia rate at $z \approx 2$ can also help to
constrain progenitor models by measuring the offset between
the peak of the cosmic SFR and the peak of the cosmic
rate of SNe~Ia.  The CANDELS observing program will extend SN~Ia rate
measurements to $z \approx 2.5$, providing a rare clue to the nature
of SN~Ia progenitors \citep{2000astro.ph..9312R}\note{Ruiz-Lapuente \&
Canal 1999}.
Prior \textit{HST} surveys for SNe~Ia at $z>1$ suggest a long delay of $\sim 3$ 
Gyr between star formation and SN~Ia production
(\citealt{2004ApJ...613..189D, 2008ApJ...673..981K, 2004ApJ...613..200S}).
\note{Dahlen et al. 2004, Kuznetsova et al. 2008; Strolger et al 2004}
However, this finding is at odds with the correlation between SN~Ia rate and 
SFR locally, which requires an explanation.  
Whether there are two channels that
produce SNe~Ia (e.g., \citealt{2005A&A...433..807M}\note{
Mannucci et al. 2005}), or lower metallicity at high redshift reduces the
efficiency of accretion \citep{2009ApJ...707.1466K}\note{
Kobayashi \& Nomoto 2009}, or there is more host extinction at high redshift than
previously believed \citep{2008MNRAS.388..829G} are hypotheses we can test
with a sample extended in redshift.  Of particular interest is WFC3's
ability to distinguish between models dominated by prompt or delayed
channels because their predicted rates diverge significantly at $z>1.5$
(Fig.~\ref{fig:sn_rates}).  Models consistent with current data predict
that our present WFC3/IR search will yield anywhere from 8 (delayed model)
to 30 (prompt model) SNe at $z>1.5$. Also, searching the volume
at $0.7<z<1$ in the IR will reduce uncertainties in
host extinction by more than a factor of two over our previous {\it
optical-only} searches with \textit{HST}\@.  This allows us to determine 
whether dust is a factor in the declining high-$z$ SN~Ia rate.

A new SN~Ia sample at $z \gtrsim 1$ can also refine the only constraints we
currently have upon the time variation of the cosmic equation of state, following
the ambitious goal of more than doubling the strength of this crucial test of
a cosmological constant by the end of {\it HST}'s life.  Such SNe~Ia 
at $1<z<1.5$, the epoch when dark matter and dark energy were vying 
for dominance over expansion (i.e., $\Omega_M \approx \Omega_{\rm DE}$), can 
be discovered in the CANDELS ACS parallel images.  These SNe~Ia 
remain incredibly valuable, as
they presently offer a rare chance to learn more about the time variation
of $w$.  At the rate \textit{HST} collected $1<z<1.5$ SNe~Ia in Cycles 11--13 
(\citealt{2004ApJ...607..665R, 2007ApJ...659...98R}), a total of 100
could be collected during the final years of \textit{HST}, more than doubling the
precision of $dw \over dz$.

\subsection {Grism Observations}\label{sec:grism}

As part of the CANDELS and CLASH SN follow-up program, we obtain
relatively deep observations with a WFC3/IR grism, usually G141, to determine
the SN type and to measure a redshift of the SN host.
While optimized for the SNe, these observations can be used for
galaxy science as well and are likely to be among the deepest grism
observations available. Eight such follow-ups are presently
budgeted; a typical depth will be $\sim 23$\,ks.  \textit{HST}
slitless grism spectra in the near-IR have low resolution but
are highly competitive with ground-based near-IR spectra due to sensitivity,
low sky background, stable flux calibration, access to spectral regions blocked by the
atmosphere, and multiplex advantage.

The orientation of the observations is typically constrained by the need
to avoid contamination of the SN spectra by the host galaxy and other
galaxies in the field and by the limited roll angles available to \textit{HST}.
We attempt to divide the observations between two orientations to
provide redundancy against inter-object contamination.  Because
the spectra subtend only $<20\arcsec$ on the detector, there is latitude
to shift the center of the field to include other sources of interest
in the grism pointing.

Grism data are just now being obtained, and it is too early to state
well-defined scientific goals.  The following are among the science 
topics of interest:

{\bf High-redshift galaxies.} 
Spectra with the G141 grism cover the Ly$\alpha$ line in galaxies at 
$8.0<z<12.5$.  The $3\sigma$ line flux limit with eight orbits of
observations is $1.5 \times 10^{-17} \ergcmsqpersec$.   Based on the
$z=6.6$ LF of Ly$\alpha$ emitters (Ouchi et
al. 2010), these spectra will be sensitive to bright objects at
$\sim 3L^*$.  The volume probed in eight grism pointings is about
$16/\phi^*$.

{\bf Low redshift H$\alpha$ science.} G141 spectra cover the H$\alpha$ + [N~II]  lines
at redshift $0.7 < z < 1.5$. The line flux limit is $1.5 \times 10^{-17} \ergcmsqpersec$.
From the \citet{2009ApJ...696..785S} LF, we expect $\gtrsim 25$ 
$\Ha$ emitters per field. With eight follow-up fields, CANDELS should find
at least twice as many $\Ha$ emitters as NICMOS found in all the years of parallel
slitless observations. About 30\% of the galaxies will be 
in the redshift range $1.3 < z < 1.5$, where the spectra also 
include [O~III] and $\Hb$, and some fraction will have detectable
$\Hb$.  

$\Ha$ and $\NII$ are blended at the resolution of the G141
grism ($R \equiv \lambda/\Delta\lambda \approx 130$). However, the $\Ha$ luminosity can be corrected for
the $\NII$ contribution \citep[e.g.,][]{1997ApJ...475..502G}. With the corrected
$\Ha$ luminosity it will be possible to (i) constrain the faint end of the $\Ha$ LF;
(ii) compare $\Ha$ luminosity and mid-IR properties of galaxies, possibly as a
function of redshift; (iii) compare the slope of the UV with the Balmer ratio $\Ha/\Hb$,
and with the integrated IR luminosity, providing constraints on extinction and
on the initial mass function; (iv) measure the size of the star forming
region; and (v) for resolved galaxies, compare $\Ha$ and UV star-formation indicators
in star-forming clumps or subregions.

{\bf Continuum and [O~III] spectroscopy at $z \approx 2$.}  The G141 spectra
cover [O~III] at $1.2<z<2.3$ and the 4000\,\AA\ break at
$1.75<z<3.1$.  These spectra will determine redshifts for galaxies 
at $z \approx 2$, where spectroscopic redshifts have been difficult to
obtain, measure $L_{\rm [O~III]}$ for AGNs and star forming 
galaxies, and allow modeling of stellar
populations using spectral features and continuum breaks.  The latter
goal is challenging because absorption lines are diluted by the low
resolution of the grism but is achievable for bright objects 
\citep[e.g.,][]{2010ApJ...718L..73V}.

\section {Summary of Program Properties and Rationale}\label{sec:requirements}

We summarize here a list of key program properties that have been
designed to address the aforementioned science goals.

\subsection{CANDELS/Wide}  
The Wide survey is the prime tool for covering large area
and sampling rare and massive objects:

$\bullet$ {\it Filters:} $H$ (F160W) is the workhorse filter
for reaching rest-frame $B$ at redshifts up to  $\sim 2.5$.  
This is augmented with $J$ (F125W) to improve photometric redshifts
near $z \approx 2$ and identification of high-redshift
galaxies at $z \gta 6$.

$\bullet$ {\it Photometric depth, structural parameters:} 
The $5\sigma$ point-source
limits in $J$ and $H$
are $\sim 27.0$\,mag in each filter individually,
sufficient to reach $L^*$ in the luminosity
function at $z \approx 7$, 
count all objects regardless of color
down to $4\times10^{9.0}$ M$_{\odot}$
($H = 26.5$\,mag)
at $z \approx 2$, and 
obtain high-quality structural information 
(concentrations, axis ratios, bulge-disk ratios, Gini/M$_{20}$) down to
$4\times10^{10}$ M$_{\odot}$ ($H = 24.0$\,mag) at $z \approx 2.5$.  

$\bullet$ {\it Area, geometry, counts, and cosmic variance:} 
A total of 0.22~deg$^2$ (including
both Deep and ERS) is imaged
in five separate fields to maximize
sample size and overcome cosmic
variance (CV).  Roughly 100 objects are expected
above $L^*$ at $z \approx 8$ (per $\Delta z = 1$) and
$\sim 300$ massive galaxies above $10^{11}$ M$_{\odot}$ at $z
\sim 2$ (per $\Delta z = 1.0$).  
Table \ref{tab:tab_cv} gives expected fractional
CV counting errors for objects with various
bias values and redshifts. If a goal is to measure 
a factor-of-two change in the
co-moving space density of rare objects
between adjacent redshift bins to $3\sigma$
accuracy, the required counting error is
$<$23\% per bin.   We show that this requirement
is more than met for bias factors $ b < 5.5$ at $z \approx
2$ (for $\Delta z = 0.75$) and $ b < 15$ at $z = 7$ 
(for $\Delta z = 1$).  These bias factors apply to
highly clustered, massive galaxies --- typical galaxies
should have even lower errors (see \S\ref{sec:cosmicvariance}).  
The Wide fields are long and thin
to decorrelate clustering on the long
axis to further reduce CV\@.  Observing the same area 
as CANDELS over a single
field would yield CV errors 
that are 50\% higher than our design.

$\bullet$ {\it Number of high-z galaxies:} A prime goal of Wide is to
return a rich sample of {\it luminous} candidate
galaxies for future follow-up.  The 0.22 deg$^2$ 
field size will contain a
total of $\sim 20$ high-$z$ 
galaxies at $z > 6.5$ to 25.6\,mag and 300 galaxies to 27\,mag 
(Table \ref{tab:numbers_of_galaxies}).

$\bullet$ {\it Number of mergers:} Assuming a 10\%
merger fraction, we expect to detect 
$\sim 150$ major mergers per $\Delta z$ bin = 0.5 
at $z \approx 2$, 
sufficient to detect a 35\% change in the merger
rate from $z = 2.5$ to $z=1.5$  to $3\sigma$ accuracy.

$\bullet$ {\it Correlation functions and environments:} 
Correlation functions are an important tool for 
estimating the halo masses of faint galaxies.
Each Wide region should be $\gta$2 correlation 
lengths across, which is 10 comoving Mpc at high redshift
(Lee et al.~2009).  The Wide 
field widths of $~\sim 8\arcmin$ span  11\,Mpc
at $z \approx 2$ and 23\,Mpc at $z \approx 8$,
meeting this requirement.   The long dimensions of 22--30\arcmin \ are
comfortably a factor of 4--5 longer.

\subsection{CANDELS/Deep}
The Deep program (in both
GOODS fields)
adds important depth for
fainter galaxies plus the extra $Y$ filter color information:

$\bullet$ {\it Filters:} The addition
of $Y$ to $J$+$H$ strengthens
the detection and redshifts
of high-$z$ galaxies and improves
photoz's for galaxies at $z \approx 2$.  
With $BViz$ added from ACS, the \Hubble\ 
filter suite covers the spectrum densely from 400--1800 nm,
providing accurate color
gradients and 
rest-frame $B$-band structural parameters
seamlessly for $z \lesssim 2.5$.

$\bullet$ {\it Photometric depth:} 
Four-orbit exposures in $Y$, $J$, and $H$
reach a 5$\sigma$ limit of $>27.7$\,mag in each filter, 
sufficient to reach 0.5$L^*$ at $z \approx 8$.  The same data provide
detailed structural parameters to $H = 24.7$\,mag
($2\times10^{10}$ M$_{\odot}$) at $z \approx 2.5$, 
covering even low-mass galaxies fading
to the red sequence.

$\bullet$ {\it Area, counts, and cosmic variance:} 
A total area of 0.033 deg$^2$ divided into two separate
regions is needed to attain sufficient sample size
and limit cosmic
variance for typical-to-small galaxies at all redshifts.
Roughly 200 galaxies are expected in the Deep regions above 
$10^{10}$ M$_{\odot}$ at $z \approx 2$ per $\Delta z = 0.5$,
and $\sim 200$ galaxies above
$0.5 L^*$ at $z \approx 8$ per $\Delta z = 1$.
Expected cosmic variance uncertainties are
20\% per $\Delta z = 1$ for very massive galaxies
at $z \approx 7$ but three times smaller for less biased (less
massive) objects (see \S\ref{sec:cosmicvariance}). 
Observing only $\sim 40$\% of each GOODS field in the Deep
survey increases CV uncertainty by only 
25\% compared to two entire GOODS fields
and saves more than a factor of two in observing time.

$\bullet$ {\it Galaxy environments:} 
With a comoving size of $\sim 3\times 5$\,Mpc at $z = 2$,
each Deep region is marginal for correlation functions but big enough for
Nth-nearest-neighbor density measurements.

\section{The CANDELS Fields}\label{sec:fields}  
 
To address our diverse set of science goals, the CANDELS survey consists
of a two-tiered Wide+Deep survey.  The CANDELS/Deep survey covers $\sim
130$ arcmin$^2$ to $\sim 10$-orbit depth within the GOODS-N and GOODS-S
Fields, while the CANDELS/Wide survey covers a total of $\sim 720$
arcmin$^2$ to $\sim 2$-orbit depth within five fields, namely GOODS-N,
GOODS-S, EGS, COSMOS, and UDS.  The use of five widely separated fields
mitigates cosmic variance (see \S\ref{sec:cosmicvariance}) and our
``wedding-cake'' approach will yield statistically robust samples of
both bright/rare and faint/common extragalactic objects.  When combined
with the existing \Hubble\ Ultra Deep Field \citep{2006AJ....132.1729B}
within GOODS-S, the CANDELS Wide and Deep surveys will document galaxy
demographics over the widest feasible ranges of mass, luminosity,
and redshift.

The five CANDELS fields were chosen because the breadth and depth of the
ancillary data available in these fields far exceeds what is available
in other regions.
Highlights include the largest spectroscopic redshift surveys (DEEP2+3,
zCOSMOS), the deepest \textit{Hubble}/ACS imaging (GOODS-S, GOODS-N), ground
optical photometry (COSMOS) and near-IR photometry (UKIDSS/UDS), the deepest
X-ray observations (GOODS-S, GOODS-N, EGS), and the deepest VLA radio
data (GOODS-N).  Furthermore, all five fields are the targets of the \Spitzer\ 
Extended Deep Survey (SEDS; Fazio et al., in prep.), which aims to cover these
fields with deep \Spitzer/IRAC 3.6\,$\mu$m and 4.5\,$\mu$m imaging to a total
depth of 12\,hr per pointing. 

Each CANDELS pointing is located at a ``sweet spot'' within the parent region
where the multi-wavelength data are best.  In this section we describe in
greater detail the data available in each of the CANDELS Deep and Wide fields
and our survey strategy in each region, including the precise location and
layout of CANDELS WFC3 and ACS imaging.  We end the section with a discussion
on how our survey strategy helps mitigate cosmic variance compared to 
simpler scenarios.  We defer the particulars of the observing strategy in each
field to \S\ref{sec:observing_strategy}.  

\subsection{CANDELS Deep Fields}

The CANDELS/Deep fields consist of the central portion of GOODS-S and
GOODS-N\@.  The primary constraint on the exact size and location of the
WFC3/IR mosaics in these fields was overlap with ancillary data. In the
following subsections we discuss the existing data in each field and present
our mosaic layouts. 

\subsubsection{The GOODS-South Field}

The GOODS-S field is a region of sky located near the southern \textit{Chandra} Deep
Field \citep{2002ApJS..139..369G} that has been targeted for some of the deepest
observations ever taken by NASA's Great Observatories \Hubble, \Chandra, and \Spitzer.
The field is centered at $\alpha$(J2000) = $03^{\rm h}$ $32^{\rm m}$ $30^{\rm s}$ and
$\delta$(J2000) = $-27^{\circ}$ $48^{\prime}$ $20^{\prime\prime}$.
The GOODS region of the field has been imaged in the optical with \textit{Hubble}/ACS in the $B, V,
i$, and $z$-bands as part of the GOODS \textit{Hubble} Treasury Program
(P.I.~Giavalisco).  This imaging covers a region
$10^{\prime}\times16^{\prime}$ in size and reaches $5\sigma$ depths of
$B=28.0, V=28.0, i=27.5$ and $z=27.3$ (extended source). 
The field has also been imaged in the mid-IR (3.6--24\,$\mu$m) as part of the
GOODS \textit{Spitzer} Legacy Program; this imaging reaches $5\sigma$ depths of
0.13, 0.22, 1.44, and 1.6\,$\mu$Jy in the IRAC 3.6, 4.5, 5.8, and 8.0\,$\mu$m
bands, respectively, and 20\,$\mu$Jy in the MIPS 24\,$\mu$m band.
Furthermore, the GOODS-S region has the deepest
\textit{Chandra} observations ever taken, which now have a total integration time
of 4\,Ms.

\ifsubmode
  \placefigure{fig:gds}
\else
  \insertfigurewide{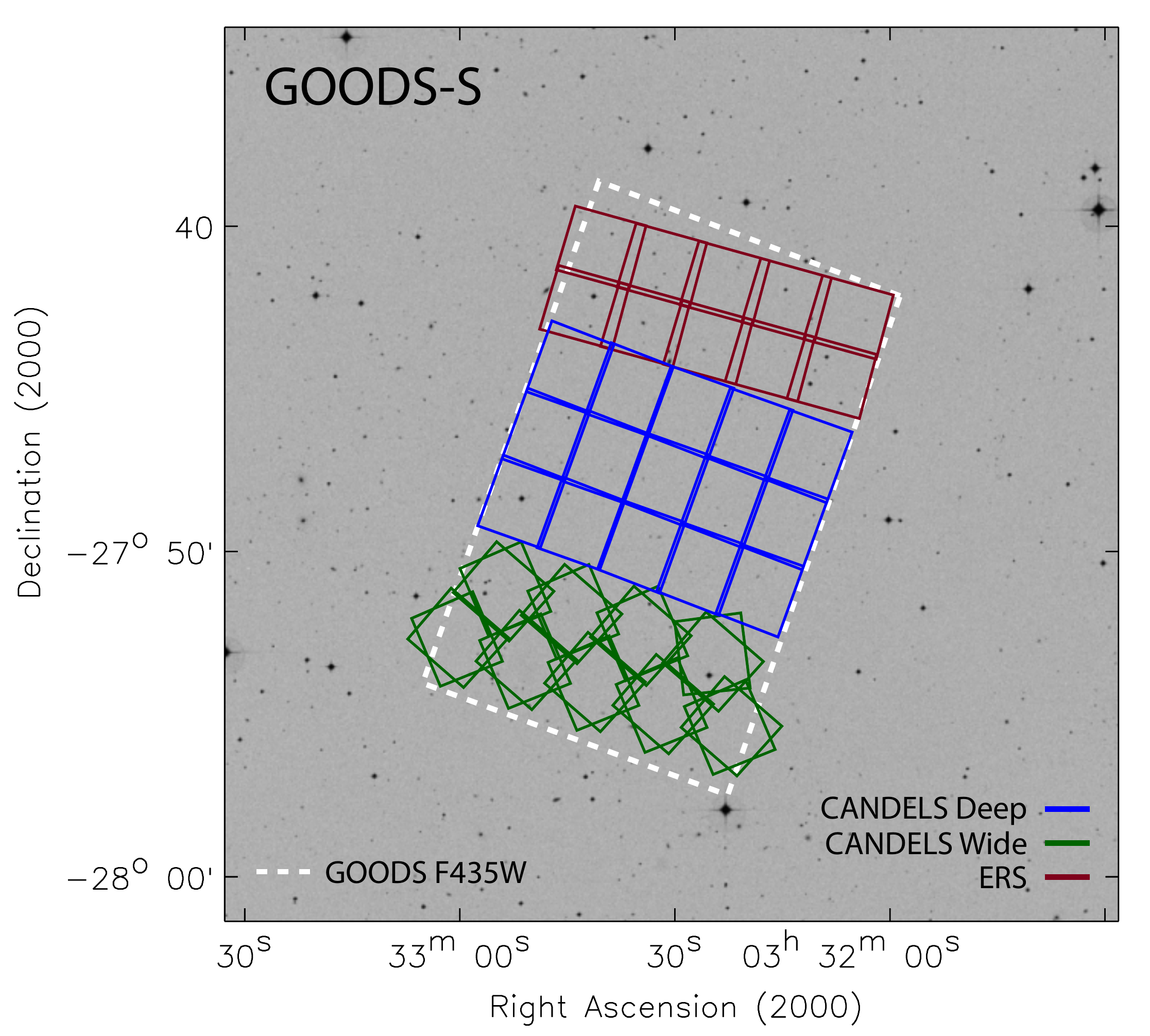}{\figgds}
\fi

Additional multi-wavelength data in the GOODS-S field include $U$-band imaging
obtained with VIMOS on the \textit{Very Large Telescope} \citep[VLT;][]{2009ApJS..183..244N},
near-IR imaging in the $Y$ band with HAWK-I \citep{2010A&A...511A..20C} and
in the $J, H,$ and $K$-bands with ISAAC \citep{2010A&A...511A..50R} on the
\textit{VLT}, additional mid-IR imaging with \textit{Spitzer}/IRAC 
(3.6--8.0\,$\mu$m) from the SEDS program and \textit{Spitzer}/MIPS (24\,$\mu$m) from the
Far-Infrared Deep Extragalactic Legacy survey (FIDEL; PI.~M.~Dickinson),  
submillimeter observations (870\,$\mu$m) taken with LABOCA on the \textit{Apex}
telescope \citep{2009ApJ...707.1201W}, and radio imaging (1.4 and 4.8\,GHz) from the
\textit{Very Large Array} \citep[\textit{VLA};][]{2008ApJS..179...71K}.  Furthermore,
the field has been the subject of numerous spectroscopic surveys \citep[e.g.,][]{2004A&A...428.1043L,2004ApJS..155..271S,2005A&A...437..883M,2005A&A...434...53V,2007A&A...465.1099R,2008A&A...482...21C,2009A&A...494..443P,2010A&A...512A..12B,2010ApJS..191..124S} 

The CANDELS/Deep WFC3/IR observations in the GOODS-S field consist of a
rectangular grid of $3\times5$ WFC3/IR tiles ($\sim 6\farcm{8}\times
10^{\prime}$) that are oriented at a position angle of $70^{\circ}$ (East of
North, see Fig.~\ref{fig:gds}). The spacing intervals of each tile
in the grid is designed to allow for maximal contiguous coverage in WFC3/IR
without introducing gaps between tiles as a result of pointing errors.
The orientation and location of the mosaic were chosen to overlap with the
deepest portion of the existing \textit{Spitzer}/IRAC imaging in the field,
which, due to an overlap strip, peaks in the center of the GOODS region and
is matched in orient and size with the CANDELS/Deep mosaic.
When combined with the CANDELS/Wide observations in the field (see \S\ref{sec:southnorthwide})
and the imaging from the ERS program
\citep[ERS;][]{2011ApJS..193...27W}, the entire
$10^{\prime}\times16^{\prime}$ GOODS region will be imaged by WFC3/IR\@.  An
overview of the resulting WFC3/IR coverage in GOODS-S from all of these observations is shown in Figure
\ref{fig:gds}.  Due to the observational requirements of our various
science programs, the actual observing strategy in GOODS-S is quite
complex.  This is described in \S\ref{sec:deepstrategy}.

\subsubsection{The GOODS-North Field}

The GOODS-N field is located near the northern \Hubble\ 
Deep Field \citep{1996AJ....112.1335W} and like its southern counterpart has been
extensively observed by NASA's Great Observatories.
The field is centered at $\alpha$(J2000) = $12^{\rm h}$ $36^{\rm m}$ $55^{\rm s}$ and
$\delta$(J2000) = $+62^{\circ}$ $14^{\prime}$ $15^{\prime\prime}$.
The GOODS region of the field has been imaged in the optical with \textit{Hubble}/ACS in the $B, V,
i$, and $z$-bands as part of the GOODS \textit{Hubble} Treasury Program.  As in
the GOODS-S field, this imaging covers a region
$10^{\prime}\times16^{\prime}$ in size and reaches $5\sigma$ depths of
$B=28.0$, $V=28.0$, $i=27.5$, and $z=27.3$\,mag (extended source). 
The field has also been imaged in the mid-IR (3.6--24\,$\mu$m) as part of the
GOODS \textit{Spitzer} Legacy Program; this imaging reaches $5\sigma$ depths of
0.58, 0.55, 1.09, and 1.20\,$\mu$Jy in the IRAC 3.6, 4.5, 5.8, and 8.0\,$\mu$m
bands, respectively, and 20\,$\mu$Jy in the MIPS 24\,$\mu$m band.
Furthermore, the field has been targeted for deep \textit{Chandra} observations
that have a total integration time of 2\,Ms \citep{2003AJ....126..539A}.
 
\ifsubmode
  \placefigure{fig:gdn}
\else
  \insertfigurewide{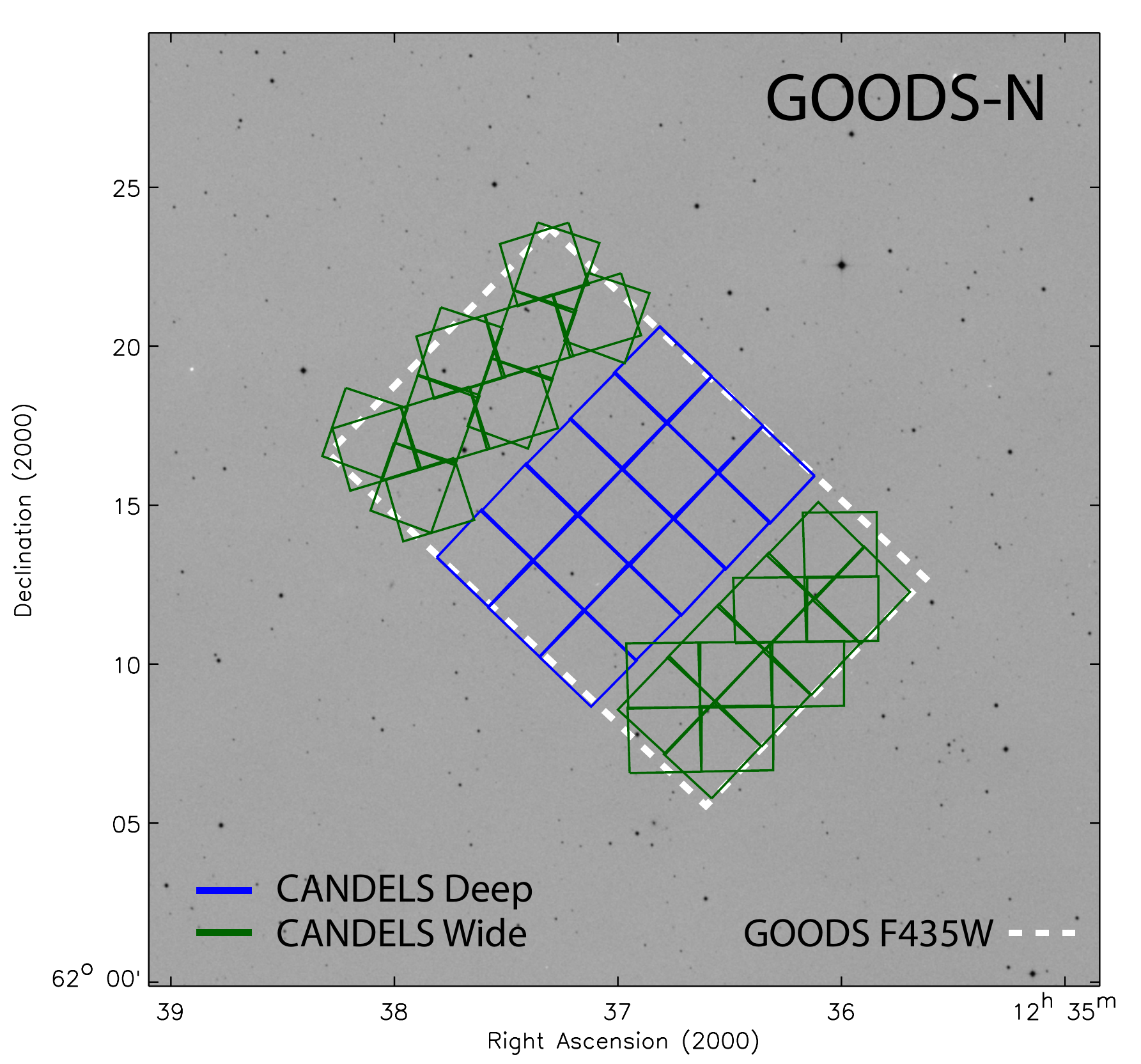}{\figgdn}
\fi

Additional multi-wavelength data in the GOODS-N field include
$U$-band imaging obtained with the MOSAIC camera on the KPNO Mayall
4m telescope \citep{2004AJ....127..180C}, near-IR imaging in the
$HK^{\prime}$-band taken with the QUIRC camera on the University of
Hawaii 2.2~m telescope \citep{2004AJ....127..180C}, additional mid-IR
imaging with \textit{Spitzer}/IRAC (3.6--8.0\,$\mu$M) from the SEDS
program and \textit{Spitzer}/MIPS (24\,$\mu$m) from the FIDEL survey
and deep radio imaging (1.4 GHz) from the \textit{Very Large Array}
\citep[\textit{VLA};][]{2010ApJS..188..178M}.  Furthermore, the field
has been the subject of numerous spectroscopic surveys
\citep[e.g.][]{2003ApJ...592..728S,2004AJ....127.3121W,2005ApJ...622L...5T,
2005ASPC..342..471S,2006ApJ...653.1004R,2008ApJ...689..687B,
2010ApJ...718..112Y,2010ApJ...714L.118D,2011ApJS..193...14C}.
 
The CANDELS/Deep WFC3/IR observations in the GOODS-N field consist of a
rectangular grid of $3\times5$ WFC3/IR tiles ($\sim 6\farcm{8}\times
10^{\prime}$) that are oriented at a position angle of $45^{\circ}$ (East of
North, see Fig.~\ref{fig:gdn}). As in GOODS-S, the orientation and
location of the WFC3/IR mosaic were chosen to overlap with the
deepest portion of the existing \textit{Spitzer}/IRAC imaging in the field,
which, due to an overlap strip, peaks in the center of the GOODS region and
is matched in orient and size with the CANDELS/Deep mosaic.
When combined with the CANDELS/Wide observations in the field 
(see \S\ref{sec:southnorthwide}),
nearly the entire $10^{\prime}\times16^{\prime}$ GOODS region will be
imaged by WFC3/IR\@.  An overview of the resulting WFC3/IR coverage in
GOODS-N from all of these observations can be seen in Figure
\ref{fig:gdn}.  The actual observing strategy in GOODS-N is quite
complex and is described in \S\ref{sec:deepstrategy}.

\subsection{CANDELS Wide Fields}\label{sec:fieldswide}

The CANDELS/Wide fields consist of COSMOS, UDS, EGS and three smaller regions
that bracket the CANDELS/Deep regions in GOODS-S and GOODS-N.
In the three primary Wide fields (COSMOS, UDS, and EGS), we have chosen a
long-thin geometry for the WFC3/IR observations in order to: i.) decorrelate
clustering on the long axis, further reducing cosmic variance, ii.)
maximize the area with overlapping and contemporaneous ACS/WFC parallel
observations on top of the WFC3/IR mosaic\footnote{Since the ACS/WFC data are
taken in parallel, they overlap different tiles than those for which WFC3/IR is
prime.}, and iii.) increase the total lag over which correlation functions
can be measured.

The precise location and orientation of the WFC3/IR mosaics on the sky were
chosen based upon the location of ancillary data in each field and upon telescope
scheduling constraints, namely roll angle visibility.  The telescope's roll angle
determines the orientation of the WFC3/IR exposures and ultimately the
placement of the ACS/WFC parallel exposures, which are offset from the prime
exposures by roughly $6^{\prime}$.  The visibility of a given roll angle is
important because each CANDELS/Wide field is observed over two epochs
separated by $\sim 50$ days in order to facilitate the search for high-$z$ SNe~Ia.  
Therefore, to ensure a mosaic is repeated at the same
orientation in each epoch, the observed roll angle must be accessible to the
telescope for at least $\sim 50$ days and across both epochs of observation.
This requirement severely limits the available observing windows.  
In the following subsections we discuss the existing
data in each of the CANDELS/Wide fields and present our mosaic layouts.

\ifsubmode
  \placefigure{fig:uds_layout}
\else
  \insertfigurewide{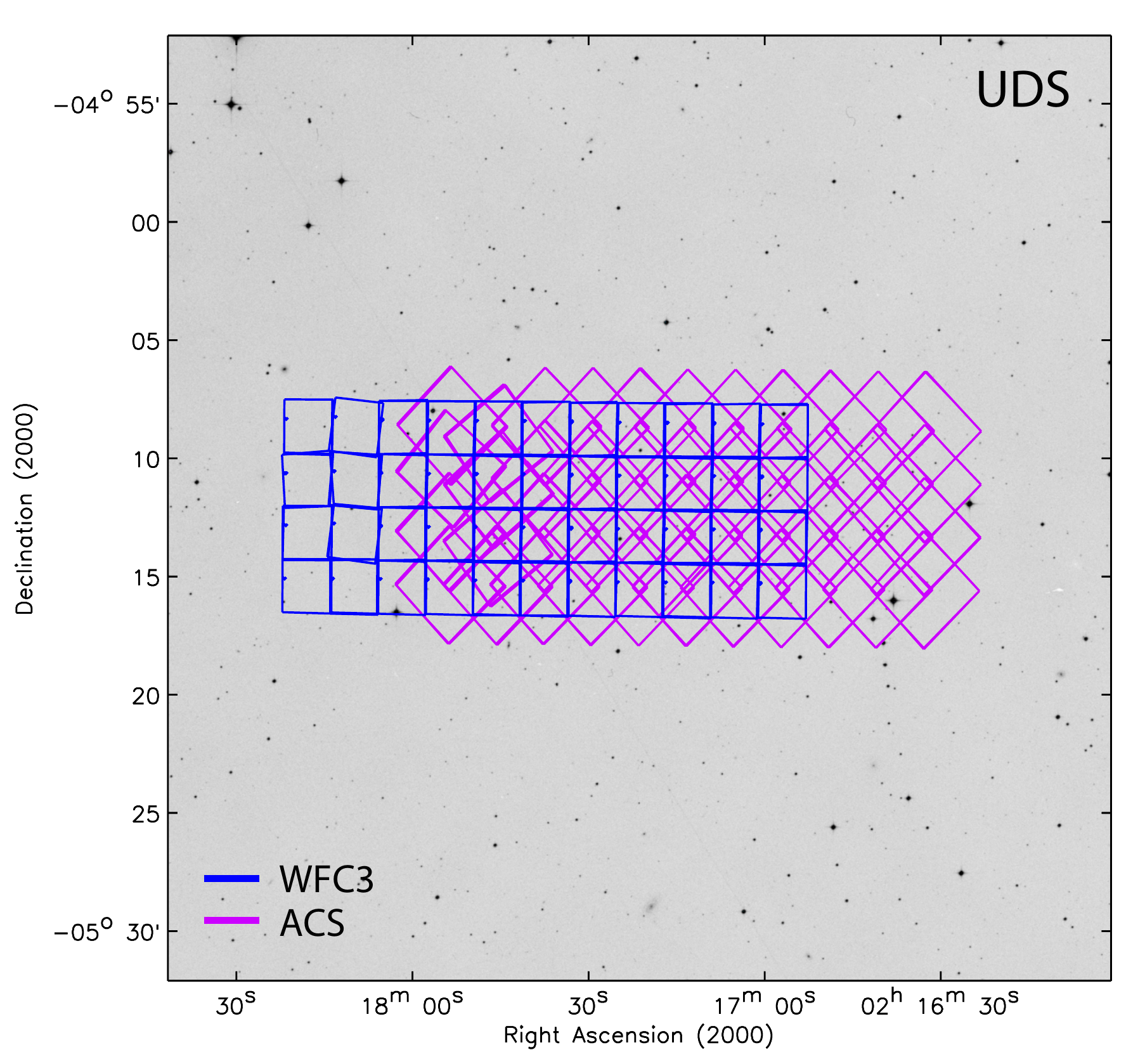}{\figudslayout}
\fi

\subsubsection{The UDS Field}

The UDS is the deepest component of the UKIRT Infrared Deep Sky Survey
\citep[UKIDSS;][]{2007MNRAS.379.1599L}.  The full UDS field covers 0.8 sq degrees on
the sky and is centered at $\alpha$(J2000) = $02^{\rm h}$ $17^{\rm m}$ $49^{\rm s}$ and
$\delta$(J2000) = $-05^{\circ}$ $12^{\prime}$ $02^{\prime\prime}$.  The field's
defining characteristic is deep \textit{J}, \textit{H}, and 
\textit{K} imaging obtained with the \textit{UKIRT} Wide Field Camera (WFCAM).
When completed in 2012, this imaging will reach $5\sigma$
depths of $J=25.2, H=24.7$, and $K=25.0$ ($2\arcsec$-diameter
aperture)\footnote{Current depths reached in the UKIDSS DR8 release are
  $J=24.9, H=24.2$, and $K=24.6$ ($2\arcsec$-diameter aperture).}.

Optical imaging in the field consists of deep \textit{Subaru}/SuprimeCam (Furusawa et
al.~2008) observations in the $B, V, R, i$, and $z$-bands, which reach $3\sigma$
depths of $B=28.4, V=27.8, R=27.7, i=27.7$ and $z=26.6$ ($2^{\prime\prime}$-diameter aperture).
Additional multi-wavelength data include $U$-band imaging
obtained with MegaCam on the \textit{Canada-France-Hawaii Telescope}
(CFHT; P.I.~O.~Almaini), mid-IR (3.6--24 $\mu$m) observation taken with IRAC and
MIPS on \textit{Spitzer} as part of the Spitzer Ultra-Deep
Survey (SpUDS; P.I.~J.~Dunlap) and the SEDS program, X-ray observations taken as
part of the \textit{Subaru/XMM-Newton} Deep Survey \citep[SXDS;][]{2004naoj.book...40S};
 and deep radio imaging (1.4\,GHz) from the \textit{VLA} \citep{2006MNRAS.372..741S}.  Furthermore,
the UDSz program (P.I.~O.~Almaini) is carrying out a large redshift survey of
the field using the VIMOS and FORS2 spectrographs on the \textit{VLT}. 

The CANDELS WFC3/IR observations in the UDS field consist of a rectangular grid of
$4\times11$ WFC3/IR tiles ($\sim 8\farcm{6}\times 23\farcm{8}$) that run
East-to-West at a position angle of $-90^{\circ}$ (see Fig.~\ref{fig:uds_layout}).
As with the CANDELS/Deep mosaics, the spacing intervals of each tile in the
grid is designed to allow for maximal contiguous coverage in WFC3/IR without
introducing gaps between tiles as a result of pointing errors.  The exposures
are all oriented so that the ACS/WFC parallels are offset along the long axis
of the mosaic, thereby producing a similar-sized mosaic overlapping the bulk
of the WFC3/IR mosaic except at its ends, where some tiles are covered only
by WFC3/IR or by ACS/WFC and not by both.

The orientation and precise location of the mosaic was chosen with the
following considerations in mind: i.) overlap with the SEDS IRAC imaging, ii.)
alignment with the VIMOS spectroscopic observations, iii.) guide star
availability, which ultimately shifted the entire mosaic $2^{\prime}$ to the
west, and iv.) telescope roll angle visibility.  The chosen position angle of
$-90^{\circ}$ East-of-North fulfills the first and second requirements and it also
corresponds to a telescope orient that can be held constant over both
observing epochs.  The layout of the mosaic and its
position within the UDS field are shown in Figure \ref{fig:uds_layout}.

\ifsubmode
  \placefigure{fig:cosmos_layout}
\else
  \insertfigurewide{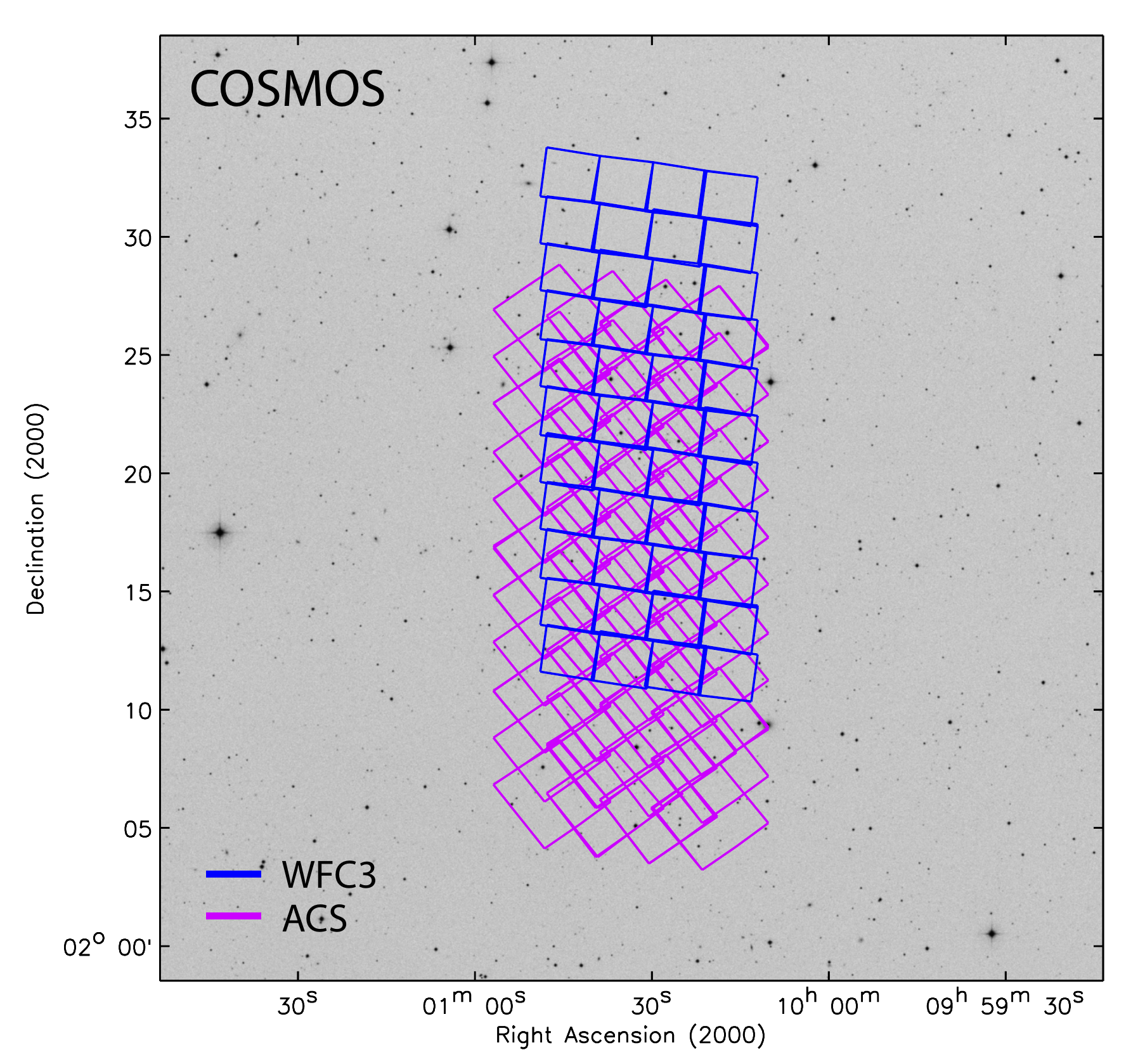}{\figcosmoslayout}
\fi

\subsubsection{The COSMOS Field}

The COSMOS field is a 2-degree region of sky surveyed with the \textit{Hubble}/ACS in the
F814W band by the COSMOS Treasury Program \citep{2007ApJS..172....1S,2007ApJS..172..196K}.  The field
is centered at $\alpha$(J2000) = $10^{\rm h}$ $00^{\rm m}$ $28^{\rm s}$ and
$\delta$(J2000) = $+02^{\circ}$ $12^{\prime}$ $21^{\prime\prime}$ and has been
extensively surveyed at a variety of wavelengths.  Ground-based optical
imaging of field includes deep \textit{Subaru}/SuprimeCam observations in the $B,
g^{\prime}, V, r^{\prime}, i^{\prime}$, and $z^{\prime}$-bands, which reach
$3\sigma$ depths of $B=27.8, g^{\prime}=27.2, V=27.1, r^{\prime}=27.2,
i^{\prime}=26.8$, and $z^{\prime}=25.9$ ($2^{\prime\prime}$-diameter
aperture).  
In the near-IR, the Ultra-VISTA project (P.I.~J.~Dunlop) is imaging the
central region of the field in the $Y$, $J$, $H$, and $K_{s}$ bands and when
completed will reach $5\sigma$ depths of $\sim 25.5$ mag.  The field is also
being observed in five medium-band filters, which cover a wavelength range
of 1--1.8$\mu$m using the NEWFIRM camera \citep[P.I.~P.~van Dokkum;][]{2009PASP..121....2V} on the Kitt Peak 4\,m telescope.

Additional multi-wavelength data in the COSMOS field include
X-ray observations with \textit{XMM-Newton} \citep{2007ApJS..172...29H}, UV
observations with GALEX \citep{2007ApJS..172..468Z}, mid-IR (3.6--24 $\mu$m)
observations with \textit{Spitzer} \citep{2007ApJS..172...86S},
submillimeter observations from the \textit{Caltech Submillimeter Observatory}
\citep[CSO;][]{2007AAS...211.8925A}, and radio observations with the \textit{VLA} \citep{2004AJ....128.1974S,2007ApJS..172...46S}. Furthermore, the zCOSMOS program \citep{2007ApJS..172...70L} is
carrying out a large spectroscopic redshift survey of the field using the
VIMOS spectrograph on the \textit{VLT}, which will ultimately yield spectra for over
35,000 galaxies out to $z \approx 2.5$. 

The CANDELS WFC3/IR observations in the COSMOS field consist of a rectangular grid of
$4\times11$ WFC3/IR tiles ($\sim 8\farcm{6}\times 23\farcm{8}$) that run
North-to-South at a position angle of $180^{\circ}$ (see Fig.~\ref{fig:cosmos_layout}).
The location of ancillary data in the COSMOS field constrained the location
and orientation of the WFC3/IR mosaic to a greater degree than in the UDS field.
The mosaic is positioned to lie within the overlap region of the mid-IR SEDS
IRAC imaging, the medium-band NEWFIRM observations, and the near-IR
Ultra-VISTA imaging. Unlike the UDS field, the telescope roll angle required
to keep the WFC3/IR exposures aligned with the long axis of the mosaic is not
visible throughout the two COSMOS observing epochs.  To find a stable
roll angle, we have rotated each WFC3/IR tile by $-8^{\circ}$ (East of North)
relative to the mosaic's position angle and used a slipped-lattice raster to
maintain the North-South orientation for the overal mosaic. Due to the larger
ACS/WFC footprint relative to WFC3/IR, the bulk of the ACS/WFC parallel
exposures still overlap the WFC3/IR mosaic even with this rotation. The
layout of the mosaic and its position within the COSMOS field are shown in
Figure \ref{fig:cosmos_layout}.

\ifsubmode
  \placefigure{fig:egs_layout}
\else
  \insertfigurewide{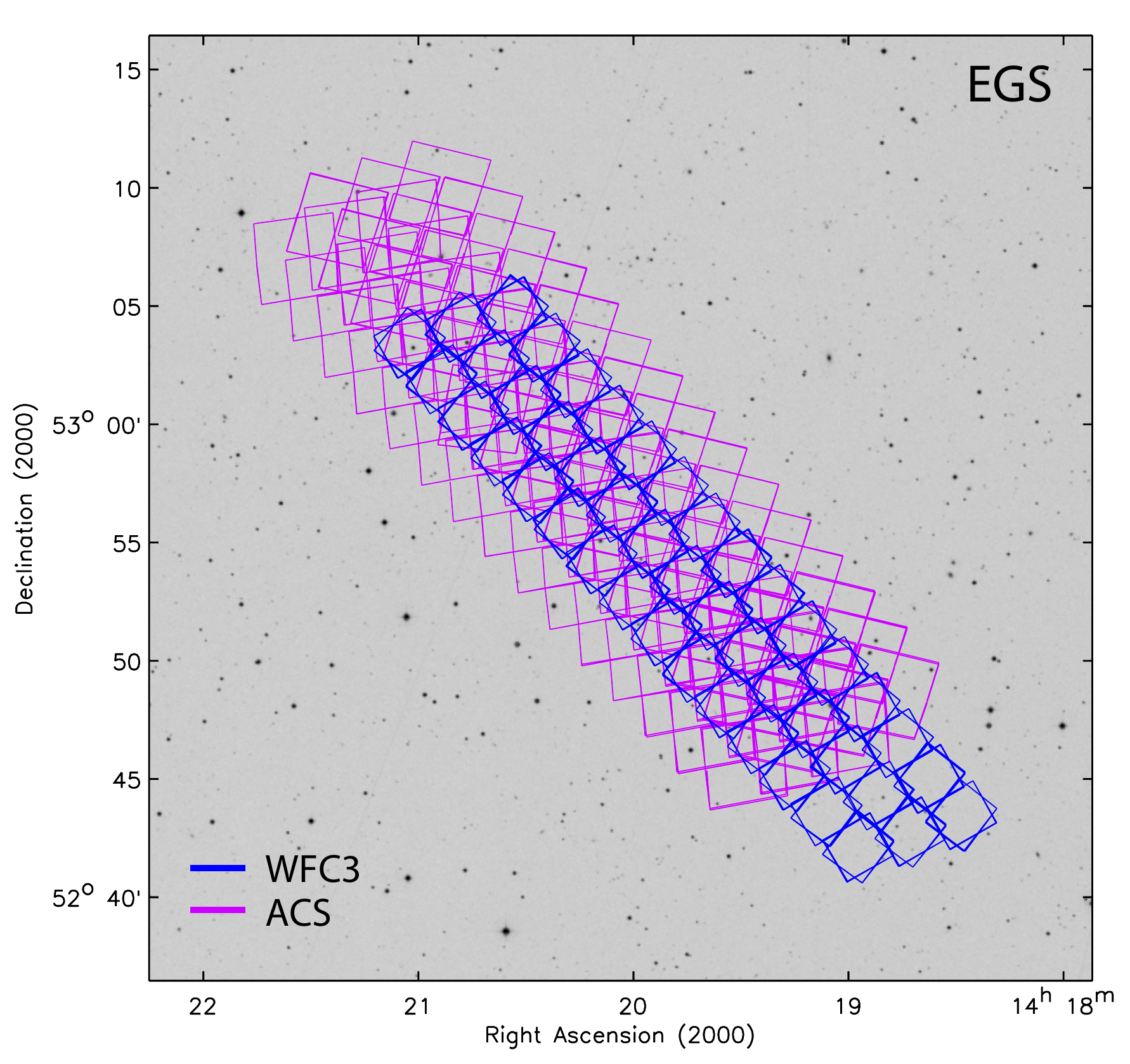}{\figegslayout}
\fi

\ifsubmode
  \placefigure{fig:egs_layout}
\else
  \insertfigurewide{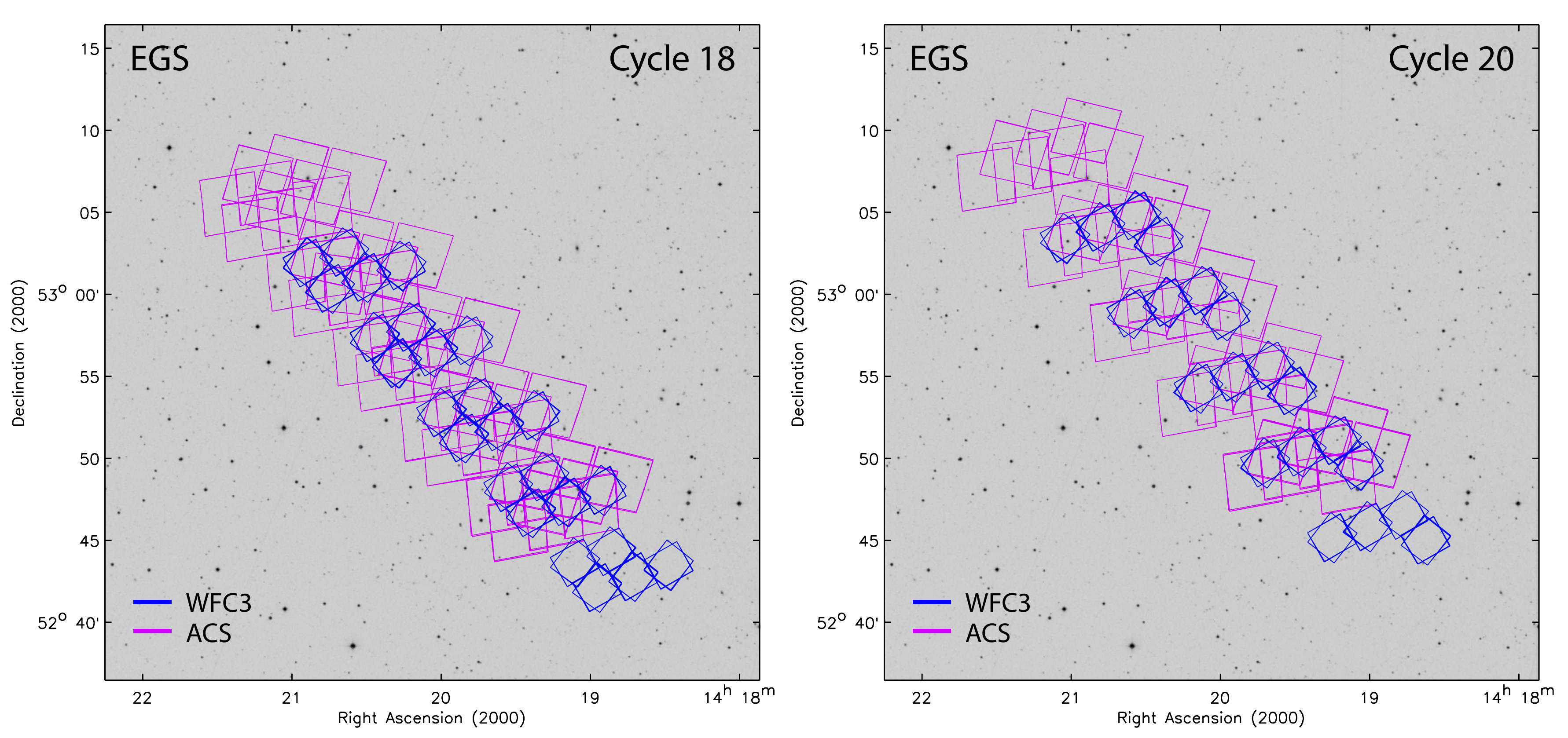}{\figegscombo}
\fi

\subsubsection{The EGS Field}

The EGS is a region of sky that has been extensively studied by the
All-wavelength Extended Groth strip International Survey \citep[AEGIS;][]{2007ApJ...660L...1D}.
The field is centered at $\alpha$(J2000) = $14^{\rm h}$ $17^{\rm m}$ $00^{\rm s}$ and
$\delta$(J2000) = $+52^{\circ}$ $30^{\prime}$ $00^{\prime\prime}$ and owes
its name to an early \textit{Hubble}/WFPC2 survey carried out by the WFPC team
\citep{2000ApJ...536...79R}.  More recently the field has been imaged with 
ACS/WFC in the F606W and F814W bands (P.I.~M.~Davis), reaching $5\sigma$
depths of 28.75 and 28.1 mag, respectively.  Ground-based optical imaging of
the field includes deep CFHT/MegaCam observations in the $u^{*}, g^{\prime},
r^{\prime}, i^{\prime}$, and $z^{\prime}$-bands, which reach $5\sigma$ depths
of $u^{*}=28.3$, $g^{\prime} \approx 27.5$, $r^{\prime}$, $i^{\prime}=26.4$, and
$z^{\prime}=26.4$ mag \citep{2008PASP..120..212G}.
The field is also being observed in five
medium-band filters, which cover a wavelength range of $1-1.8\mu$m, using the
NEWFIRM camera \citep[P.I.~P.~van Dokkum;][]{2009PASP..121....2V} on the Kitt Peak
4\,m telescope. 

Additional multi-wavelength data in the EGS include $B, R,$ and $I$-band imaging
obtained with the CFHT 12K mosaic camera
\citep{2001ASPC..232..398C,2004ApJ...609..525C}, near-IR imaging in the $J$
and $K_{s}$ bands taken with Wide-Field 
Infrared Camera (WIRC) on the \textit{Palomar} 5m telescope, mid-IR (3.6--24 $\mu$m)
observations taken with IRAC and MIPS on \textit{Spitzer} as
part of the GTO program 8 and the SEDS program, far-IR ($70\mu$m)
observations taken with \textit{Spitzer}/MIPS as part of the FIDEL survey, X-ray observations
taken with \textit{Chandra} \citep{2009ApJS..180..102L}, UV observations taken with
\textit{GALEX}, and radio imaging (1.4 and 4.8\,GHz) from the \textit{VLA}
\citep{2006AJ....132.2159W,2007ApJ...660L..77I}.  Furthermore, the DEEP2 and DEEP3
programs \citep[][, Newman et al.~2011, in prep.]{2003SPIE.4834..161D} have carried
out an extensive spectroscopic survey of the EGS using the Deep Imaging
Multi-Object Spectrograph \citep[DEIMOS;][]{2003SPIE.4841.1657F} on the Keck II 10m telescope. 

The CANDELS WFC3/IR observations in the EGS consist of a rectangular grid of
$3\times15$ WFC3/IR tiles ($\sim 6\farcm{7}\times 30\farcm{6}$) that is
oriented at a position angle of $42^{\circ}$ (see Fig.~\ref{fig:egs_layout}). 
As in the COSMOS field, the location of ancillary data in the EGS heavily
constrains the location and orientation of the WFC3/IR mosaic.  The existing
\textit{Hubble}/ACS imaging in the field is roughly $10^{\prime}$ wide and is oriented at a
position angle of $40^{\circ}$.  The CANDELS mosaic is positioned to lie within these
existing data and to maximize overlap with the mid-IR SEDS imaging, the
far-IR FIDEL imaging, and the deepest portion of the \textit{Chandra} observations in the
field. The EGS also has the strictest scheduling constraints; there exists no roll angle
for the EGS that can be held fixed throughout both observing epochs.  
Therefore, we have devised a layout in which each WFC3/IR tile is rotated
$+11^{\circ}$ (East-of-North) during the first imaging epoch and
$-11^{\circ}$ during the second.  This results in a larger portion of the ACS/WFC
parallel exposures falling outside the WFC3/IR mosaic than in the UDS and
COSMOS fields, but due to the larger ACS/WFC footprint, the mosaic remains
nearly entirely covered by the ACS/WFC optical imaging.  The
layout of the mosaic and its position within the EGS are shown in
Figure \ref{fig:egs_layout}.

\subsubsection{The GOODS South and North Fields}\label{sec:southnorthwide}

The Wide-depth CANDELS WFC3/IR observations in the GOODS-S and GOODS-N
fields consist of three regions (one in GOODS-S and two in GOODS-N)
that flank the CANDELS/Deep regions in each field.  The location and
orientation of these regions are shown in Figures \ref{fig:gds} and \ref{fig:gdn}. 
In GOODS-S this Wide-depth ``flanking field'' contains a total of nine WFC3/IR
tiles oriented to place most of their ACS/WFC parallel exposures into the
CANDELS/Deep region.  In GOODS-N, the two flanking fields contain eight WFC3/IR
tiles each and are again oriented such that their ACS/WFC parallel exposures
fall within the CANDELS/Deep region.  A second flanking field in GOODS-S is
effectively provided by the existing WFC3/IR observations from the ERS
program, which are adjacent and to the North of the CANDELS/Deep region. 
When combined with the CANDELS/Deep observations and the ERS imaging, these
flanking fields will provide WFC3/IR coverage over nearly the entire
$10^{\prime}\times16^{\prime}$ extents of the two GOODS regions.

\subsection{Mitigating Cosmic Variance}\label{sec:cosmicvariance}
  
%
\vspace{-0.28in}                                                      
\begin{center}
\tabletypesize{\scriptsize}
\begin{deluxetable}{cccc}
\setcounter{table}{4}
\tablewidth{0pt}
\tablecaption{Cosmic Variance Estimates\label{tab:tab_cv}}
\tablecolumns{4}
\tablehead{\colhead{CANDELS Field} & \colhead{Area} & \colhead{$CV_{z=2}$} & \colhead{$CV_{z=7}$} } 
\startdata                                              
GOODS-North (Deep)  & $6\farcm{8}\times 10\farcm{0}$ &  0.476 $(b/5.5)$  &  0.578 $(b/15)$ \\ 
GOODS-South (Deep)  & $6\farcm{8}\times 10\farcm{0}$ &  0.476 $(b/5.5)$  &  0.578 $(b/15)$ \\ 
GOODS-North (Wide)  & $9\farcm{7}\times 16\farcm{4}$ &  0.404 $(b/5.5)$  &  0.469 $(b/15)$ \\ 
GOODS-South (Wide)  & $9\farcm{9}\times 14\farcm{9}$ &  0.412 $(b/5.5)$  &  0.481 $(b/15)$ \\ 
COSMOS              & $9\farcm{0}\times 22\farcm{4}$ &  0.375 $(b/5.5)$  &  0.429 $(b/15)$ \\
UDS                 & $9\farcm{4}\times 22\farcm{0}$ &  0.354 $(b/5.5)$  &  0.405 $(b/15)$ \\ 
EGS                 & $6\farcm{4}\times 30\farcm{5}$ &  0.375 $(b/5.5)$  &  0.430 $(b/15)$ \\
\hline \\
Deep Total  & 0.038 sq deg                   &  0.337 $(b/5.5)$  &  0.409 $(b/15)$ \\
Wide Total  & 0.290 sq deg                    &  0.171 $(b/5.5)$  &  0.197 $(b/15)$ \\
\vspace{-0.075in}                                                      
\enddata
\tablecomments{A single $32\farcm{4}\times 32\farcm{4}$ field covering the
  same total area as the Wide program results in $CV_{z=2} = 0.250$ $(b/5.5)$ and
  $CV_{z=7} = 0.259$ $(b/15)$.}
\end{deluxetable}
\end{center}
\clearpage

The primary reason CANDELS has focused on five different fields, rather than
one or two, is to minimize the impact of the variation in the mean density of
matter (and hence also of galaxies of some class) from one region to another,
commonly referred to as ``sample variance'' or ``cosmic variance''.  This effect
can be comparable to or larger than Poisson variance, and hence will have
large impact on the determination of the abundance of a class of objects
(e.g., the normalization of the LF).  The impact of cosmic
variance is most simply reduced by surveying a larger volume of the Universe;
however, that is expensive in terms of telescope time.  Given a
fixed survey area, it will be minimized by dividing that area amongst
multiple widely-separated fields, as the density fluctuations in those fields
will be statistically independent;  for a single field, cosmic variance is
reduced if the field has a long, narrow geometry as opposed to a square one,
as that maximizes the typical distance between objects and ensures greater
independence between different portions of the field (i.e., samples more
quasi-independent volumes; cf.~\citealt{2002ApJ...564..567N}).  Cosmic variance is
also a driver for the geometry of individual CANDELS/Wide fields; we have
made them as elongated as possible while maintaining overlap with other
relevant datasets and retaining the ability to measure correlation functions
accurately. 

We have used the public QUICKCV code \citep{2002ApJ...564..567N} to predict the
impact of cosmic variance for representative CANDELS samples, and compared to
other simple scenarios.  The fractional uncertainty in the abundance of some
class of object will scale as its linear bias, $b$.  We consider two fiducial
scenarios: counting the abundance of galaxies at $z = 1.75-2.25$ with stellar
mass $10^{11} - 10^{11.5}$, using the results of \citet{2003SPIE.4841.1657F} to
predict their bias\footnote{We find $b \approx 5.5$, equivalent to a correlation
  length in comoving coordinates of $r_0 = 12.5$ $h^{-1}$ Mpc, assuming
  a correlation function of form $\xi(r) = (r/r_0)^{-\gamma}$ with slope
  $\gamma=1.8$.}, or counting the abundance of a sample of galaxies at
$z=6.5-7.5$ with the same clustering properties (in comoving coordinates) as
the $z=2$ sample, corresponding to a bias of 15 compared to dark matter at
the same redshift.  These samples represent approximate upper limits to the
impact of cosmic variance; any more biased sample at the same redshifts
should be extremely rare and dominated by Poisson uncertainties.  The results
of these calculations are given in Table \ref{tab:tab_cv}.\footnote{We have assumed an
  amplitude of dark matter fluctuations today of $\sigma_8=0.85$; the numbers
  presented will scale as $(\sigma_8/0.85)$, for a sample of fixed bias.} 

Hence, the net cosmic variance for the CANDELS/Wide survey should be roughly
$17\%$ $(b/5.5)$ for a sample at $z=1.75-2.25$, or $20\%$ $(b/15)$ for a sample
at $z=6.5-7.5$.  Even for more typical galaxies with a large-scale structure
bias one-third as large as our fiducial samples, cosmic variance should
contribute significantly to errors so long as the sample contains more than
$\sim 100$ objects over the CANDELS survey.  These numbers are significant,
but compare very favorably to a square (or nearly so) survey that covers a
single field, such as the COSMOS survey.
A single-field survey of the same total area as CANDELS (i.e.~a $32.4 \arcmin
\times 32.4 \arcmin$ field) would have errors from cosmic variance of
25\% $(b/5.5)$ at $z \approx 2$ or 26\% $(b/15)$ at $z \approx 7$, almost 50\% larger than for our chosen geometry. 
To match the cosmic variance uncertainties from a CANDELS sample, a
single field would have to be $55.5 \arcmin \times 55.5 \arcmin$ across,
corresponding to an area (and hence time to survey) 2.94$\times$ larger than
CANDELS.    

\ifsubmode
  \placefigure{fig:gdsepa}
\else
   \insertfigurewide{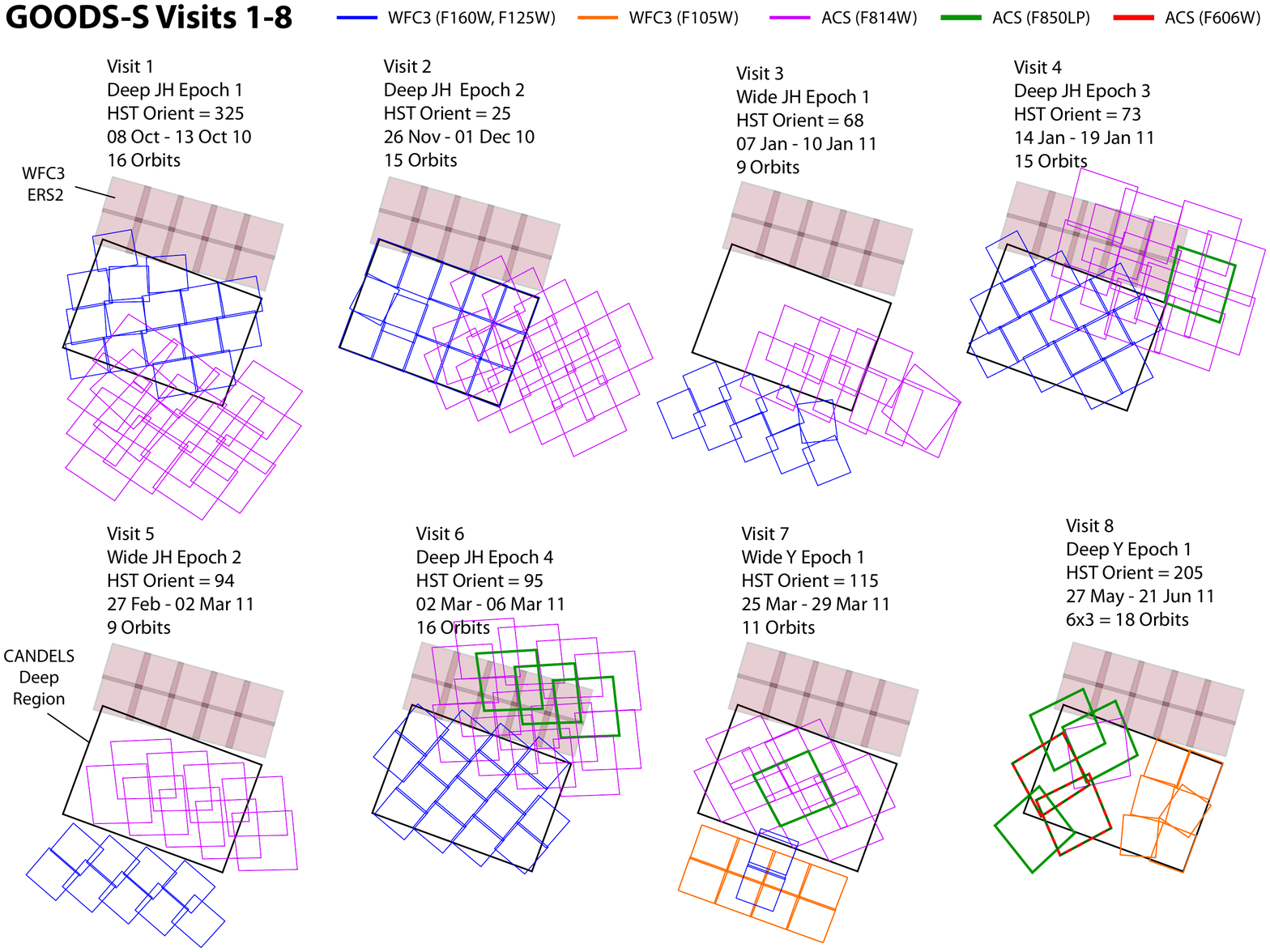}{\figgdsepa}
\fi

\ifsubmode
  \placefigure{fig:gdsepb}
\else
   \insertfigurewide{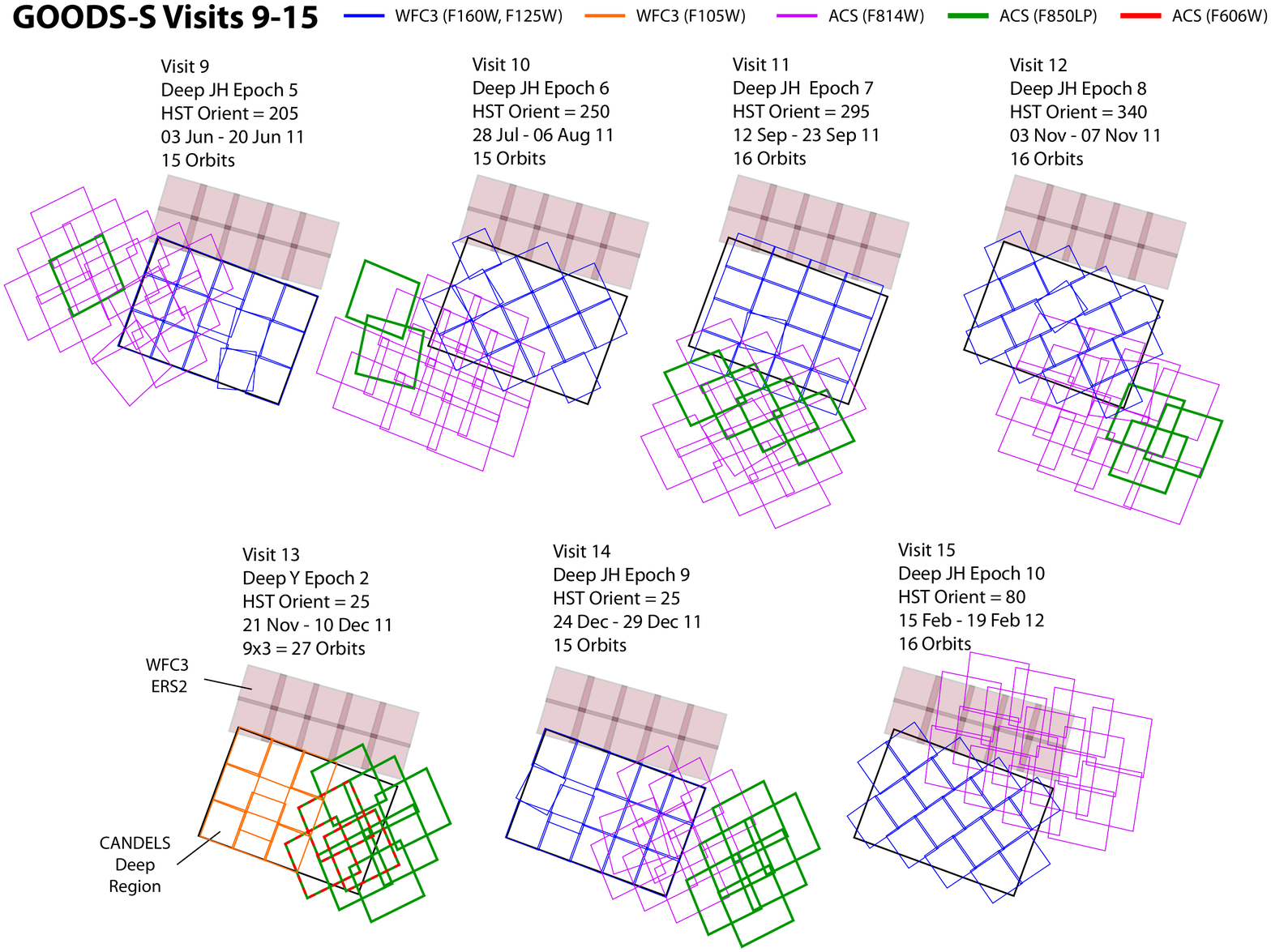}{\figgdsepb}
\fi

\section {Observing Strategy}\label{sec:observing_strategy}

The CANDELS observing strategy is very tightly constrained by
several competing science goals and by the scheduling limitations of
\textit{HST}\@.  Because CANDELS is an \textit{HST} Multi-cycle Treasury
Program, it will be observed over the three years of \textit{HST}
Cycles 18, 19, and 20.  At the time of writing, only the Cycle 18 and
19 observations have officially been planned through to the so-called
``Phase II'' stage.  In this section, we describe the observing strategy
in detail, with particular attention to the constraints that have shaped
our decisions.

The Cycle 18 observations include both epochs of the full UDS field, both
epochs of the first half of the EGS field (see Figs.~\ref{fig:egs_layout}
and \ref{fig:egs_combo}), and several epochs of the GOODS-S field.
The latest specifics of these observations may be obtained from the STScI
website as Programs 12060 (non-SN-searched GOODS-S), 12061 and 12062
(SN-searched GOODS-S), 12063 (first-half EGS), and 12064 (UDS).  Program
12062 includes observations that extend into Cycle 19.  The COSMOS
and GOODS-N observations for Cycle 19 were planned through the Phase II
stage in mid-2011.  The COSMOS observations have been assigned Program
12440, and the GOODS-N observations have been assigned Programs 12442
(non-SN-searched GOODS-N), 12443, 12444, and 12445 (successive thirds of
the SN-searched GOODS-N).  The Cycle 20 observations will comprise the
latter portion of GOODS-N (Programs 12444, 12445, and roughly half of
12442) and the second-half of EGS (Program 12063), whose specifics will
have been largely pre-determined by our earlier observations of these
fields in Cycles 19 and 18, respectively.  The scheduling constraints
upon all CANDELS fields are already sufficiently tight that we can
describe the expected layouts and timings with reasonable certainty.

As in the previous section, the detailed description 
of the observing strategy is best split between the 
CANDELS/Wide fields and the CANDELS/Deep fields.

\subsection{Wide-Field Strategy}\label{sec:widestrategy}
In the CANDELS/Wide fields, the goal is to cover
a large area at two-orbit depth while maintaining
the ability to search for SNe in the near-IR
with WFC3.

The SN search requires that we repeatedly observe a 
given location in WFC3 F125W and F160W 
with time separation(s) of $\sim 52$ days.  
Given that each of the CANDELS/Wide areas comprises dozens
of WFC3/IR fields of view, the minimum efficient dwell time 
is one full orbit per pointing
per epoch.  The minimum SN search is a
pair of epochs separated by 52\,d, which gives us a total
of two-orbit depth with WFC3/IR in the Wide fields.

It is crucial that we have nearly contemporaneous 
observations in an optical filter to discriminate the type
(and the photometric redshift) of a WFC3-detected
SN\@.  We have therefore adopted a long and narrow geometry for 
the CANDELS/Wide fields, such that ACS parallel observations will 
overlap the majority of the WFC3/IR coverage (see Figs.~\ref{fig:uds_layout},
\ref{fig:cosmos_layout}, and \ref{fig:egs_layout}).  

For the CANDELS UDS,
our desired long-axis position angle (East--West) fortuitously corresponds
to a telescope orientation that can be held constant across
both epochs.  The same is nearly true for the COSMOS field,
where the desired North--South alignment is within several
degrees of a stable \textit{HST} orientation and permits a slipped-lattice 
WFC3/IR raster (see Fig.~\ref{fig:cosmos_layout}).  The difference between
this arrangement and an orthogonal raster is small enough 
that ACS parallel fields still overlap the WFC3/IR footprint, 
as desired for the SN search.

For the EGS field, there is no \textit{HST} orientation near our
desired Northeast--Southwest alignment that can be held
for the entire duration of observing the full field at both
epochs.  We therefore observe half the field in one campaign
(both epochs) in Cycle 18, and the other half
in another campaign (also both epochs) in Cycle 20.  Furthermore, in each 
EGS campaign the first epoch is observed at a position angle
$11\arcdeg$ counterclockwise of the desired alignment; the
second epoch at $11\arcdeg$ clockwise of 
the desired alignment.  The rotations are necessary because of our inability to maintain 
a fixed roll angle of \textit{HST} across the $\sim 8.5$ weeks spanned
by the two epochs.  As with the COSMOS field, the displacements
away from the EGS tiling orientation are not so large that the 
ACS parallels miss the WFC3/IR footprint (see Fig.~\ref{fig:egs_layout}).

Some care must be taken in dividing the full EGS
tiling so that WFC3/IR and ACS are overlapping in the respective halves.
It is unavoidable, when dividing the EGS tilings into two campaigns,
that there will be less contemporaneous overlap of WFC3/IR and ACS than if
we could observe the entire field in a single campaign.  Our solution,
which we believe minimizes this overlap loss, is the tiling scheme 
shown in Figure \ref{fig:egs_combo}.  The key element is to divide the
mosaic into bands separated by $\sim 6\arcmin$, matching the offset
between the contemporaneous fields of view of WFC3/IR and ACS.

Because of the large offset between these fields of view,
the first three rows of WFC3/IR pointings along the
long axis of the CANDELS/Wide fields will inevitably lack ACS parallel coverage.
As a corollary, the ACS parallel coverage of the CANDELS/Wide fields
will extend several arcminutes past the WFC3/IR footprint at the other end
of the long axis.  Lacking parallel ACS coverage in the first three
rows of WFC3/IR pointings, we have elected to use WFC3/UVIS in order to
obtain the contemporaneous optical coverage needed for SN discrimination.
This added UVIS exposure, which can be taken in the ``white-light''
filter F350LP for maximum throughput, foreshortens the WFC3/IR exposure
depth in these first three rows by 410\,s per epoch.  The UDS and COSMOS
fields, which are $4\times11$ tilings, therefore have 12 tiles with
reduced IR exposure.  The EGS field, which is a $3\times15$
tiling, has nine tiles with reduced IR exposure.

To even sensitivity in AB magnitudes, we seek to obtain a 2:1 ratio of exposure
time in the WFC3/IR filters F160W and F125W\@.  We also desire four WFC3/IR
exposures per epoch, using the same filter and exposure combinations in 
each epoch, to make a well-dithered mosaic for SN 
detection (see Koekemoer et al. 2011 for details).  We therefore 
strive at each WFC3/IR pointing to obtain two long F160W exposures and 
two short F125W exposures.  Pointings that require an additional WFC3/UVIS
exposure (see above) will have an F160W:F125W exposure ratio closer
to 1:1.  This variation in exposure ratio is necessary because the F125W 
exposures are already near the minimum duration to accommodate the \textit{HST} 
buffer dump of the prior ACS (and sometimes WFC3) exposure in parallel 
with the current exposure.  Shorter F125W exposure time would trigger a 
serial (rather than parallel) buffer dump, in effect precluding the exposure.

For the ACS parallel exposures in the CANDELS/Wide fields, the need for
maximally efficient discovery of high-$z$ dropout galaxies pushes toward a
2:1 ratio of exposure in the filters F814W and F606W\@.  In the EGS and 
COSMOS fields, which contain legacy ACS data in these filters (see 
\S\ref{sec:fieldswide} for details), the goal is to adjust the CANDELS exposure ratios such that 
the cumulative exposure is 2:1 for F814W:F606W\@.  Because the ACS
field of view is very nearly twice that of WFC3/IR, our abutting WFC3/IR
raster results in most of the ACS parallel footprint having double
the coverage, or effectively two-orbit depth per epoch.  
The two epochs thus provide a total of four-orbit depth of ACS parallel
coverage divided between F814W and F606W.  

In the CANDELS UDS field, where there exists no prior 
full-field coverage with ACS, each epoch yields 2/3-orbit F606W 
and 4/3-orbit F814W\@.  In the CANDELS EGS field, which contains 
prior F606W and F814W images totaling one-orbit depth apiece 
(Program 10134; P.I.: M.~Davis), we take no F606W data in 
the first epoch, and split between F606W and F814W in the second
epoch, to achieve approximate 2- and 4-orbit total depth in
F606W and F814W\@.  In the CANDELS COSMOS field, where there exists
one orbit of F814W, we allocate the new ACS parallels in a 1:2
ratio of F606W:F814W in the first epoch, and a 2:1 ratio in the
second epoch, resulting in a total depth of 2- and 3-orbits in
F606W and F814W. 
  
Figures \ref{fig:udsexpmaps}a-e show the planned exposure maps for the CANDELS
UDS field in the filters F350LP (WFC3/UVIS), F606W (ACS/WFC), 
F814W (ACS/WFC), F125W (WFC3/IR), and F160W (WFC3/IR).

\ifsubmode
\placefigure{fig:udsexpmaps}
\else
\begin{figure*}[ht]
\plottwo{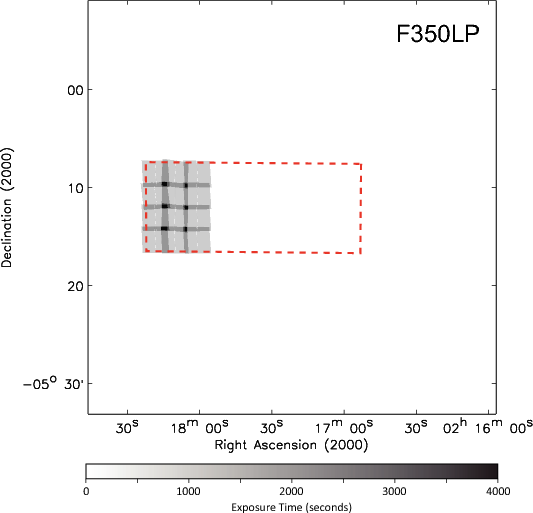}{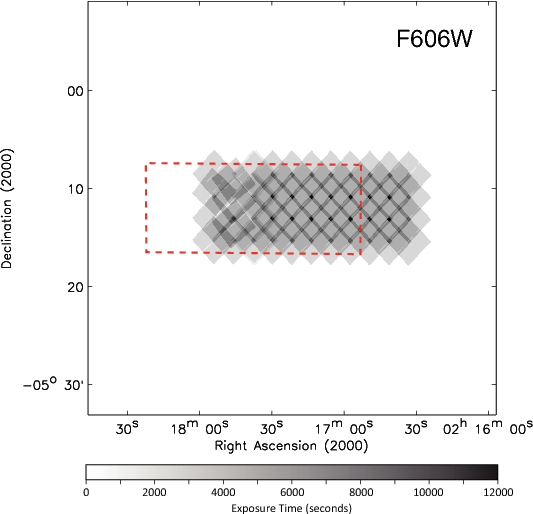}
\plottwo{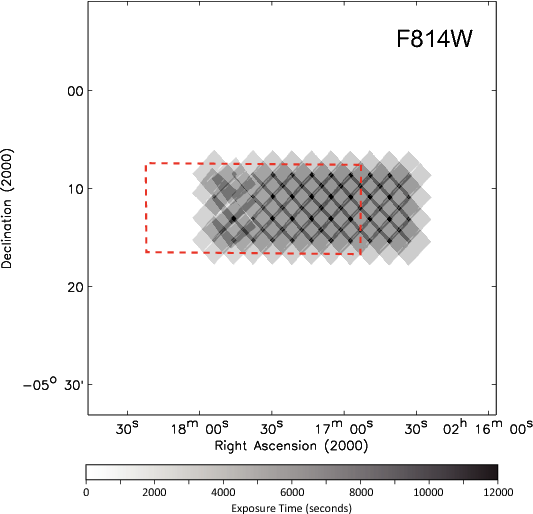}{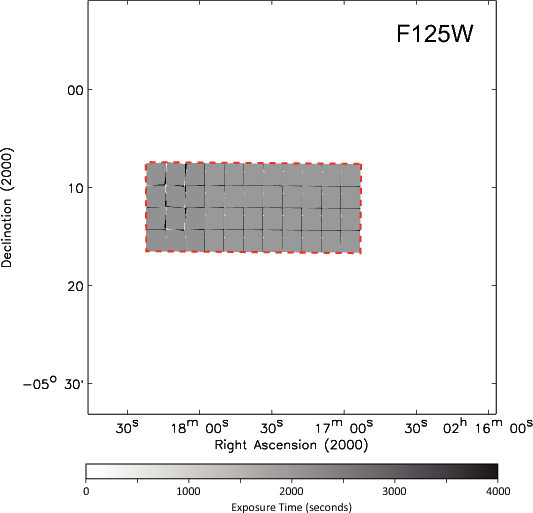}
\plottwo{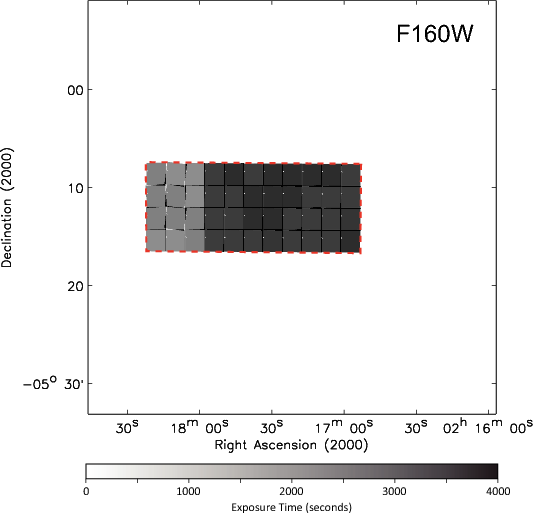}{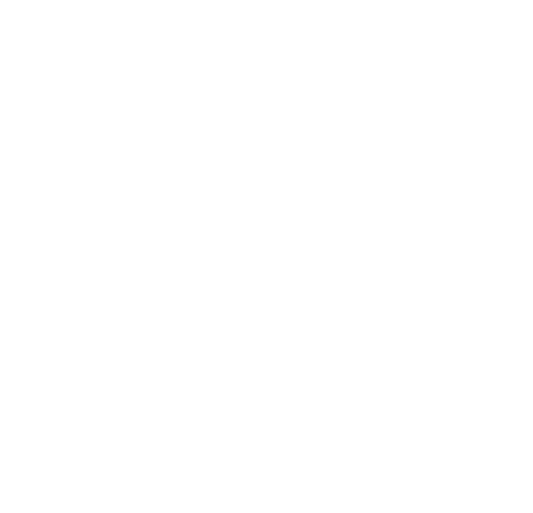}
\caption{\figudsexpmaps}
\end{figure*}
\fi

Figures \ref{fig:egsexpmaps}a-e show the planned exposure maps for the CANDELS
EGS field in the filters F350LP (WFC3/UVIS), F606W (ACS/WFC), 
F814W (ACS/WFC), F125W (WFC3/IR), and F160W (WFC3/IR).  The 
ACS exposure maps include the legacy EGS imaging from Program 10134.

\ifsubmode
\placefigure{fig:egsexpmaps}
\else
\begin{figure*}[ht]
\plottwo{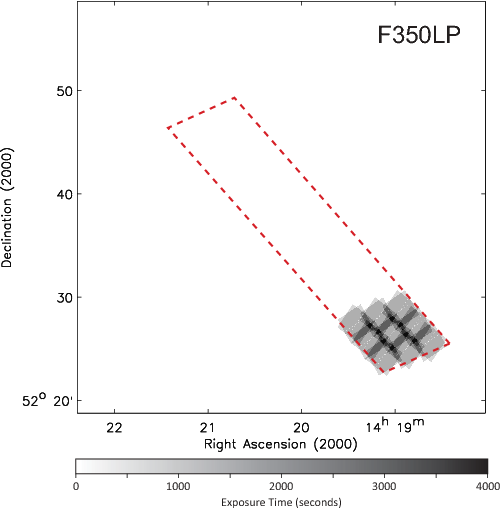}{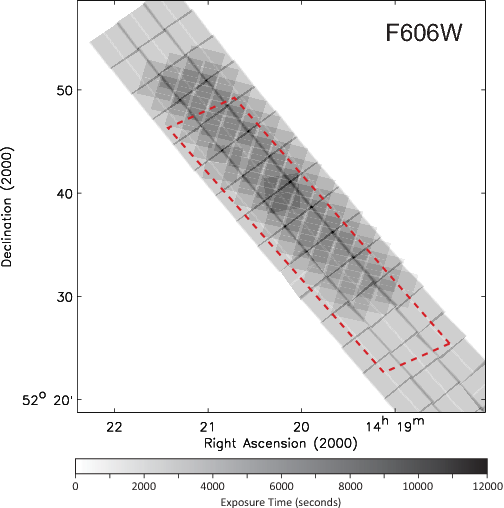}
\plottwo{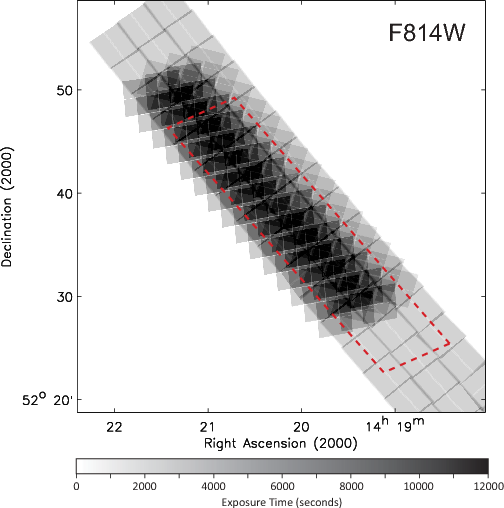}{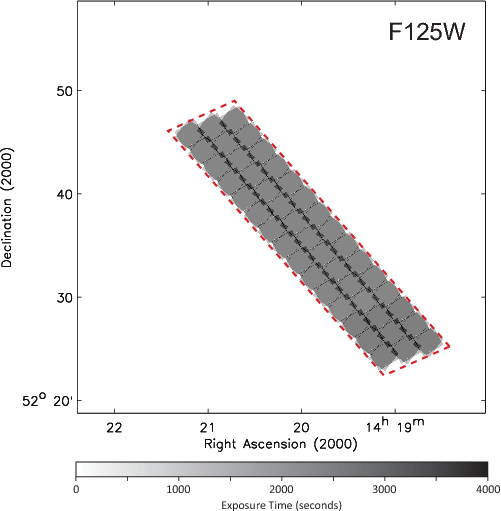}
\plottwo{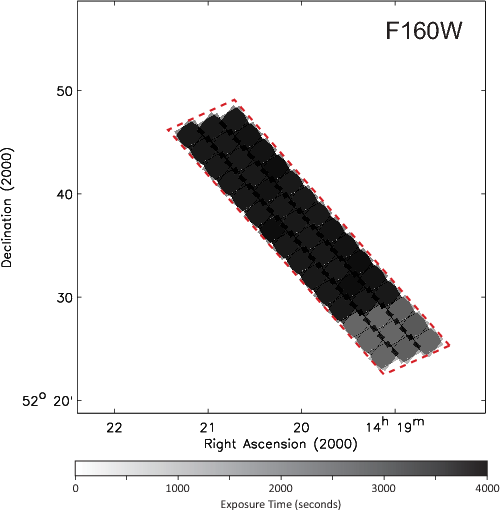}{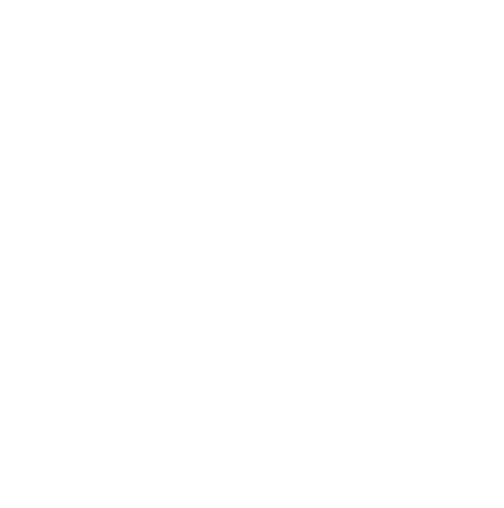}
\caption{\figegsexpmaps}
\end{figure*}
\fi

Figures \ref{fig:cosexpmaps}a-e show the planned exposure maps for the CANDELS
COSMOS field in the filters F350LP (WFC3/UVIS), F606W (ACS/WFC), 
F814W (ACS/WFC), F125W (WFC3/IR), and F160W (WFC3/IR).  The 
ACS exposure maps include the legacy COSMOS imaging from Programs 9822
and 10092.

\ifsubmode
\placefigure{fig:cosmosexpmaps}
\else
\begin{figure*}[ht]
\plottwo{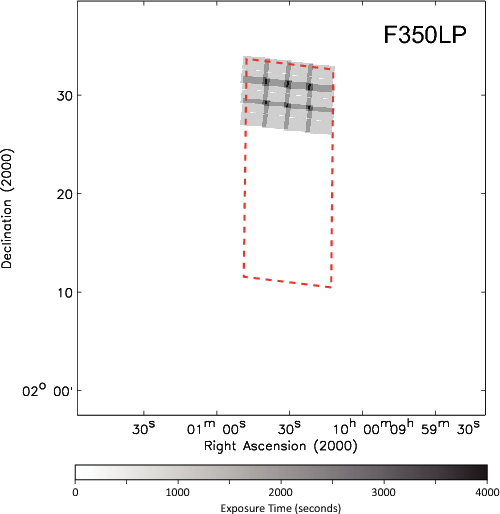}{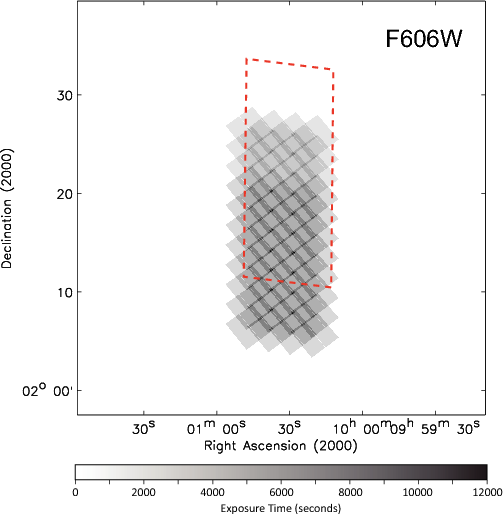}
\plottwo{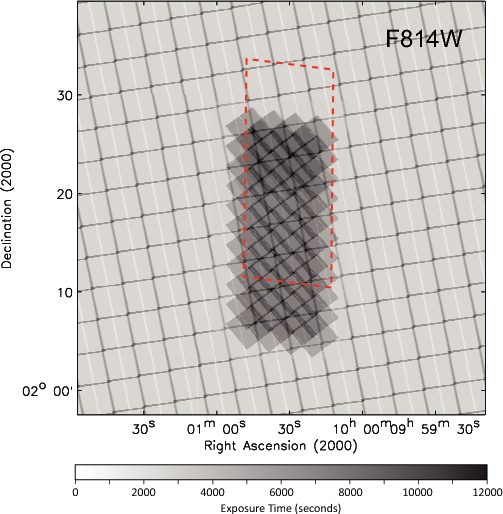}{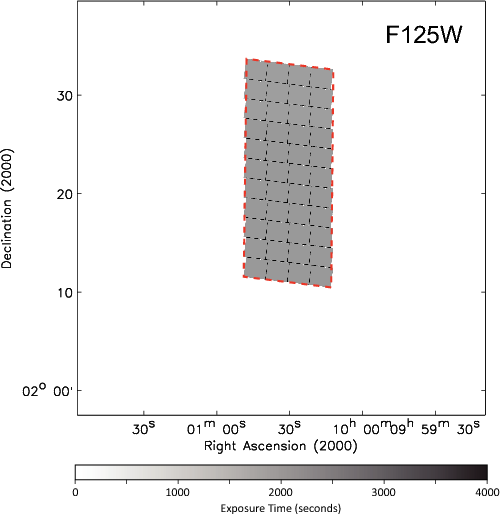}
\plottwo{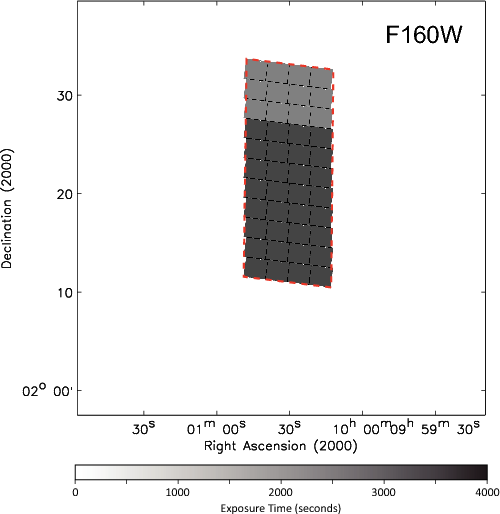}{EGS_blank.png}
\caption{\figcosmosexpmaps}
\end{figure*}
\fi

Figures \ref{fig:widead}a-c show the curves of area versus exposure 
depth for each of the ACS and WFC3 filters that will comprise CANDELS UDS, EGS,
and COSMOS fields.  Each of these area vs.~depth plots only consider the
area within the footprint of the CANDELS F160W images and includes the
legacy ACS data.

\ifsubmode
\placefigure{fig:widead}
\else
\begin{figure*}[ht]
\plottwo{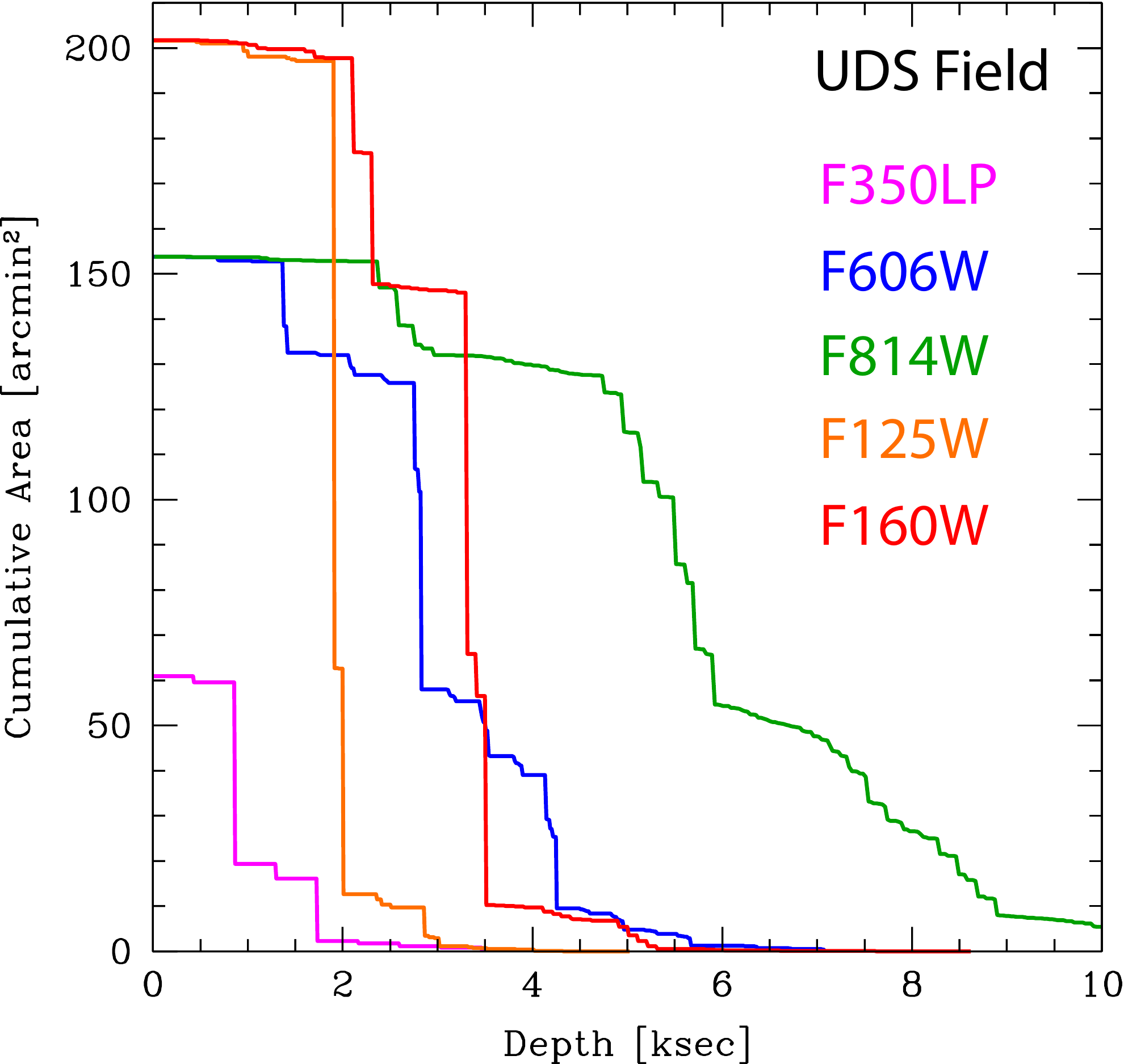}{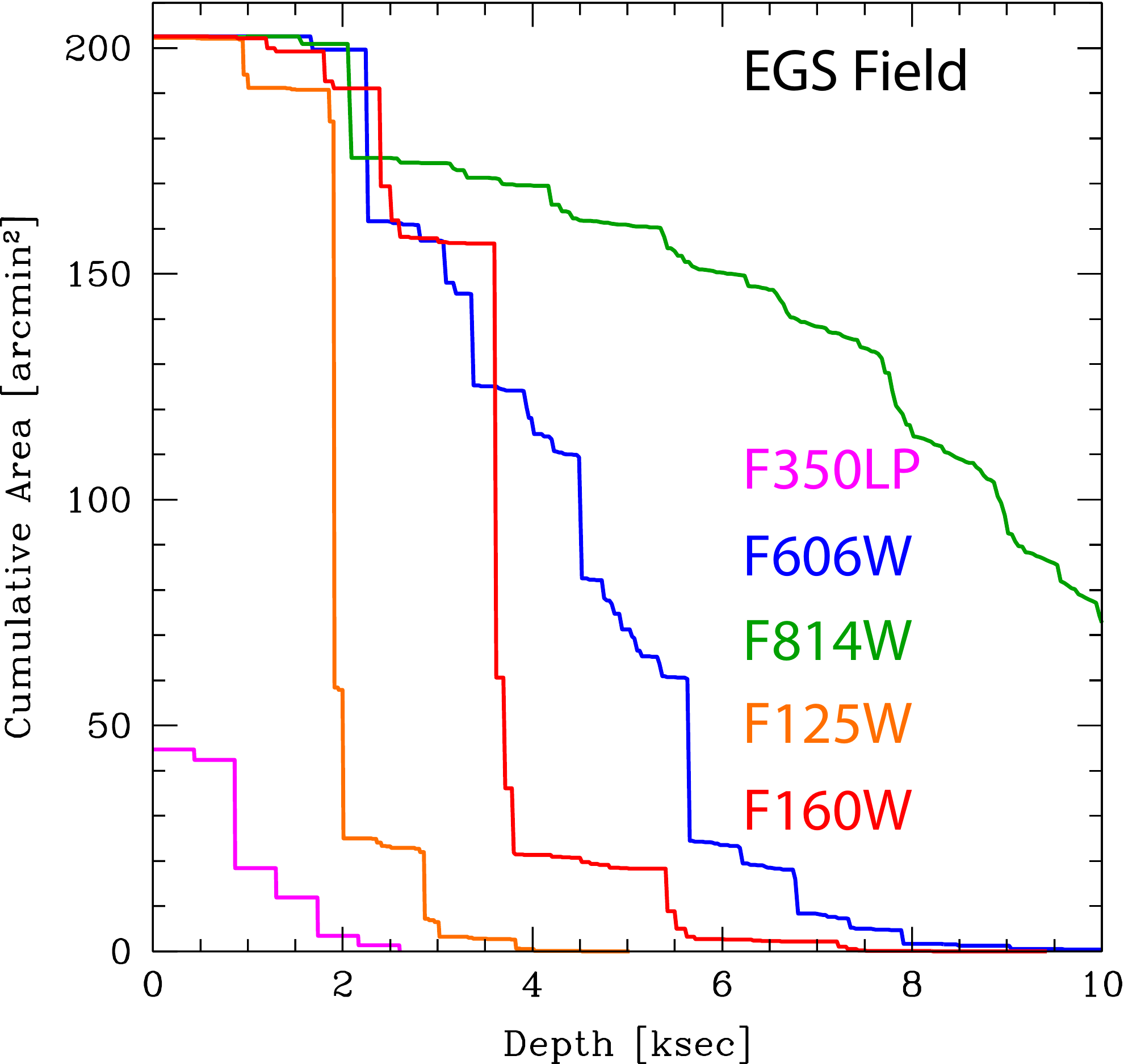}
\plottwo{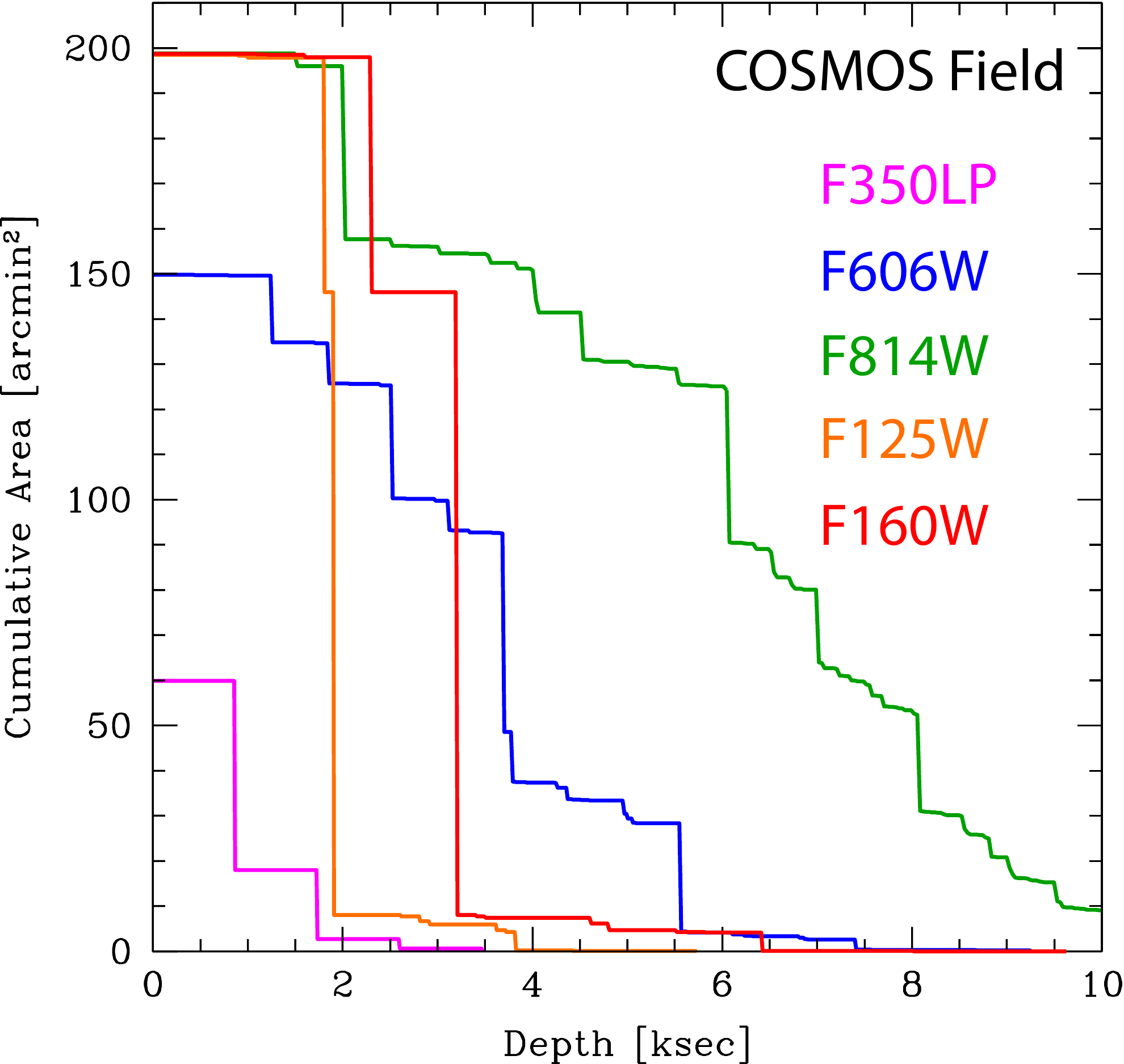}{blank_areadepth.pdf}
\caption{\figwidead}
\end{figure*}
\fi

\subsection{Deep-field Strategy}\label{sec:deepstrategy}

In the CANDELS/Deep fields, the primary observational goal
is to obtain WFC3 and ACS imaging across moderately large regions 
that are substantially deeper than the CANDELS/Wide fields.  Unlike the
CANDELS/Wide fields, completed in two epochs, the CANDELS/Deep fields are 
imaged for ten SN-searched epochs in F125W and F160W, as well
as additional epochs in F105W ($Y$-band) that are not SN-searched.
The target is 10\,ks total exposure in these three WFC3 filters.

To enable the robust identification of extreme-redshift galaxies
in the CANDELS/Deep fields, we also design the observing plan to attain at 
least 25\,ks of ACS F814W exposure across the entire area covered
by the deep WFC3/IR images.  The majority
of this F814W depth is obtained from ACS parallel exposures during the $Y$-band 
epochs (e.g., Fig.~\ref{fig:gdsepa}, Visit 8; Fig.~\ref{fig:gdsepb}, Visit 13).
 
Our orbit allocation permits us to cover roughly 40\% of the
$\sim 10\arcmin \times 16\arcmin$ GOODS ACS regions for CANDELS/Deep.  We
have chosen a placement that contains the zones of deepest GOODS \textit{Spitzer}
coverage: $>45$ hr exposure time with IRAC\@ (3.6$\mu$m--8$\mu$m).
Recognizing that the remainder of the GOODS ACS regions are still
extremely valuable for moderate-redshift science in conjunction with WFC3,
we image the GOODS areas outside the CANDELS/Deep fields with a two-epoch
observing strategy similar to the CANDELS/Wide fields above.  Because these
flanking fields have superior multicolor ACS imaging as compared with the
CANDELS/Wide fields, we also image them with F105W ($Y$-band) to complete
the \textit{HST} wavelength coverage of the entire GOODS area from $B$-band (F435W)
to $H$-band (F160W).  Judicious scheduling of these CANDELS/Deep flanking
fields allows us to place additional ACS parallel F814W exposures into
the CANDELS/Deep fields (e.g., Fig.~\ref{fig:gdsepa}, Visits 3, 5, and 7).

In the CANDELS GOODS-S field, we have designed our
observing strategy to be complementary to the existing 
observations by the WFC3 ERS\@.  
Those ERS observations match quite closely what we would
desire for a northwest 
flanking field of CANDELS/Deep, and to depths in
F125W and F160W that actually exceed the CANDELS/Wide target.  
We therefore observe only the southeast flanking field 
in GOODS-S\@.  We note that the ERS observations
in $Y$-band were taken with F098M rather than F105W\@.  We choose
F105W in the GOODS-S flanking field for consistency with
the CANDELS/Deep fields, the GOODS-N flanking fields, and deep
F105W imaging in GOODS-S from the HUDF09 program.  We note
that the ERS F098M exposure time is roughly twice that of
the flanking-field F105W, so the $Y$-band sensitivity across ERS will
be at least that of the flanking fields despite the narrower
bandpass.

In CANDELS GOODS-N, where there exists no ancillary WFC3
imaging, we will observe both flanking fields identically in F105W,
F125W, and F160W\@.  During the GOODS-N Deep epochs, we will also
take advantage of CVZ opportunities
to image the Deep region in UV with WFC3/UVIS (F275W and F336W).
Because these UV exposures will be read-noise limited, we have chosen
the $2\times2$-binned mode.  This binning yields $\sim 0.5$\,mag
additional sensitivity at only modest expense in angular
resolution for the drizzled UV mosaics, given our sub-pixel dithering.
To facilitate the selection of Lyman-break drop-outs at $z \approx 2$
and to increase the sensitivity to LyC radiation at $z \approx 2.5$ (see
\S\ref{sec:uv}), the goal is a 2:1 ratio of exposure time for F275W:F336W\@.
We expect extended-source sensitivities of $\sim 27.9$ and $\sim 27.7$\,mag
in F275W and F336W, respectively.  Caveats include uncertainties
in the durations of the narrow windows for CVZ scheduling, and in the
degradation of the UVIS detectors' charge transfer efficiency ---
important for these low-background UV exposures.

During the ten epochs of the CANDELS/Deep campaigns, the ACS
parallel observations pivot with the orientation of \textit{HST}
throughout the year, sometimes landing outside the larger 
GOODS ACS box entirely (see Figs.~11
and 12).  In GOODS-S, this extended region has shallow ACS
coverage in F606W and F850LP from the GEMS program.  Because
the area coverage of ACS/WFC is $\sim 2$ times that of WFC3/IR,
we have approximately two-orbit depth in ACS parallel coverage
at each Deep epoch.  We have developed a strategy for filter 
selection among these far-flung ACS parallels according to the 
decision tree:
\begin{itemize}
\item
If the ACS field intersects the CANDELS/Deep: first priority 
is to exceed 25\,ks depth in F814W; second priority is
one orbit of F850LP for variability study (with GOODS and PANS); 
third priority is
additional F606W or F814W as appropriate to reach a 1:2 ratio
of exposure in those two filters.
\item
If the ACS field intersects the CANDELS/Deep flanking fields: 
first priority is to exceed two orbits of F814W (the F606W 
depth across the GOODS fields exceeds 3 orbits); second priority 
is one orbit of F850LP for variability study (with GOODS and
PANS); third priority is
additional F606W or F814W as appropriate to reach a 1:2 ratio
of exposure in those two filters.
\item
If the ACS field is entirely outside the CANDELS+ERS WFC3: first 
priority in GOODS-S is one orbit of F850LP for variability 
study (with GEMS); otherwise we allocate between 
F606W and F814W as appropriate to reach 
a 1:2 ratio of exposure in those two filters.  The full coverage
of our CANDELS GOODS-S ACS parallels includes at least 1-orbit-depth 
F606W from the GEMS program.  There is very little ancillary ACS imaging
in F606W, F814W, or F850LP just outside the footprint of GOODS-N.
\end{itemize}

Because the CANDELS GOODS epochs individually cover compact
regions with WFC3, we are unable to use the ACS parallels for
contemporaneous optical discrimination of distant SNe as we 
can with the CANDELS/Wide fields.  Therefore, all of our CANDELS GOODS
SN search epochs include a short WFC3/UVIS F350LP exposure at each
pointing.  This configuration is nearly identical to that used
for the small portions of the CANDELS/Wide fields lacking ACS parallel
coverage.  The $Y$-band epochs, which are not used for SN~Ia searching,
are taken without F350LP exposures.  In the Deep regions, the F105W
is also obtained with multi-orbit visits, reducing \textit{HST} guide-star
acquisition overhead.  Both situations result in more F105W exposure 
per orbit than we achieve for F125W and F160W\@.  Our three-orbit visits 
for the Deep F105W obtain nearly as much exposure time as F125W and 
F160W attain in their ten shared-orbit visits. 

Figures \ref{fig:gdsloexpmaps}a-f show the planned 
exposure maps for the CANDELS
GOODS-S field in the bluer filters: F275W (WFC3/UVIS), F336W (WFC3/UVIS),
F350LP (WFC3/UVIS), F435W (ACS/WFC), F606W (ACS/WFC), and F775W (ACS/WFC). 
Figures \ref{fig:gdshiexpmaps}a-f show the planned 
exposure maps for the CANDELS GOODS-S field in the redder 
filters: F814W (ACS/WFC3), F850LP (ACS/WFC),
F098M (WFC3/IR), F105W (WFC3/IR), F125W (WFC3/IR), and F160W (WFC3/IR). 
The exposure maps include all available ACS and WFC3 legacy imaging 
in this region matching the combined GOODS and CANDELS filter sets.  
In particular, the F275W and F336W exposures come entirely from the 
ERS (Program 11359) and are shown because CANDELS will also be 
obtaining exposures in these filters in the GOODS-N field.
The F098M imaging in the ERS is also shown, as this will be used as
a proxy for the F105W imaging that CANDELS is employing elsewhere in the 
field.  We also note that the F775W data is entirely from legacy 
observations, principally GOODS and PANS.

\ifsubmode
\placefigure{fig:gdsloexpmaps}
\placefigure{fig:gdshiexpmaps}
\else
\begin{figure*}[ht]
\plottwo{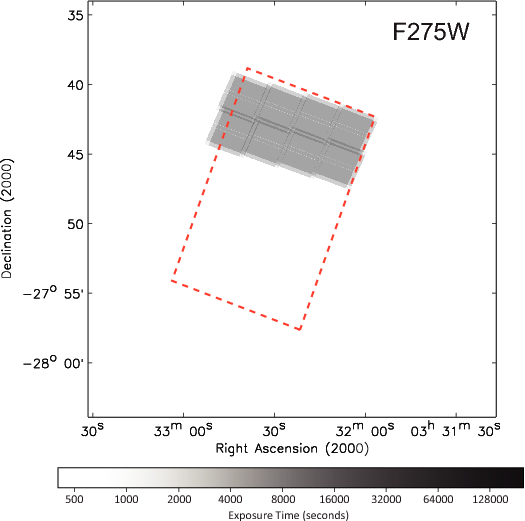}{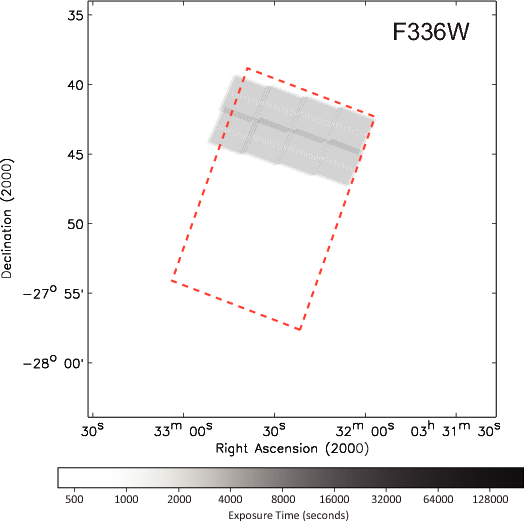}
\plottwo{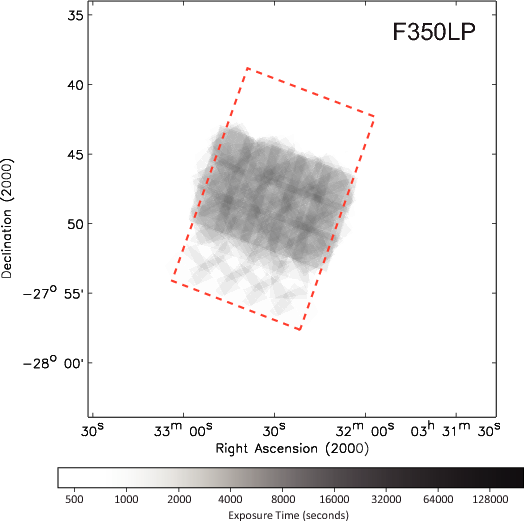}{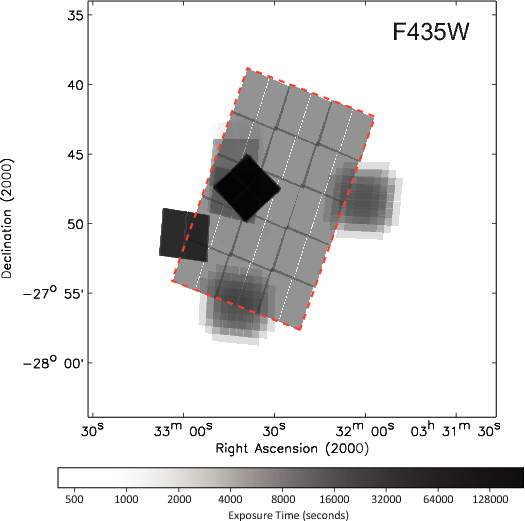}
\plottwo{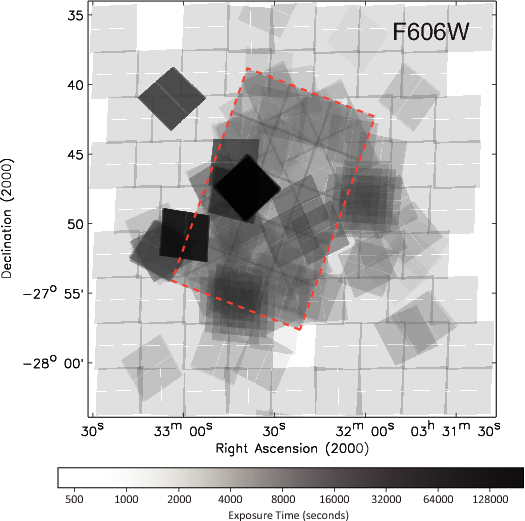}{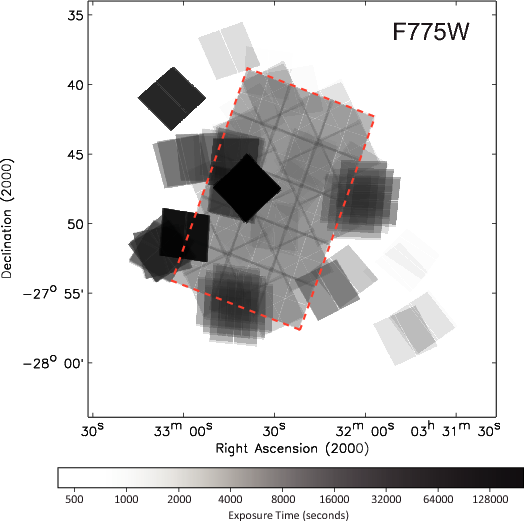}
\caption{\figgdsloexpmaps}
\end{figure*}

\begin{figure*}[ht]
\plottwo{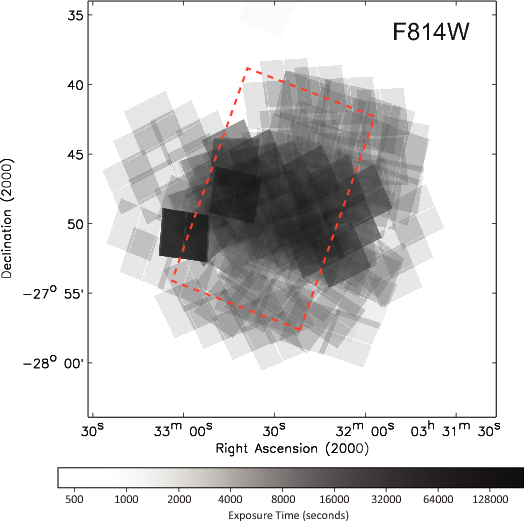}{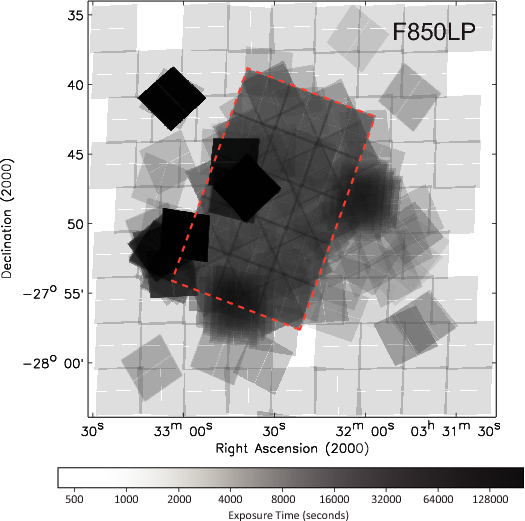}
\plottwo{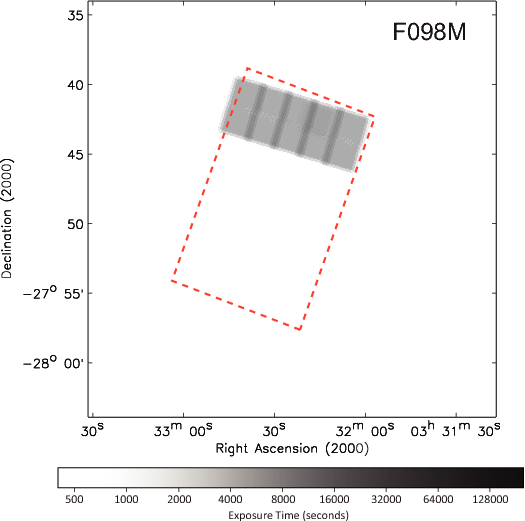}{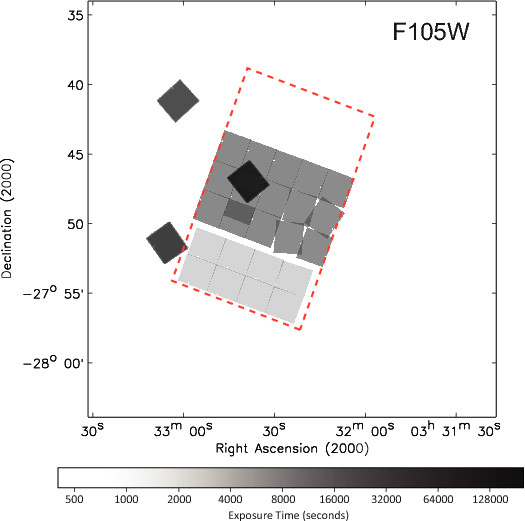}
\plottwo{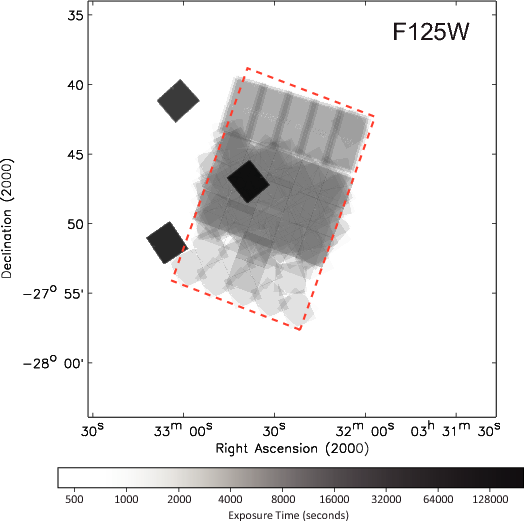}{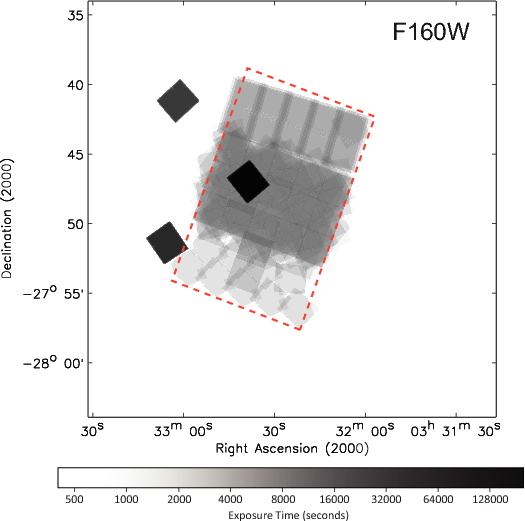}
\caption{\figgdshiexpmaps}
\end{figure*}
\fi

Figures \ref{fig:gdsad}a,b show the curves of area versus exposure 
depth for each of the ACS and WFC3 filters that will comprise the 
CANDELS GOODS-S field.  Each of these area vs.~depth plots  
considers only the area within the footprint that is the union of the ERS 
and the CANDELS images but does include all legacy ACS and WFC3 data.
Note that the F098M and F105W coverages are almost entirely non-overlapping,
and thus their area-depth curves may be approximately stacked to represent
the $Y$-band counterpart to the available $J$ and $H$ images.
\ifsubmode
\placefigure{fig:gdsad}
\else
\begin{figure*}[ht]
\plottwo{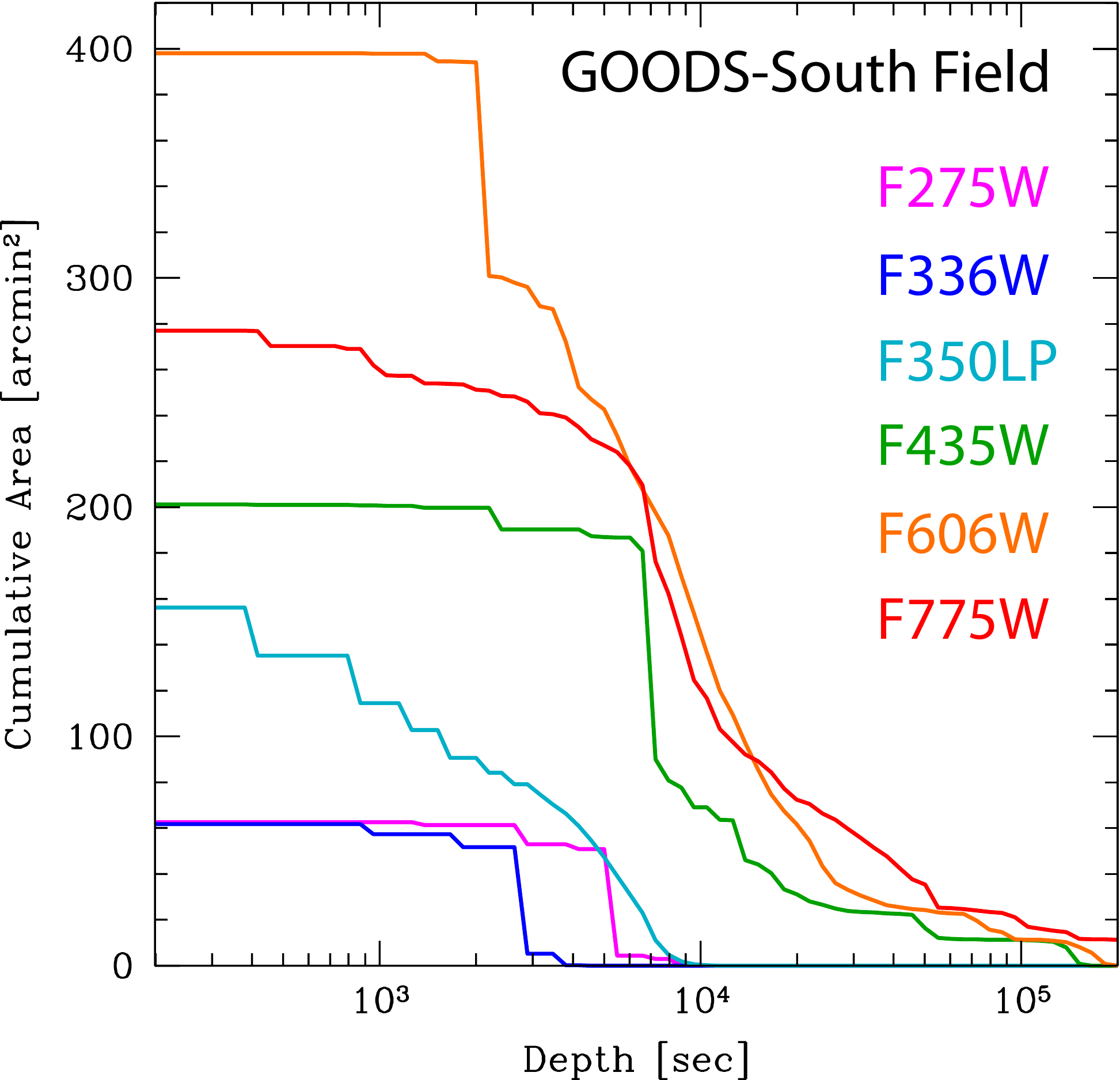}{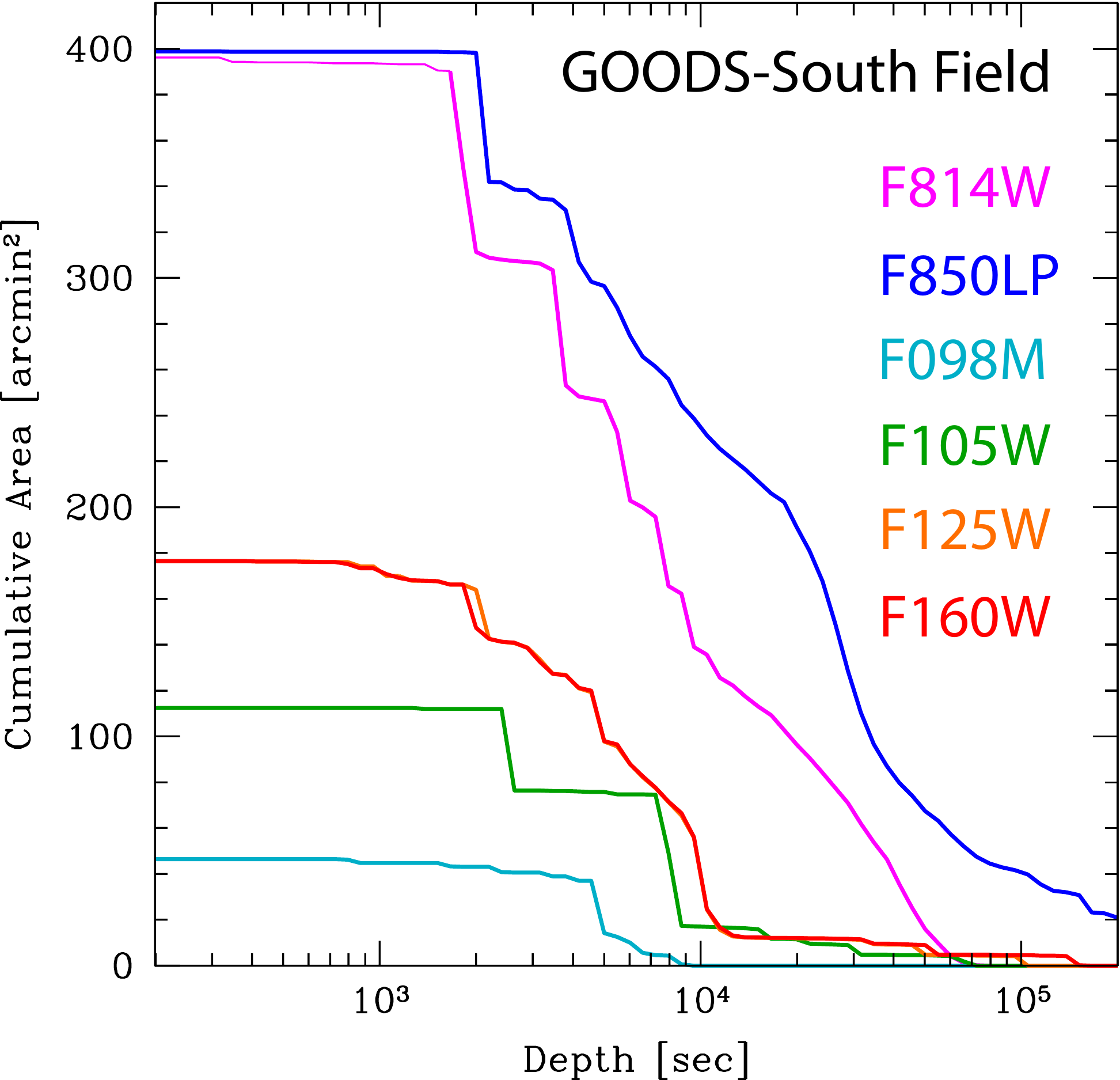}
\caption{\figgdsad}
\end{figure*}
\fi

Figures \ref{fig:gdnloexpmaps}a-f show the planned 
exposure maps for the CANDELS
GOODS-N field in the bluer filters: F275W (WFC3/UVIS), F336W (WFC3/UVIS),
F350LP (WFC3/UVIS), F435W (ACS/WFC), F606W (ACS/WFC), and F775W (ACS/WFC). 
We note that the original GOODS+PANS F775W coverage in GOODS-N has
been substantially augmented by the ACS parallel exposures of the large
WFC3/IR grism survey by Weiner et al.~(Program 11600).  This additional
F775W imaging is reflected in Figure~\ref{fig:gdnloexpmaps}f and in the 
F775W area vs.~depth curve to follow.
Figures \ref{fig:gdnhiexpmaps}a-e show the planned 
exposure maps for the CANDELS GOODS-N field in the redder 
filters: F814W (ACS/WFC3), F850LP (ACS/WFC),
F105W (WFC3/IR), F125W (WFC3/IR), and F160W (WFC3/IR). 
The exposure maps include all available ACS legacy imaging 
in this region matching the combined GOODS and CANDELS filter sets.  
At the time of writing, the filter assignments for the CANDELS
ACS parallel exposures have not been finalized.  We currently show
all ACS parallel exposures as F814W (see Fig.~\ref{fig:gdnhiexpmaps}), 
but a small fraction of these exposures will instead be taken with
F606W and F850LP, as we have done in the CANDELS GOODS-S.

\ifsubmode
\placefigure{fig:gdnloexpmaps}
\placefigure{fig:gdnhiexpmaps}
\else
\begin{figure*}[ht]
\plottwo{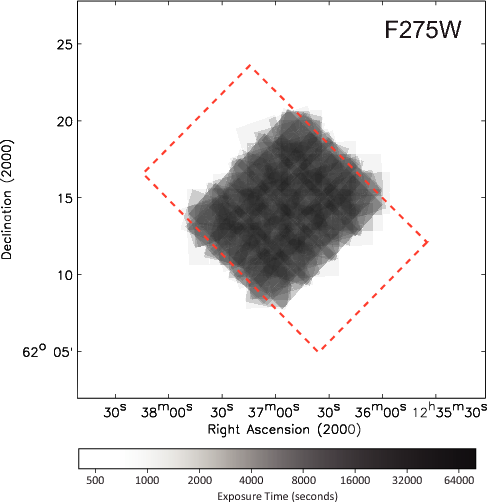}{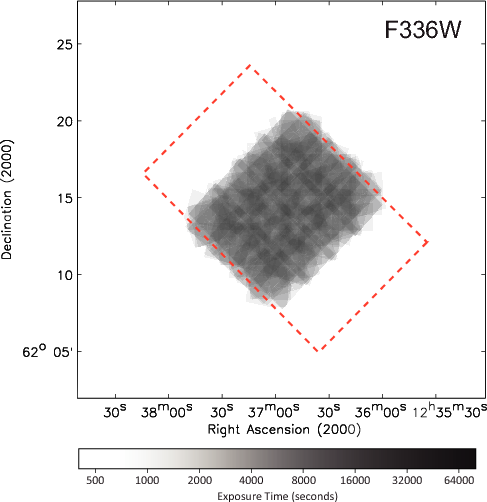}
\plottwo{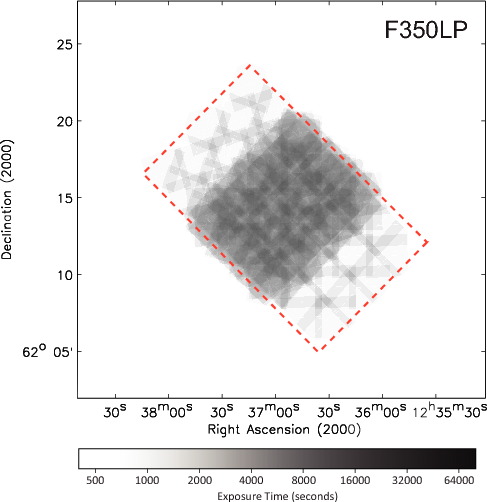}{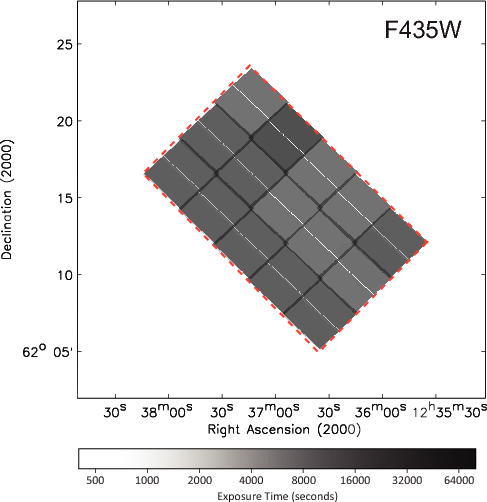}
\plottwo{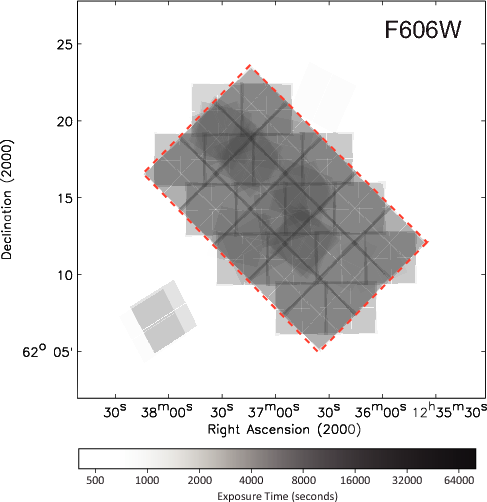}{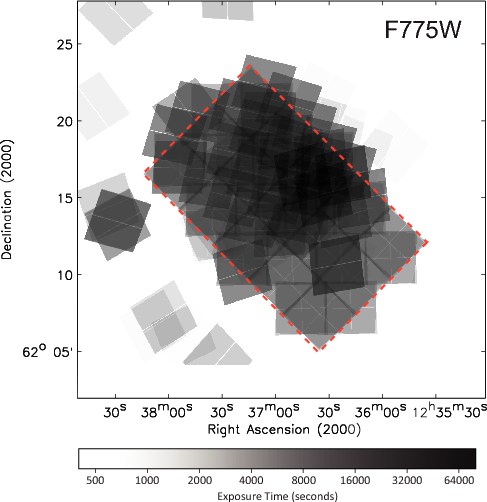}
\caption{\figgdnloexpmaps}
\end{figure*}

\begin{figure*}[ht]
\plottwo{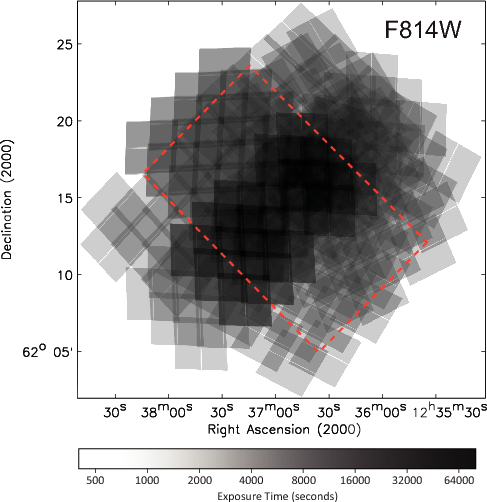}{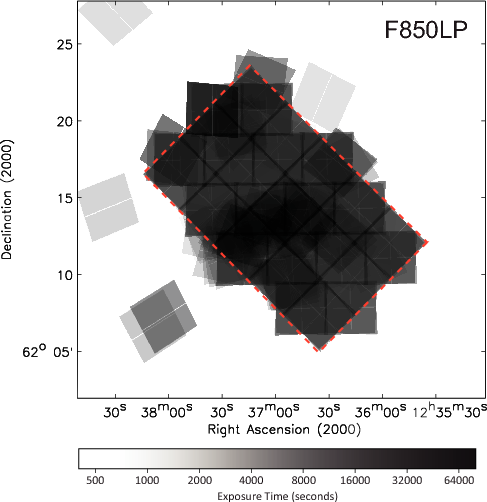}
\plottwo{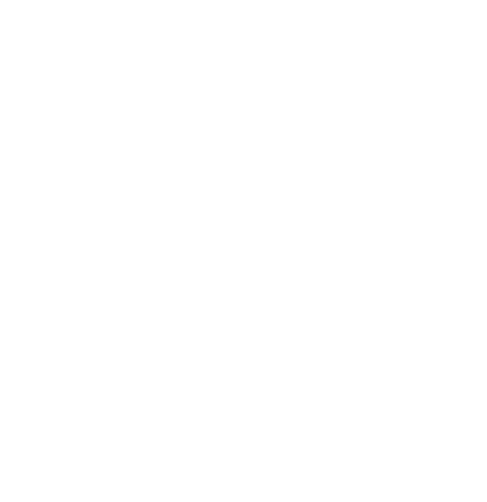}{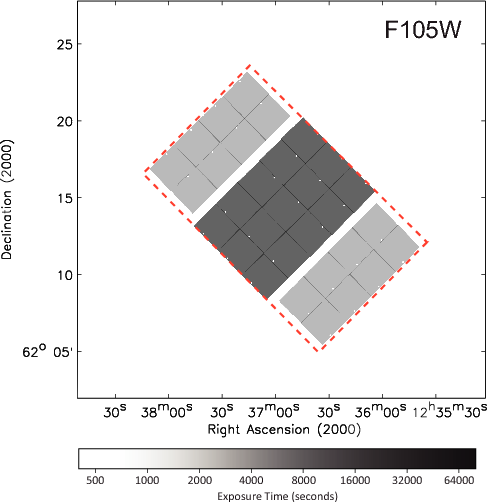}
\plottwo{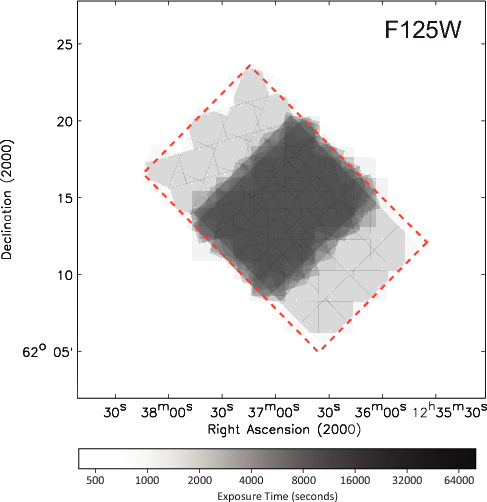}{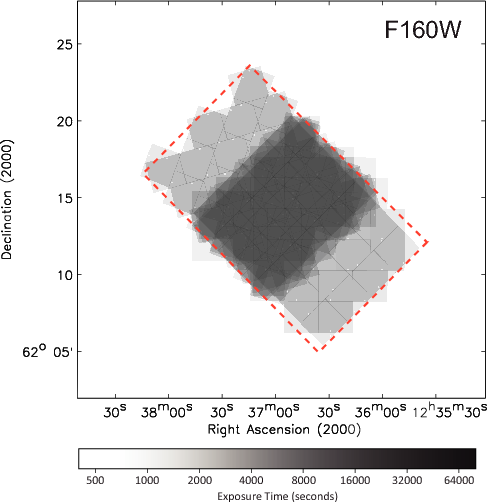}
\caption{\figgdnhiexpmaps}
\end{figure*}
\fi

Figures \ref{fig:gdnad}a,b show the curves of area versus exposure 
depth for each of the ACS and WFC3 filters that will comprise the 
CANDELS GOODS-N field.  Each of these area vs.~depth plots  
considers only the area within the footprint of 
CANDELS images but includes all legacy ACS imaging within that region.  
The curves for F606W
and F850LP will be slightly augmented at the expense of F814W, once
the GOODS-N filter assignments have been finalized.
\ifsubmode
\placefigure{fig:gdnad}
\else
\begin{figure*}[ht]
\plottwo{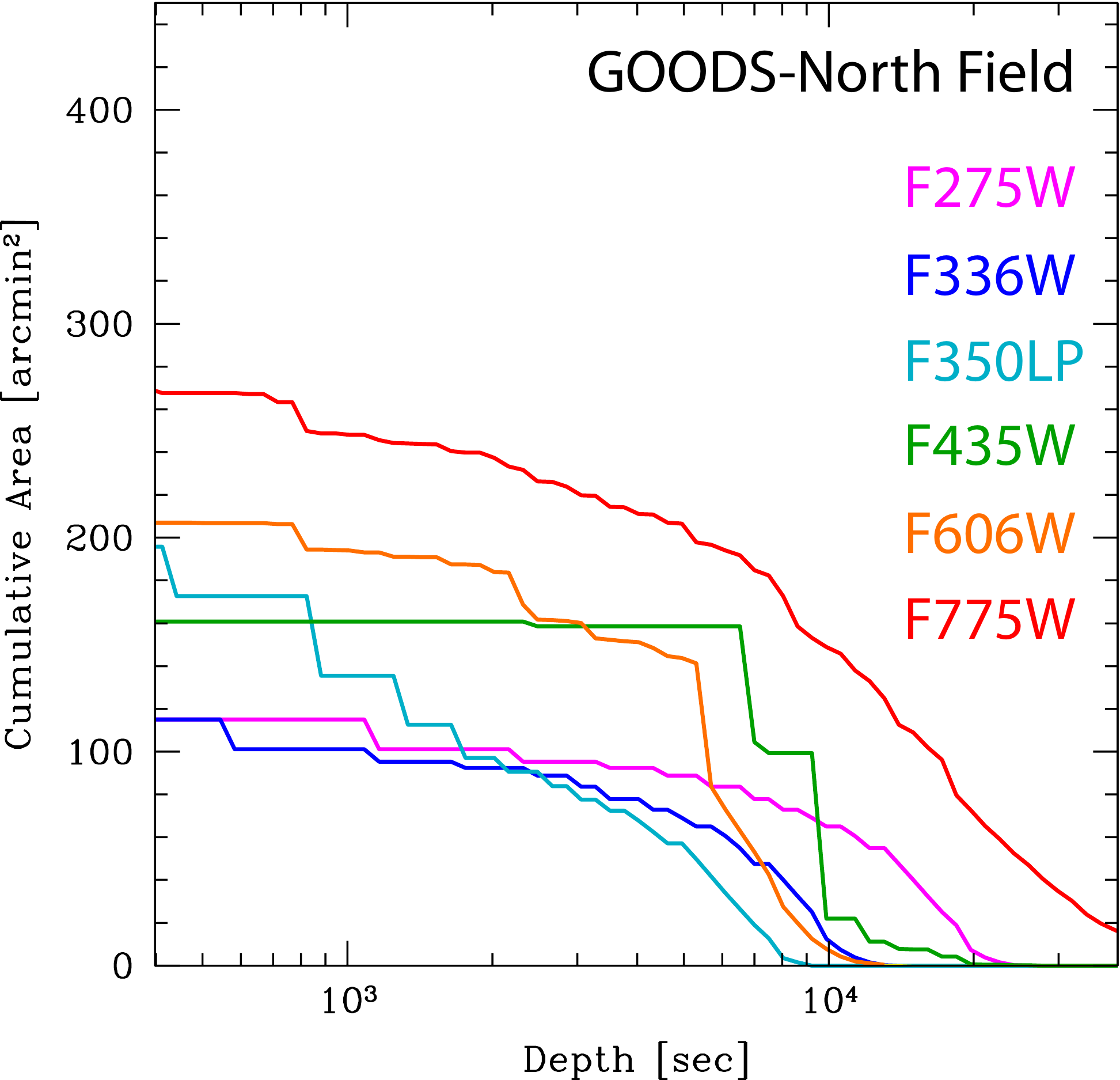}{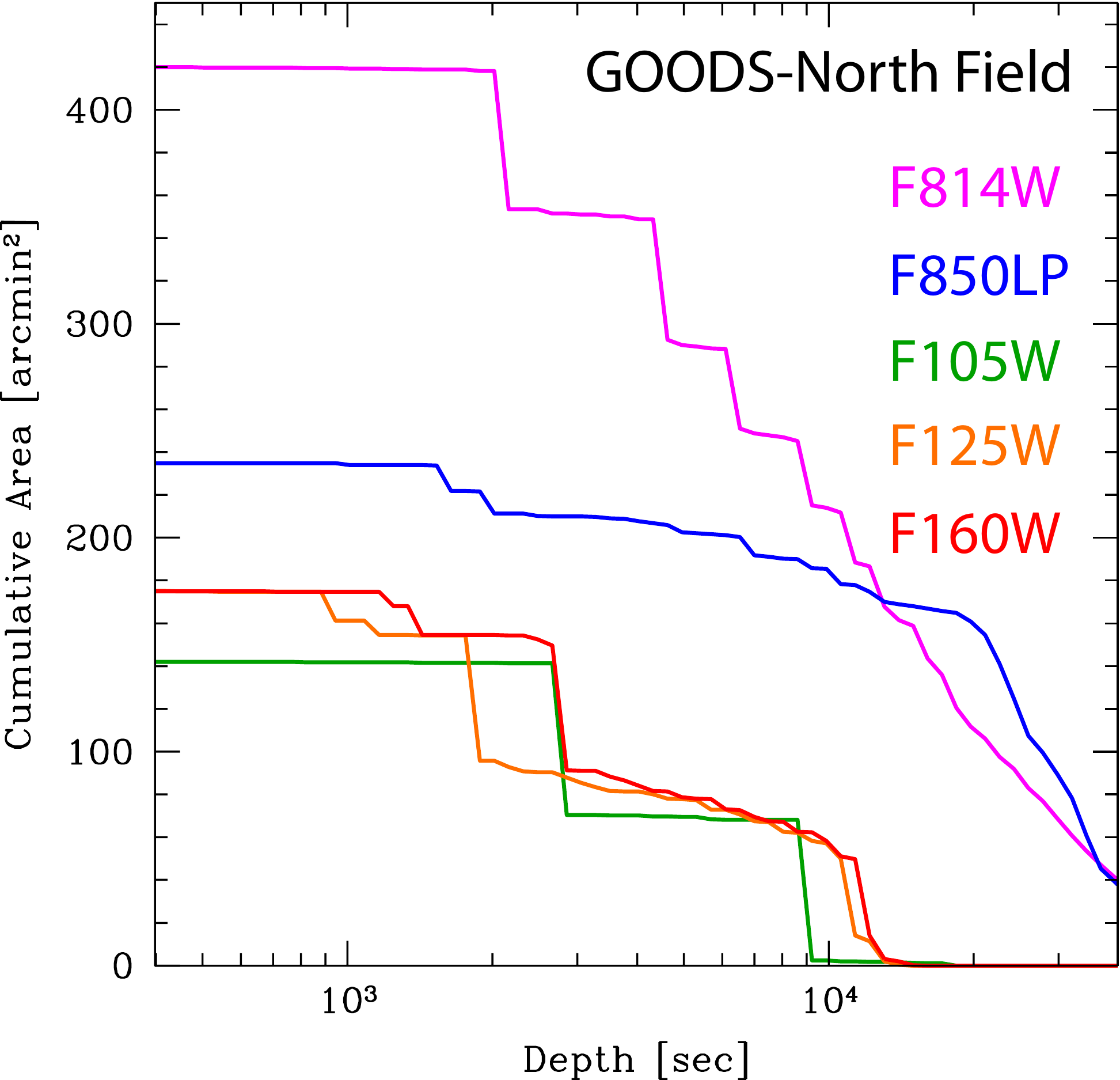}
\caption{\figgdnad}
\end{figure*}
\fi

The aggregate areas, \textit{HST} exposure times, and filter 
sensitivities of the Wide and Deep programs are summarized 
in Table \ref{tab:areadepth}.  
Given the heterogeneous ancillary \textit{HST} data in CANDELS, the exposure 
depths vary substantially from field to field and within each field
for both Wide and Deep programs (see Figs.~\ref{fig:widead}, 
\ref{fig:gdsad}, and \ref{fig:gdnad}).  We therefore quote an approximate
lower limit to the filter sensitivities in Table \ref{tab:areadepth}.
Point-source limiting magnitudes correspond to the ``optimal'' signal-to-noise
estimate provided by the \textit{HST} exposure time calculator.
Extended-source limiting magnitudes correspond to
$5\times$ the photometric error within a $0.2$ arcsec$^2$ aperture. 
Note that the GOODS-S ERS field is entirely legacy \textit{HST} imaging 
except for CANDELS ACS parallels to the GOODS-S Deep.

\clearpage
%
\ifsubmode 
  \vspace{-1in}
\fi
\begin{deluxetable}{lrcc}
\ifsubmode 
  \tabletypesize{\scriptsize}
\fi
\tablewidth{0pc}
\tablecolumns{4}
\tablecaption{CANDELS Areas, Exposure, and $5\sigma$ Sensitivities \label{tab:areadepth}}
\tablehead{
\colhead{HST} & \colhead{Exposure} &   
\colhead{Pt.-Src.~$5\sigma$} & \colhead{Extended $5\sigma$} \\
\colhead{Filter}   & \colhead{[s]} & \colhead{[mag]} & \colhead{[mag]} 
}
\startdata
\multicolumn{4}{c}{\rule{0pt}{12pt}\underline{\textit{CANDELS/Deep Program}}}\\
\multicolumn{4}{c}{\rule{0pt}{10pt}GOODS-South Field ($\sim 7\arcmin\times 10\arcmin$)}\\
F435W & 7200 & 28.7 & 27.7 \\
F606W & 10600 & 29.3 & 28.3 \\
F775W & 8600 & 28.5 & 27.6 \\
F814W & 31900 & 29.4 & 28.4 \\
F850LP & 28600 & 28.5 & 27.6 \\
F105W & 8100 & 28.2 & 28.1 \\
F125W & 9100 & 27.9 & 27.9 \\
F160W & 9200 & 27.6 & 27.6 \\
\multicolumn{4}{c}{\rule{0pt}{8pt}GOODS-North Field ($\sim 7\arcmin\times 10\arcmin$)}\\
F275W & 13500 & 29.1 & 27.9 \\
F336W & 6300 & 28.9 & 27.7 \\
F435W & 7200 & 28.7 & 27.7 \\
F606W & 8000 & 29.1 & 28.2 \\
F775W & 18600 & 28.9 & 28.0 \\
F814W & 32000 & 29.4 & 28.4 \\
F850LP & 30000 & 28.5 & 27.6 \\
F105W & 8900 & 28.3 & 28.2 \\
F125W & 10600 & 28.0 & 28.0 \\
F160W & 11200 & 27.7 & 27.8 \\
\multicolumn{4}{c}{\rule{0pt}{12pt}\underline{\textit{CANDELS/Wide Program}}}\\
\multicolumn{4}{c}{\rule{0pt}{10pt}UDS Field ($\sim 9\arcmin\times 24\arcmin$)}\\
F606W & 2800 & 28.4 & 27.5 \\
F814W & 5700 & 28.4 & 27.5 \\
F125W & 1900 & 27.0 & 27.0 \\
F160W & 3300 & 27.1 & 27.1 \\
\multicolumn{4}{c}{\rule{0pt}{8pt}EGS Field ($\sim 7\arcmin\times 26\arcmin$)}\\
F606W & 5700 & 28.9 & 28.0 \\
F814W & 10000 & 28.8 & 27.9 \\
F125W & 1900 & 27.2 & 27.1 \\
F160W & 3600 & 27.3 & 27.4 \\
\multicolumn{4}{c}{\rule{0pt}{8pt}COSMOS Field ($\sim 9\arcmin\times 24\arcmin$)}\\
F606W & 3300 & 28.5 & 27.6 \\
F814W & 6900 & 28.5 & 27.6 \\
F125W & 1900 & 27.0 & 27.0 \\
F160W & 3200 & 27.1 & 27.1 \\
\ifsubmode
  \tablebreak
\fi
\multicolumn{4}{c}{\rule{0pt}{8pt}GOODS-South Flanking ($\sim 4\arcmin\times 10\arcmin$)}\\
F435W & 7200 & 28.7 & 27.8 \\
F606W & 8600 & 29.2 & 28.2 \\
F775W & 7000 & 28.4 & 27.4 \\
F814W & 7500 & 28.6 & 27.7 \\
F850LP & 30000 & 28.5 & 27.6 \\
F105W & 2700 & 27.6 & 27.5 \\
F125W & 2100 & 27.1 & 27.1 \\
F160W & 2100 & 26.8 & 26.8 \\
\multicolumn{4}{c}{\rule{0pt}{8pt}GOODS-South ERS ($\sim 4\arcmin\times 9\arcmin$)}\\
F225W & 5700 & 27.7 & 26.6 \\
F275W & 5700 & 27.8 & 26.7 \\
F336W & 2800 & 27.7 & 26.6 \\
F435W & 7200 & 28.7 & 27.8 \\
F606W & 5500 & 28.9 & 28.0 \\
F775W & 7000 & 28.4 & 27.5 \\
F814W & 10000 & 28.7 & 27.8 \\
F850LP & 17200 & 28.2 & 27.3 \\
F098M & 5000 & 27.5 & 27.4 \\
F125W & 5000 & 27.6 & 27.5 \\
F160W & 5000 & 27.3 & 27.3 \\
\multicolumn{4}{c}{\rule{0pt}{8pt}GOODS-North Flanking (2 @ $\sim 4\arcmin\times 9\arcmin$)}\\
F435W & 7200 & 28.7 & 27.7 \\
F606W & 5600 & 28.9 & 28.0 \\
F775W & 8100 & 28.4 & 27.5 \\
F814W & 12000 & 28.8 & 27.9 \\
F850LP & 25000 & 28.4 & 27.5 \\
F105W & 2900 & 27.7 & 27.5 \\
F125W & 1900 & 27.1 & 27.0 \\
F160W & 2800 & 27.0 & 27.0 
\enddata
\end{deluxetable}
\clearpage

\subsection{Scheduling of Observations}\label{sec:sched}
We summarize the currently planned timetable of CANDELS epochs in
Figure \ref{fig:epochs}, including dates of execution, orbits per epoch,
\textit{HST} orientation, and a five-character alphanumeric descriptor that will
prefix the given epoch's CANDELS data releases (see Koekemoer et al.~2011
for details).  The entries are color-coded by field, in chronological
order with GOODS-S (cyan), UDS (orange), EGS (blue), COSMOS (yellow),
and GOODS-N (magenta).  The very first entry is a single test orbit
to verify sensitivities and dithering pattern, taken in the GOODS-S
region (at the northwest corner of the ERS) and included in the Program
12061 PhaseII description.  The portions of the CANDELS program that
have been assigned \textit{HST} Program IDs at the time of writing are listed
in the last column of the timetable.

With ten SN~Ia-search epochs spaced every $\sim 52$ days, each CANDELS
Deep campaign must execute steadily across nearly 1.5 years.  To avoid 
overlap of epochs among the different CANDELS fields during the
three-year execution of the program, we started
the GOODS-S observations early in Cycle 18 and plan to 
begin the GOODS-N observations almost immediately after
GOODS-S completes, in the latter part of Cycle 19.  The last
GOODS-N epoch is planned to complete near the end of Cycle 20.
The slightly longer duration of the GOODS-S program is due to the
couple of months in Spring 2011 when the field is too close to the
sun for \textit{HST} observations, and the inter-epoch gap must be extended
well beyond the 52\,d optimal for SN~Ia-searching.  GOODS-N, on the
other hand, can be observed year-round.

In the GOODS-N field, the $\sim 52$ day cadence for optimal detection
of extreme-redshift SNe~Ia fortuitously matches the intervals between
CVZ opportunities.  Our intention is to align the CANDELS GOODS-N
observing schedule with the CVZ cadence in order to obtain WFC3/UVIS
F275W and F336W exposures in the Deep region during the bright-earth portion
of the \textit{HST} orbits, at no additional cost in orbits to the program.

The CANDELS/Wide campaigns, whose tightly-constrained 
\textit{HST} orientations must be held fixed for $>\!60$d spans, 
are scheduled to avoid overlap with the CANDELS/Deep epochs, and to 
distribute the CANDELS \textit{HST} orbit allocation as evenly as
possible across the full three years of the Multi-Cycle Treasury Program.

In all cases we have tried to compress the epochs into as short
an interval as reasonably schedulable with \textit{HST}, and to perform
all observations in a given epoch at identical \textit{HST} orientation.
It is inevitable that the large WFC3/IR rasters include a small fraction
of tiles with no suitable \textit{HST} guide-stars at exactly the desired orientation.
In the rare cases that the nearest guide-star orientation to our 
desired epoch's value is unobservable during an epoch's scheduled window, 
or when a guide-star failure occurs in the course of program 
execution and the pointing must be re-observed, the corresponding tile is by
necessity observed outside the window listed in the timetable.

The up-to-date Phase II-level description of the CANDELS \textit{HST} observations,
including variances from the plans described herein because of
guide-star failures, scheduling incompatibilities, etc., may be
obtained by querying the corresponding Program ID status on the STScI
website\footnote{http://www.stsci.edu/hst/scheduling/program\_information}.

\ifsubmode
  \placefigure{fig:epochs}
\else
\begin{figure*}[ht]
  \includegraphics[width=\textwidth]{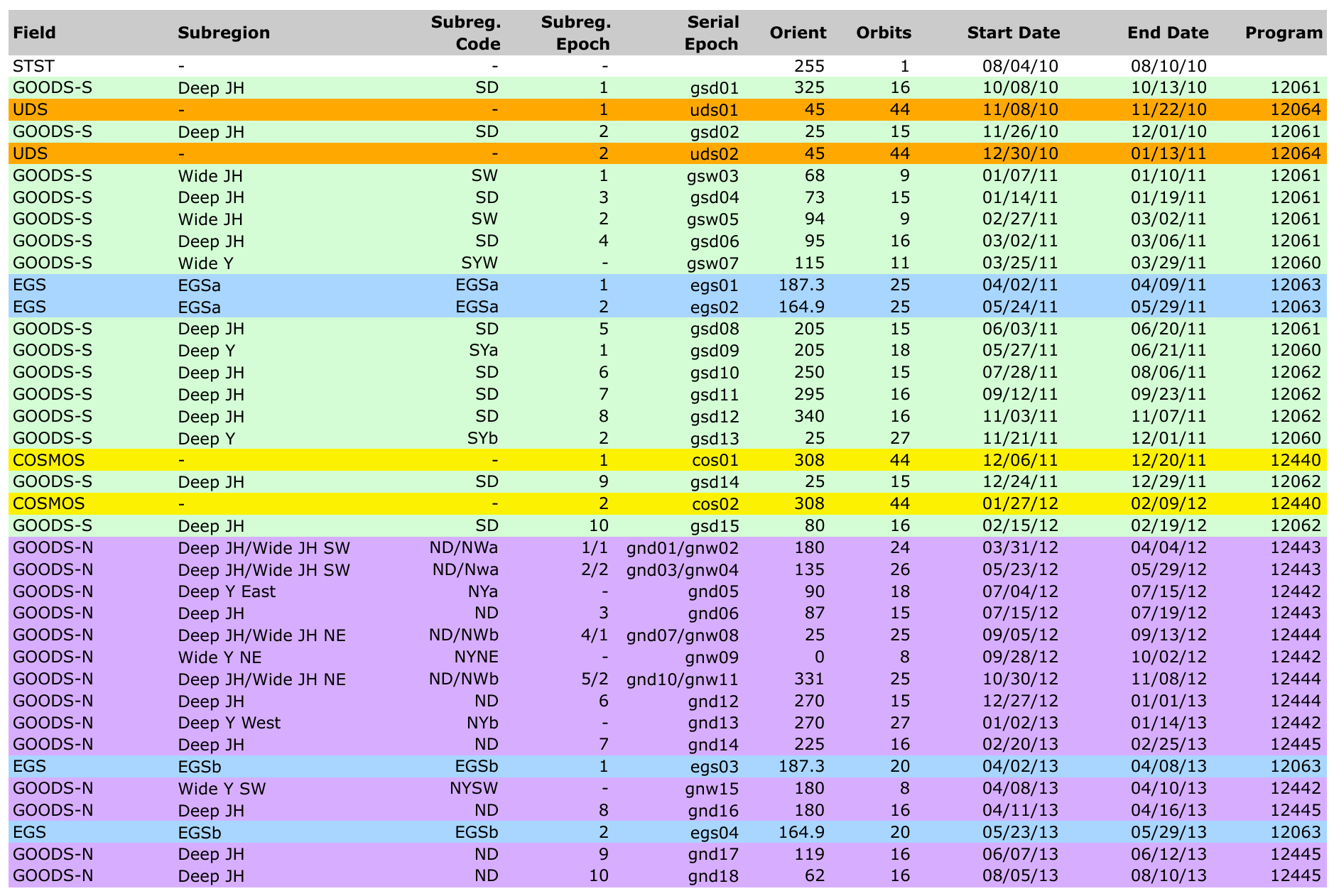}
\caption{\figepochs}
\end{figure*}
\fi

\ifsubmode
\else
\clearpage
\fi

\section{Summary}\label{sec:discussion}

We have presented an overview of the Cosmic Assembly Near-infrared
Extragalactic Legacy Survey (CANDELS), an ongoing \textit{HST} imaging survey of
five pre-eminent extragalactic deep fields, allocated approximately ten
percent of \textit{HST}'s available observing time for three years.  We intend this
overview to serve as a reference for the anticipated wealth of science
enabled by this survey, underway since Fall 2010.  We have enumerated
the CANDELS collaboration's principal science goals that have guided the
survey design, but we also emphasize that CANDELS is an \textit{HST} Treasury
survey with strong commitment to rapid dissemination of data products
for use by the wider astronomical community.  Our science goals may be
roughly segregated into the topics of ``Cosmic Dawn'' ($z\gtrsim6$),
``Cosmic High Noon'' ($z \approx 2$), SN cosmology, and moderate-$z$
galaxies' far-UV\@.  We refer the reader to Table \ref{tab:science_goals}
for an outline of these science goals.

We have described the specifics of our observing strategy
for achieving the aforementioned science goals, in as complete detail as
possible for a survey still in its early stages and not yet fully evolved
from conception to implementation.  We have presented justifications
for the five chosen CANDELS fields, with attention to the ancillary data
that are a crucial complement to achieve our goals.  We have also provided
our best estimate of the CANDELS \textit{HST} schedule of observations, aggregate
exposure maps, and limiting sensitivities.  These sensitivity estimates,
while useful benchmarks for our survey design, will undoubtedly vary
somewhat from the delivered mosaics' sensitivities, which we will duly
report in our forthcoming papers on CANDELS data releases.

The \textit{HST} images we have received to date have been completely
processed through
the CANDELS pipeline, described in the companion paper by Koekemoer et
al.~(2011).  There we present the details of the calibration procedures
and image processing, as well as the resulting public-release mosaic
data products for each field.  In brief, we create a set of single-epoch
combined mosaics in all the filters that are obtained in each epoch
separately, as well as a set of combined mosaics that include all the data
so far obtained on a given field.  We perform a series of specialized
calibration steps that include improved treatment of detector warm
pixels and persistence in the WFC3/IR images, and correction for bias
striping, amplifier crosstalk, and CTE degradation in the ACS images.
Full details may be found in Koekemoer et al.~(2011), as well as
the public CANDELS website\footnote{http://candels.ucolick.org}.
Three months after the observations of each CANDELS
epoch are completed (see Fig.~\ref{fig:epochs}), we release
the calibrated mosaic images to the public via the STScI
archive\footnote{http://archive.stsci.edu/prepds/candels/}.  At the time
of writing, we have already released the first several GOODS-S epochs,
the full UDS campaign, and the first EGS campaign (covering over half 
that field).  
We strongly encourage the astronomical
community to make use of CANDELS to advance their own research.

\section {Acknowledgments}\label{sec:ack}

We would like to thank our Program Coordinators, Tricia Royle and
Beth Perriello, along with the rest of the \Hubble\ planning team,
for their efforts to schedule this challenging program. The WFC3 team
has made substantial contributions to the program by calibrating and
characterizing the instrument, and have provided much useful advice.
Rychard Bouwens gave helpful input on the observing strategy for the
CANDELS/Deep survey.  John Mackenty suggested using $2 \times 2$ on-chip
binning for the UV observations, which significantly improves the
signal-to-noise ratio of those observations.  The CANDELS observations
would not have been possible without the contributions of hundreds
of other individuals to the \Hubble\ missions and the development and
installation of new instruments.  

Support for {\it HST} Programs GO-12060 and GO-12099
was provided by NASA through grants from the Space Telescope Science
Institute, which is operated by the Association of Universities for
Research in Astronomy, Inc., under NASA contract NAS5-26555.
A.~Fontana acknowledges support from agreement ASI-INAF I/009/10/0.

{\it Facilities:} \facility{HST (WFC3)}

\bibliographystyle{apj}
\bibliography{apjmnemonic,manuscript}

\ifsubmode
\insertfigend{rosario_udf_ih_montage.eps}{\figmontage}
\insertfigend{mclure_z7z8lf_v2.eps}{\figlbglf}
\insertfigend{limiting_mags.eps}{\figlimits}
\insertfigend{CANDELS-EBL.eps}{\figebl}
\insertfigend{wuyts_comp_zphot.H25crop.eps}{\figzphot}
\insertfigend{rosario_figA_ih_montage_v3.eps}{\figlargemontage}
\insertfigend{lotz_combo_v2.eps}{\figmorph}
\insertfigend{lotz_galaxymerger.eps}{\figmerger}
\insertfigend{uv.eps}{\figuv}
\insertfigend{SN_CANDELS_fig1.eps}{\figsndndz}
\newpage
\epsscale{0.8}
\begin{figure*}[ht]
\plotone{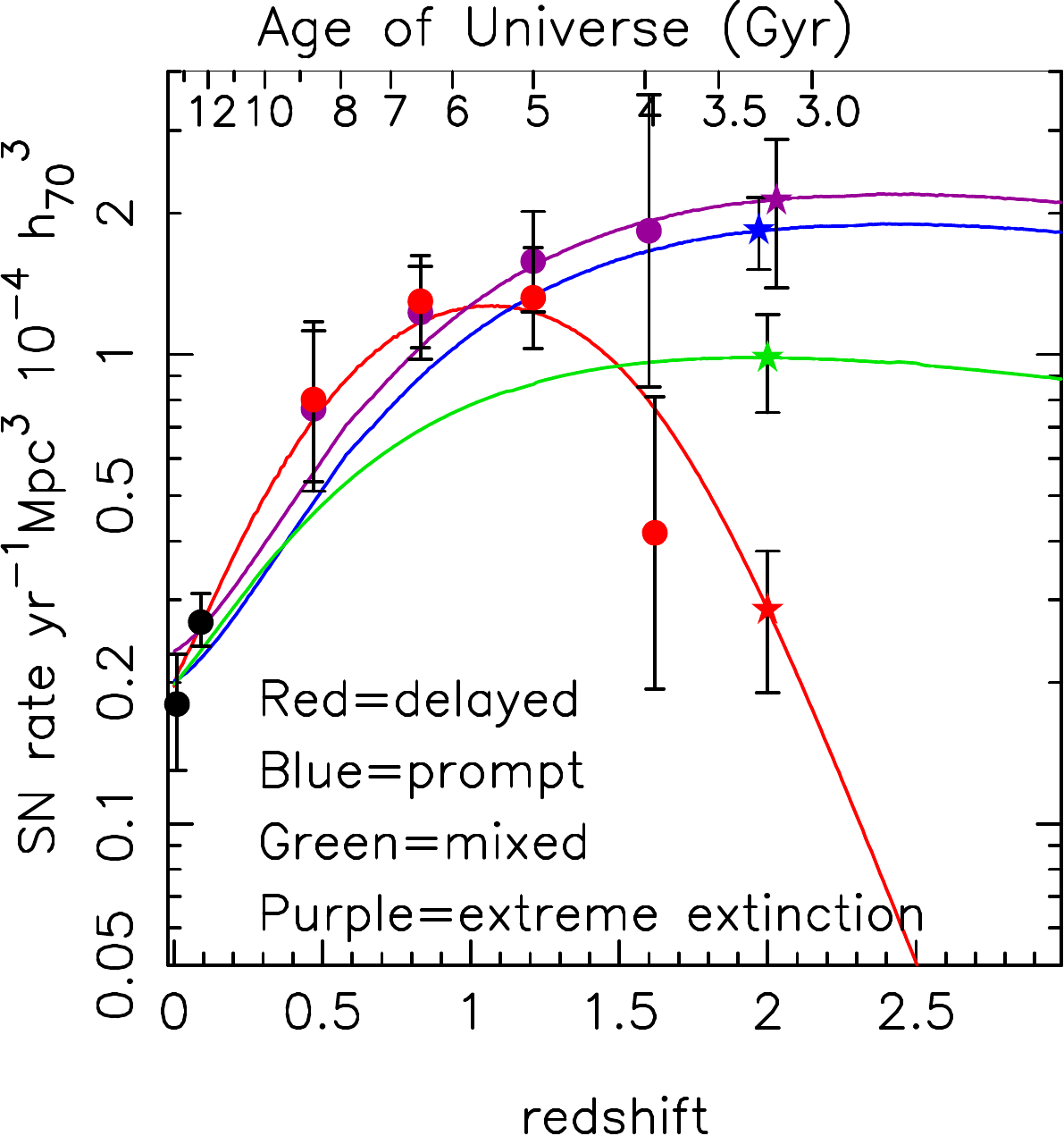}
\caption{\figsnrates}
\end{figure*}
\clearpage
\insertfigend{gdssimp.eps}{\figgds}
\insertfigend{gdnsimp.eps}{\figgdn}
\insertfigend{uds_layout.eps}{\figudslayout}
\insertfigend{cosmos_layout.eps}{\figcosmoslayout}
\insertfigend{EGS_layout.eps}{\figegslayout}
\insertfigend{EGS_layout_epochs.eps}{\figegscombo}
\insertfigend{GOODSS_visit_summary1.eps}{\figgdsepa}
\insertfigend{GOODSS_visit_summary2.eps}{\figgdsepb}
\newpage
\epsscale{0.8}
\begin{figure*}[ht]
\plottwo{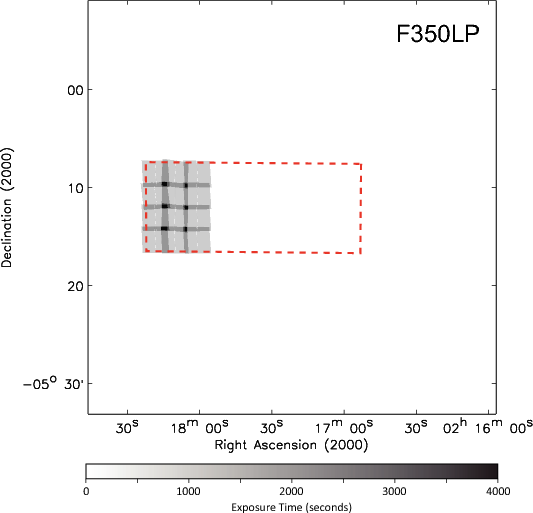}{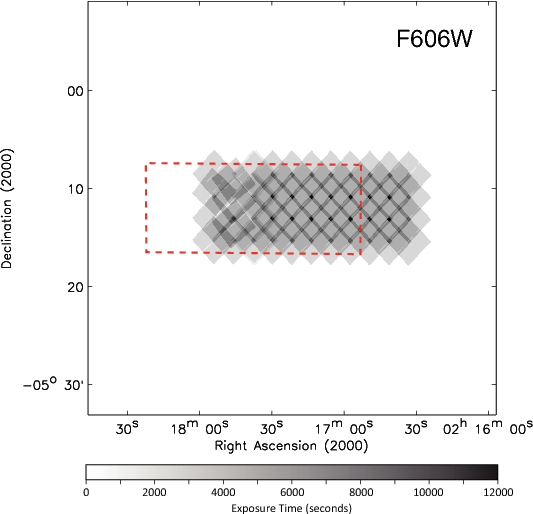}
\plottwo{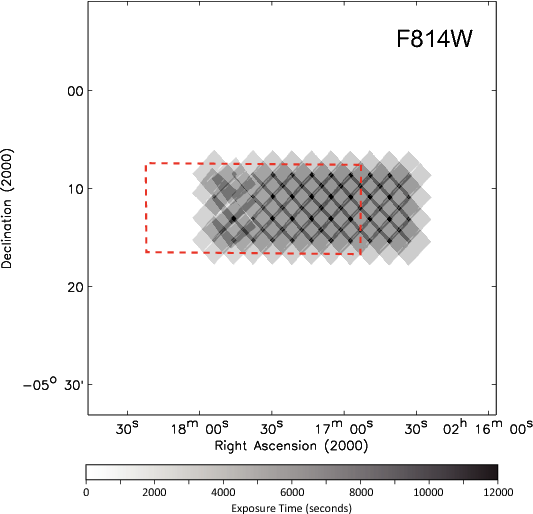}{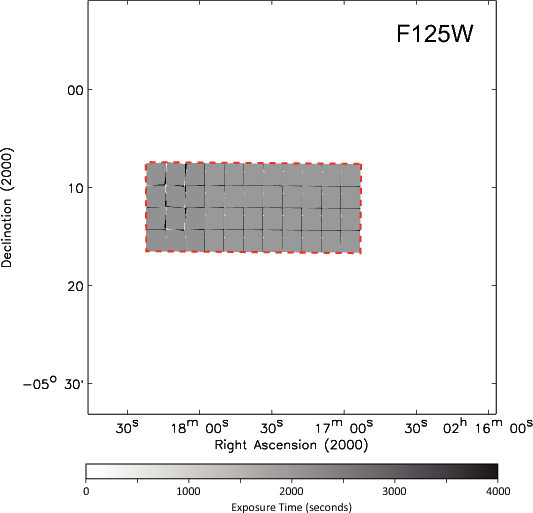}
\plottwo{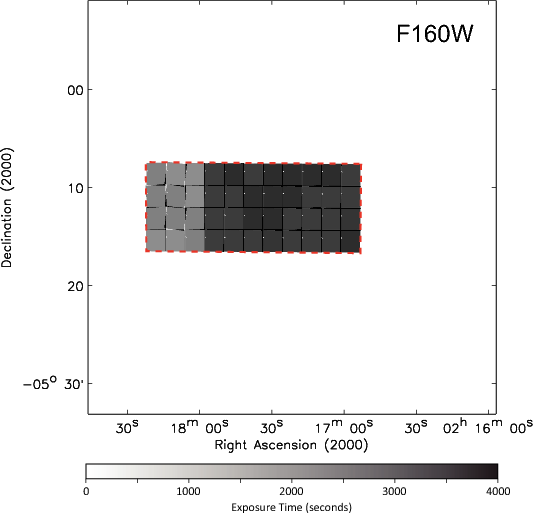}{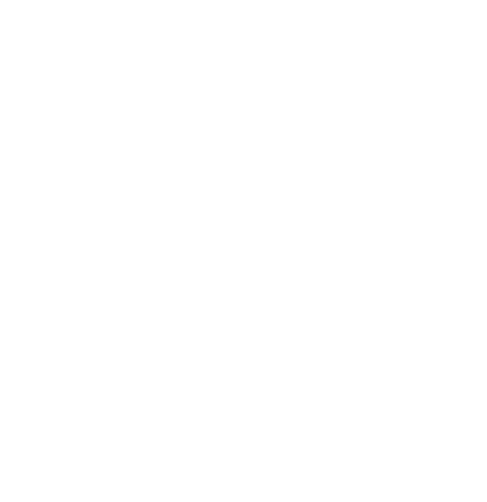}
\caption{\figudsexpmaps}
\end{figure*}
\newpage
\begin{figure*}[ht]
\plottwo{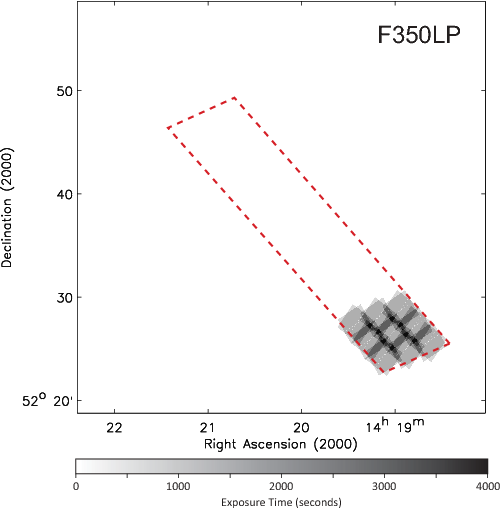}{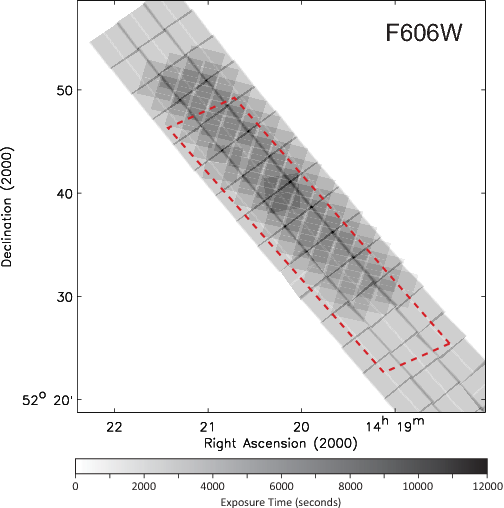}
\plottwo{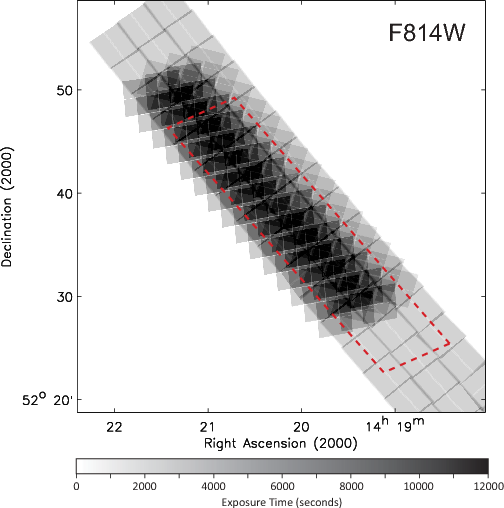}{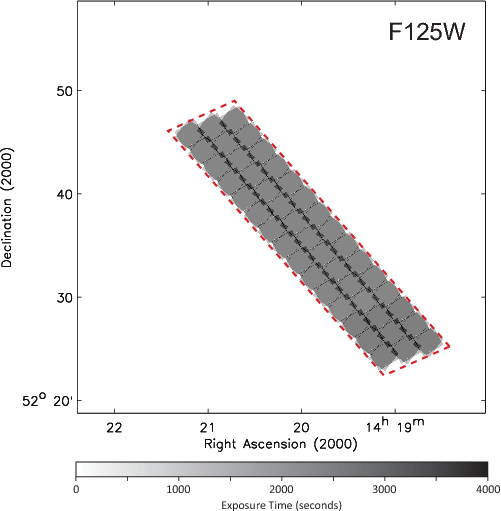}
\plottwo{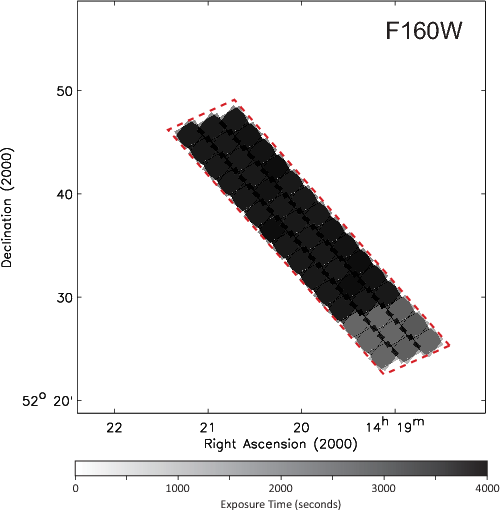}{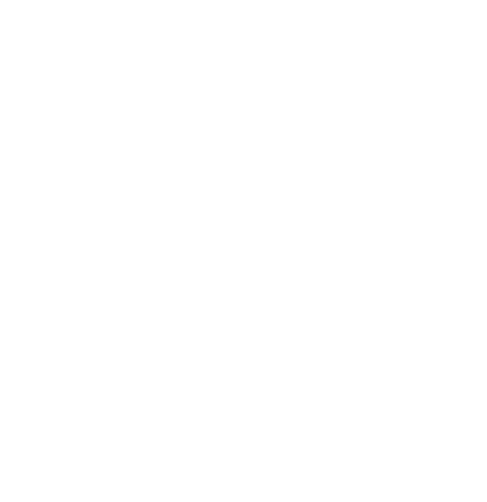}
\caption{\figegsexpmaps}
\end{figure*}
\newpage
\begin{figure*}[ht]
\plottwo{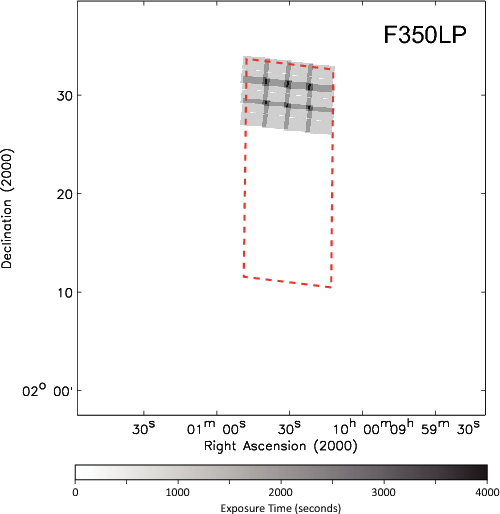}{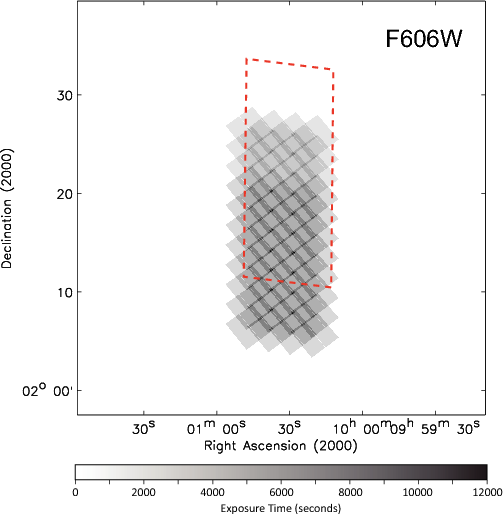}
\plottwo{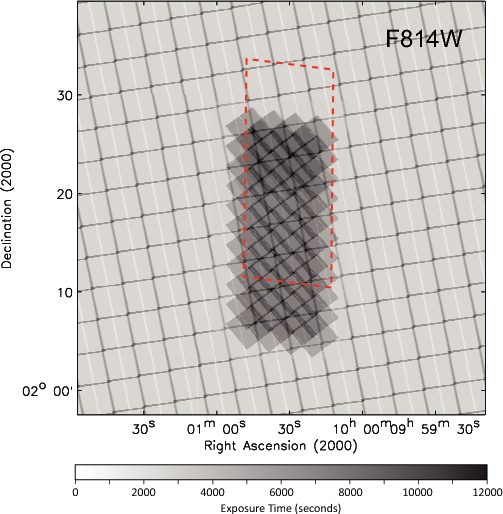}{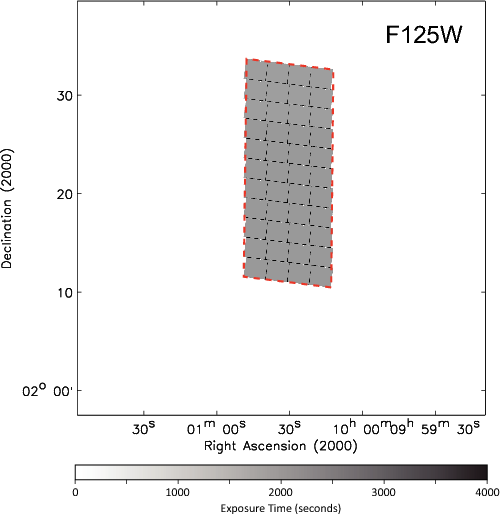}
\plottwo{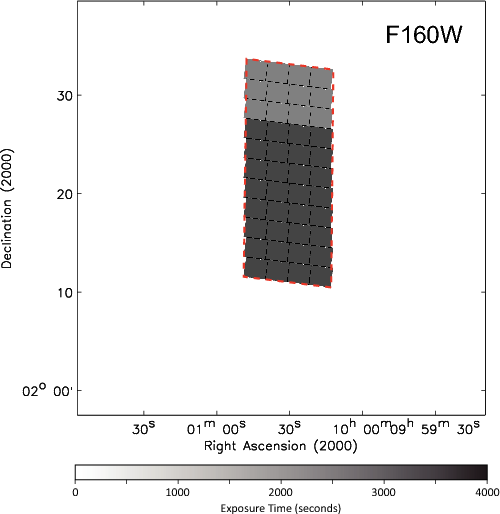}{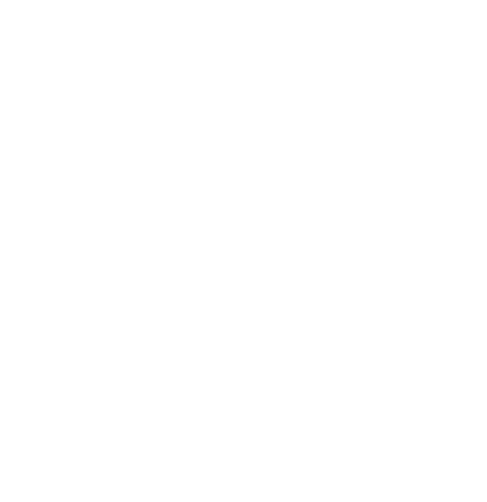}
\caption{\figcosmosexpmaps}
\end{figure*}
\newpage
\begin{figure*}[ht]
\plottwo{uds_areadepth2.eps}{egs_areadepth2.eps}
\plottwo{cos_areadepth2.eps}{blank_areadepth.eps}
\caption{\figwidead}
\end{figure*}
\newpage
\begin{figure*}[ht]
\plottwo{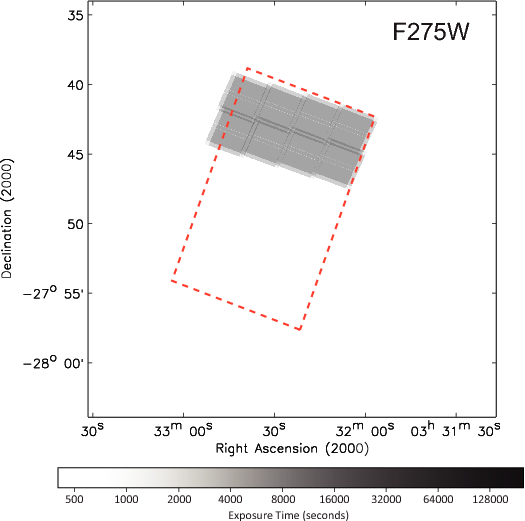}{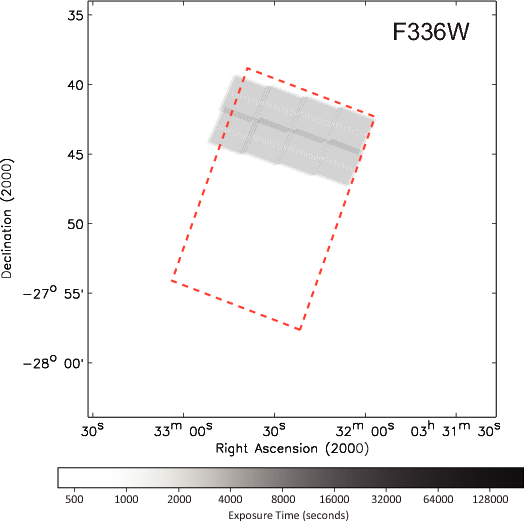}
\plottwo{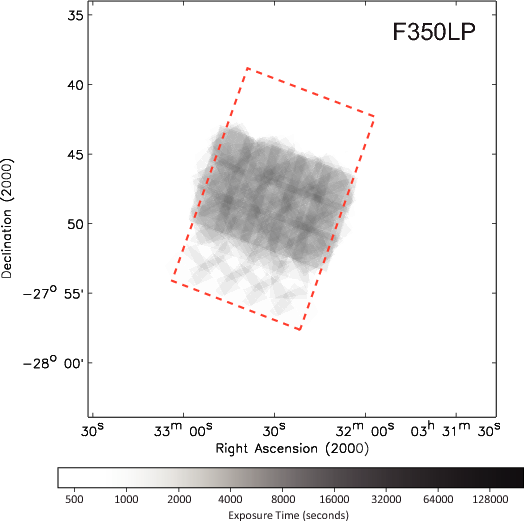}{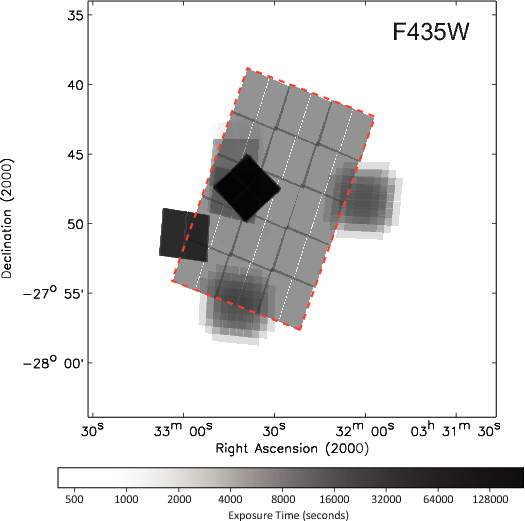}
\plottwo{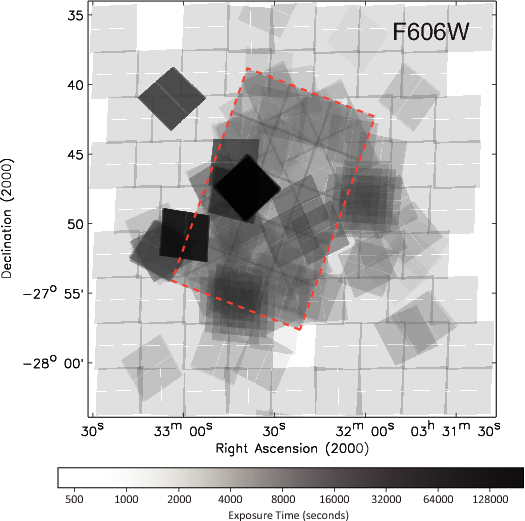}{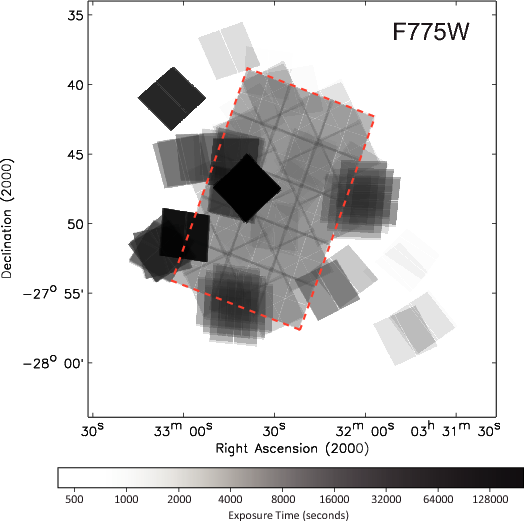}
\caption{\figgdsloexpmaps}
\end{figure*}
\newpage
\begin{figure*}[ht]
\plottwo{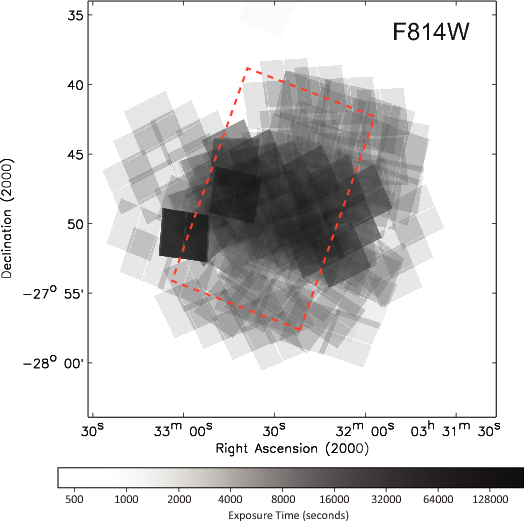}{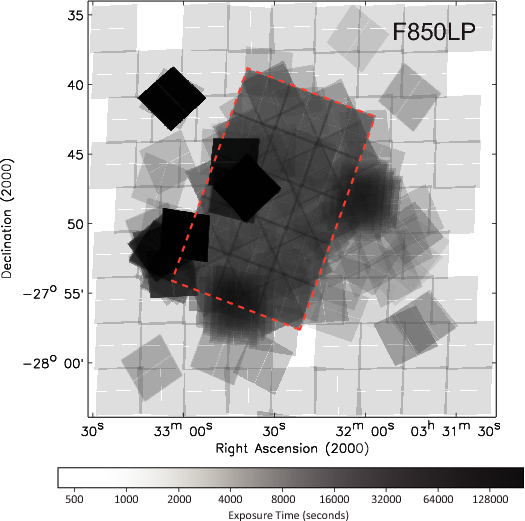}
\plottwo{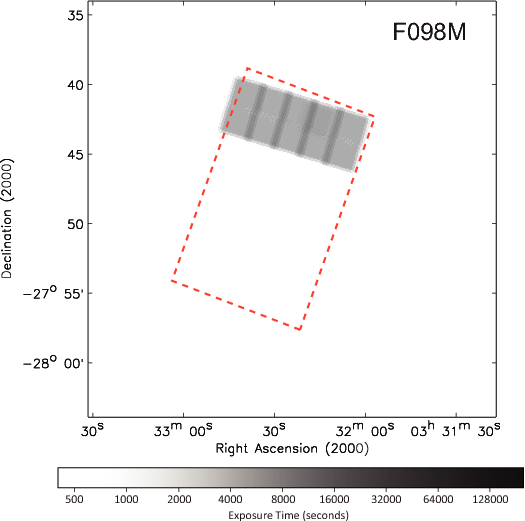}{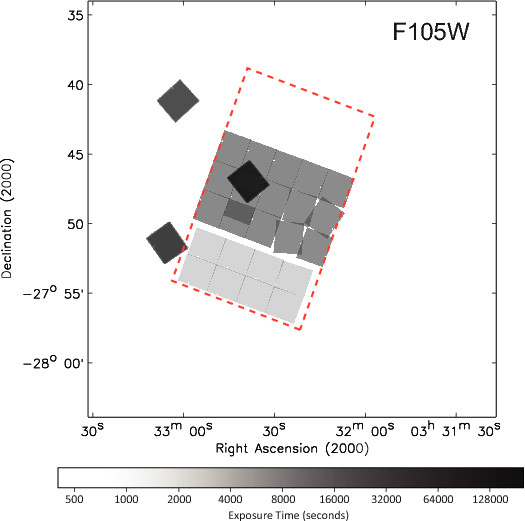}
\plottwo{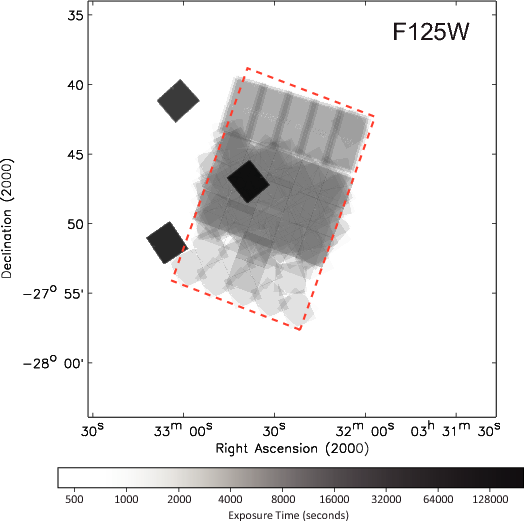}{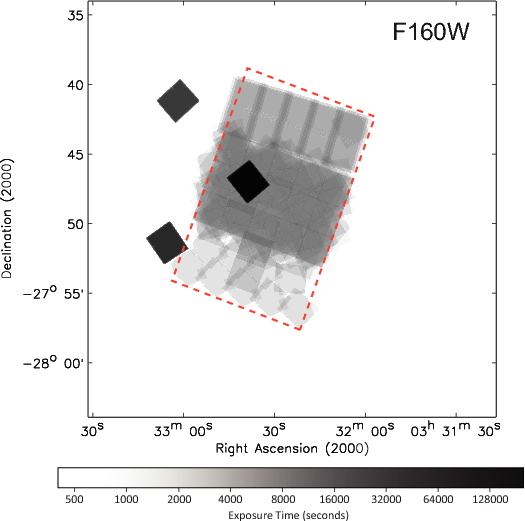}
\caption{\figgdshiexpmaps}
\end{figure*}
\newpage
\begin{figure*}[ht]
\plottwo{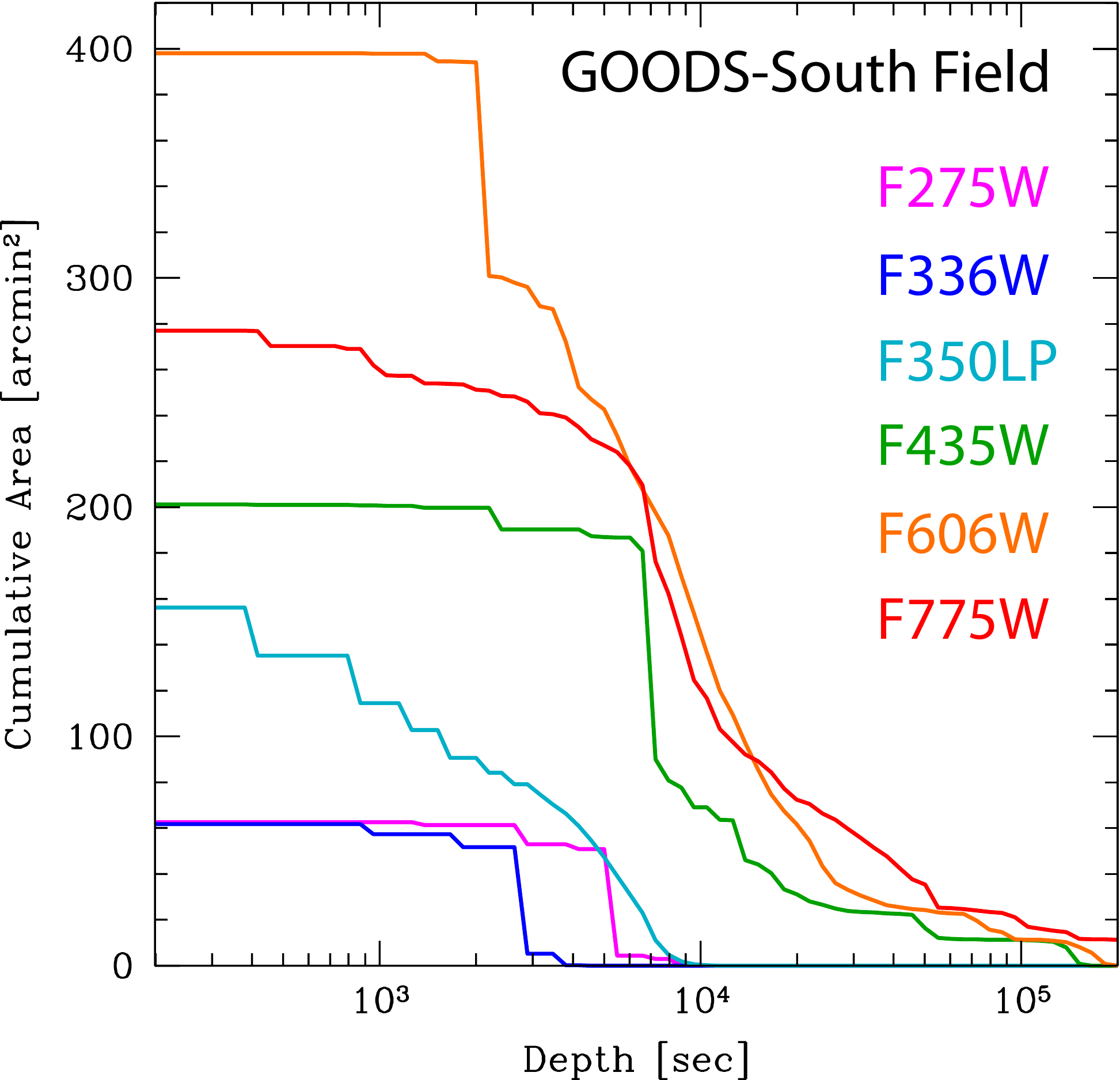}{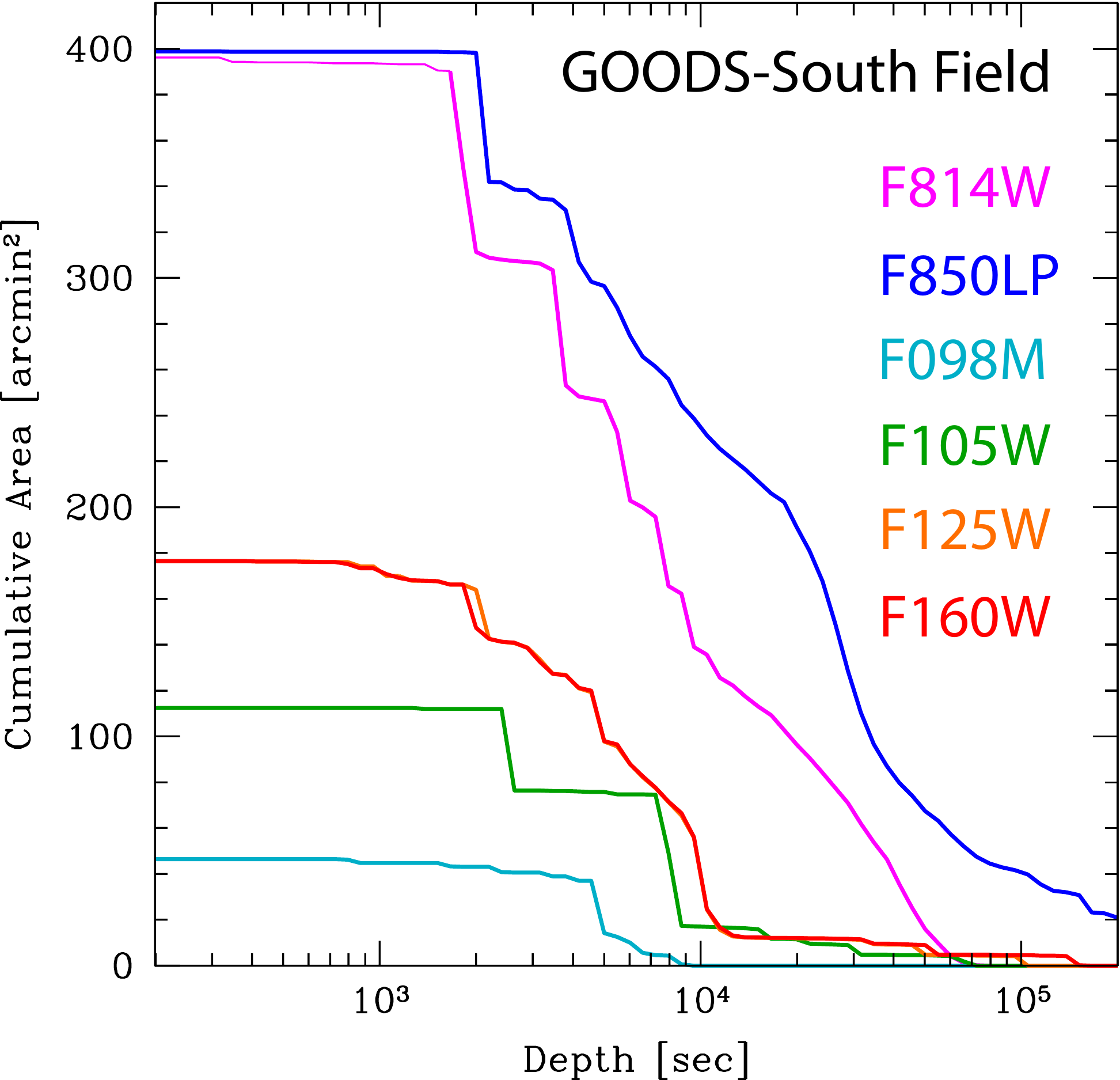}
\caption{\figgdsad}
\end{figure*}
\newpage
\begin{figure*}[ht]
\plottwo{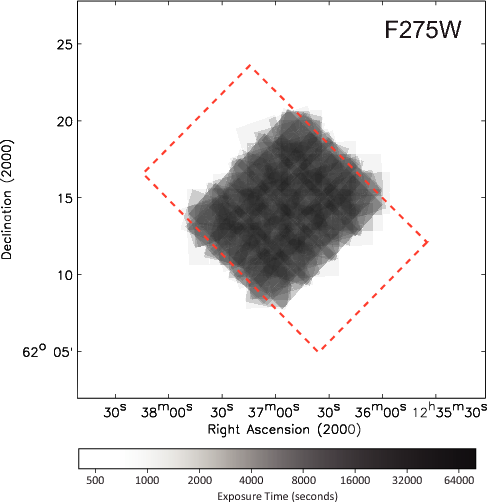}{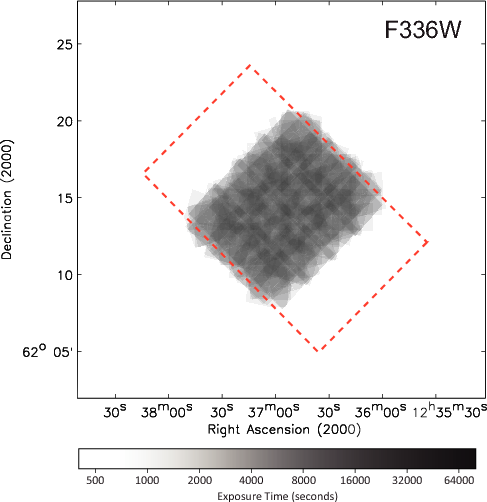}
\plottwo{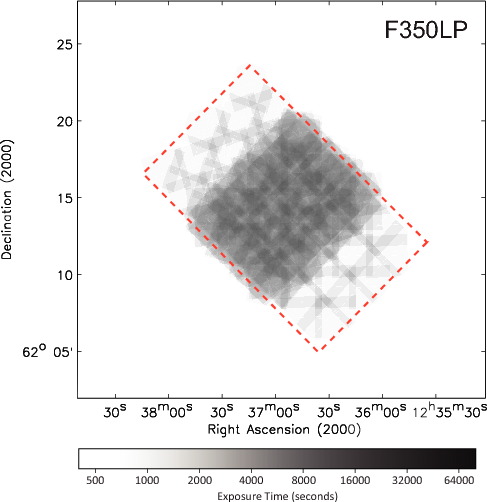}{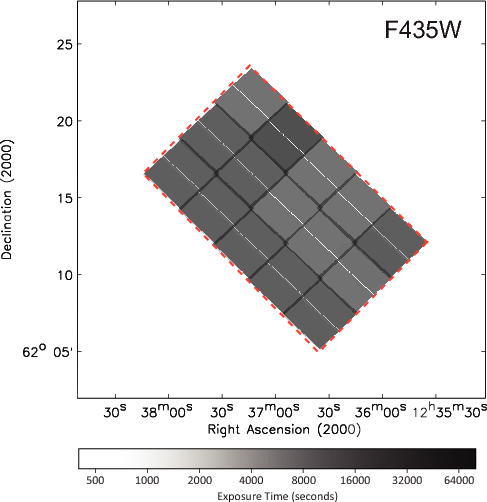}
\plottwo{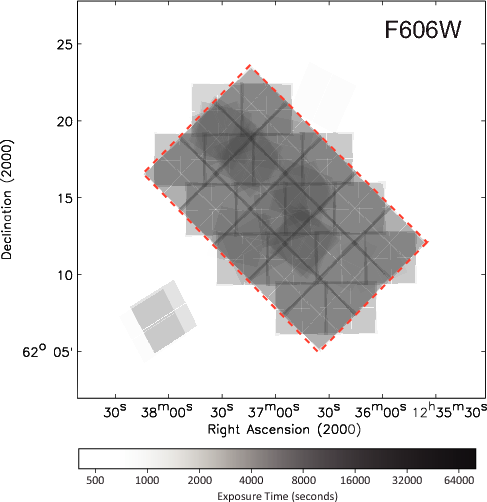}{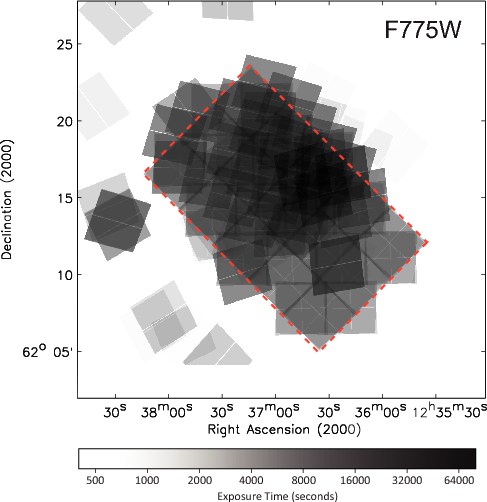}
\caption{\figgdnloexpmaps}
\end{figure*}
\newpage
\begin{figure*}[ht]
\plottwo{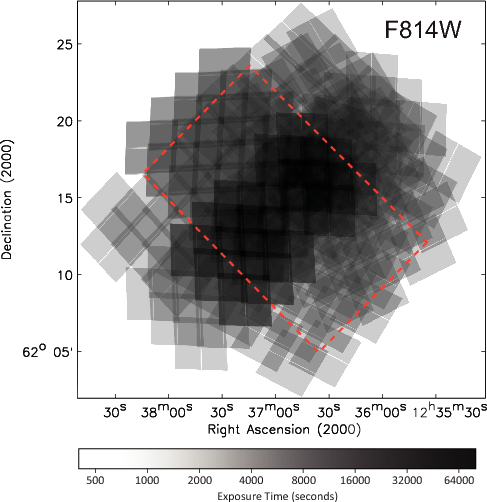}{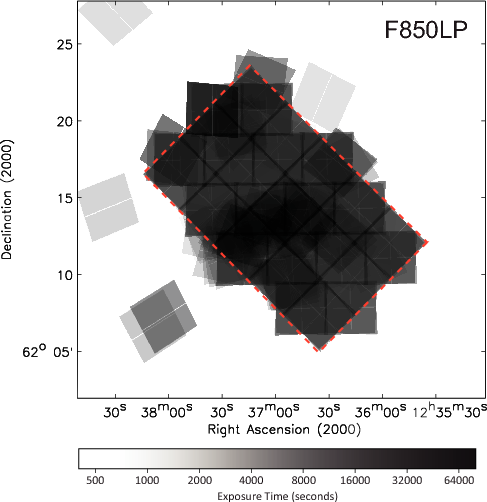}
\plottwo{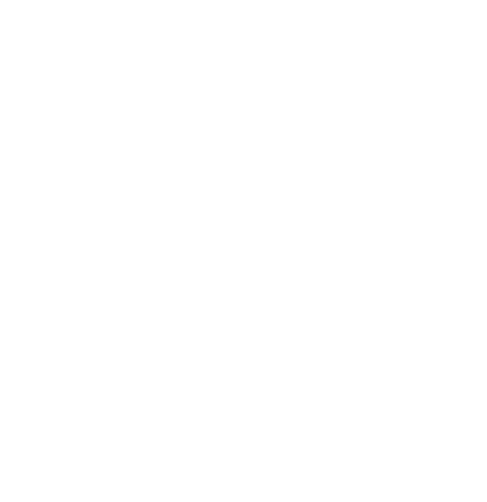}{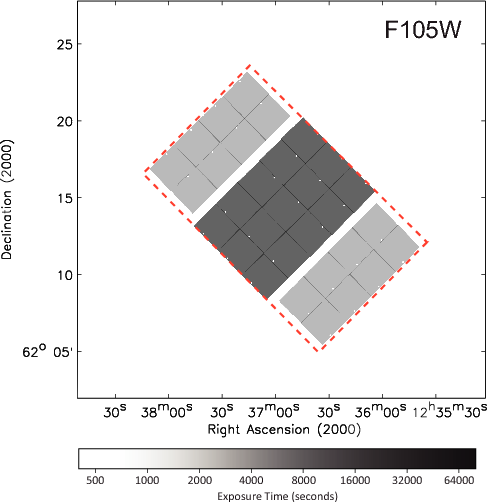}
\plottwo{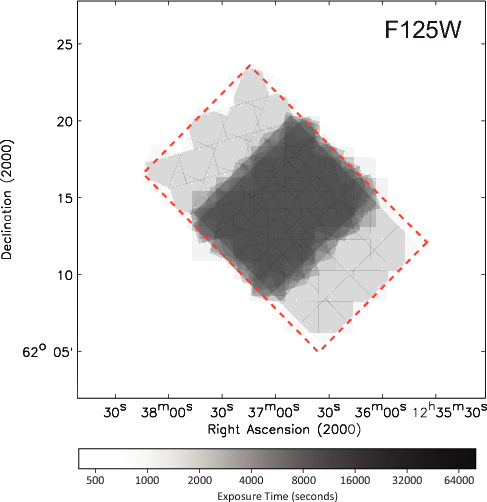}{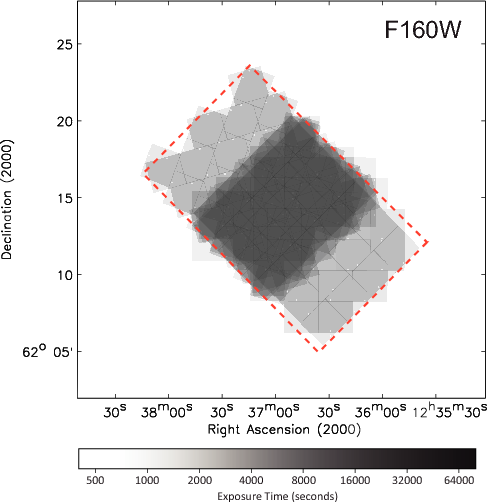}
\caption{\figgdnhiexpmaps}
\end{figure*}
\newpage
\begin{figure*}[ht]
\plottwo{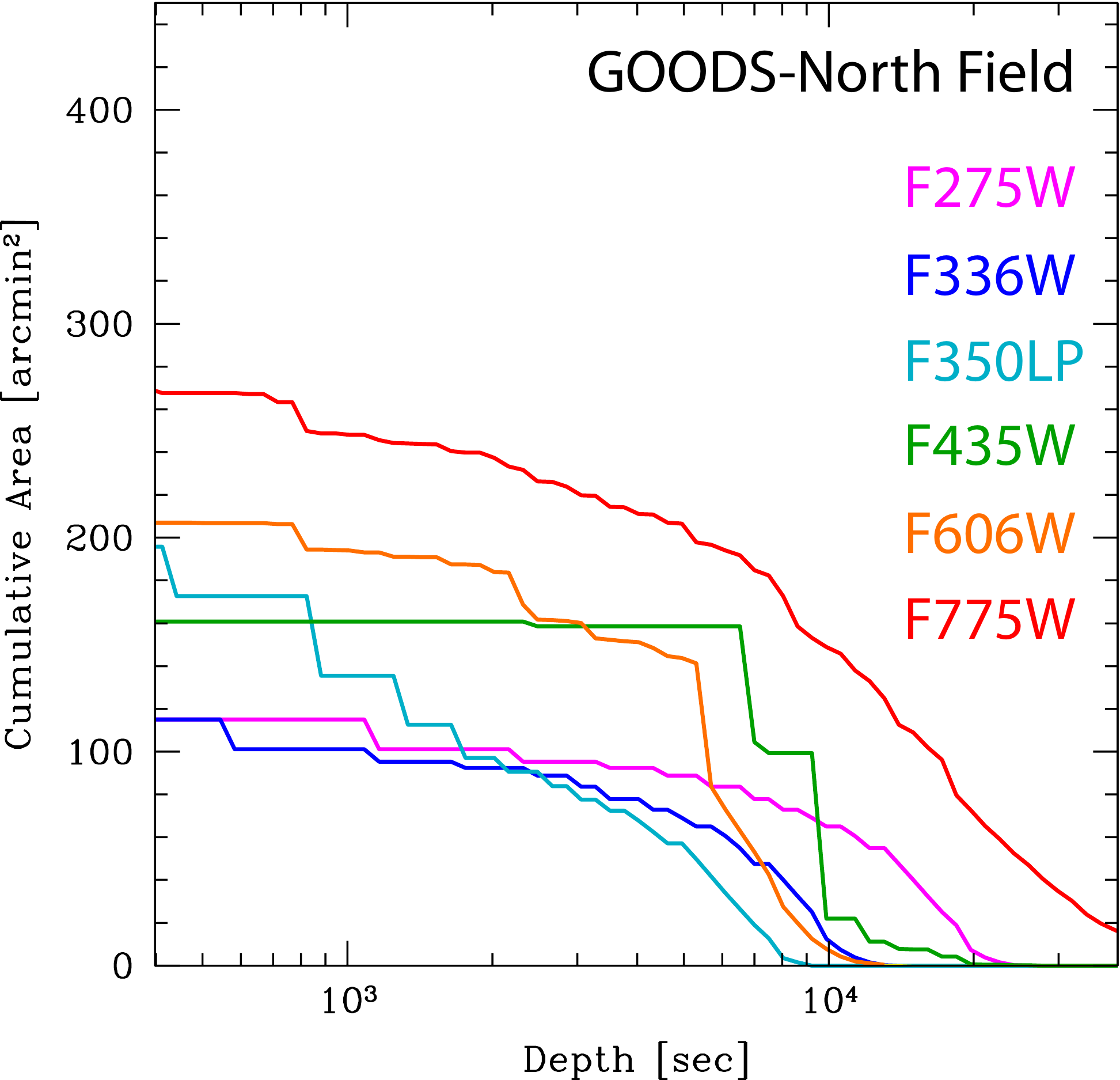}{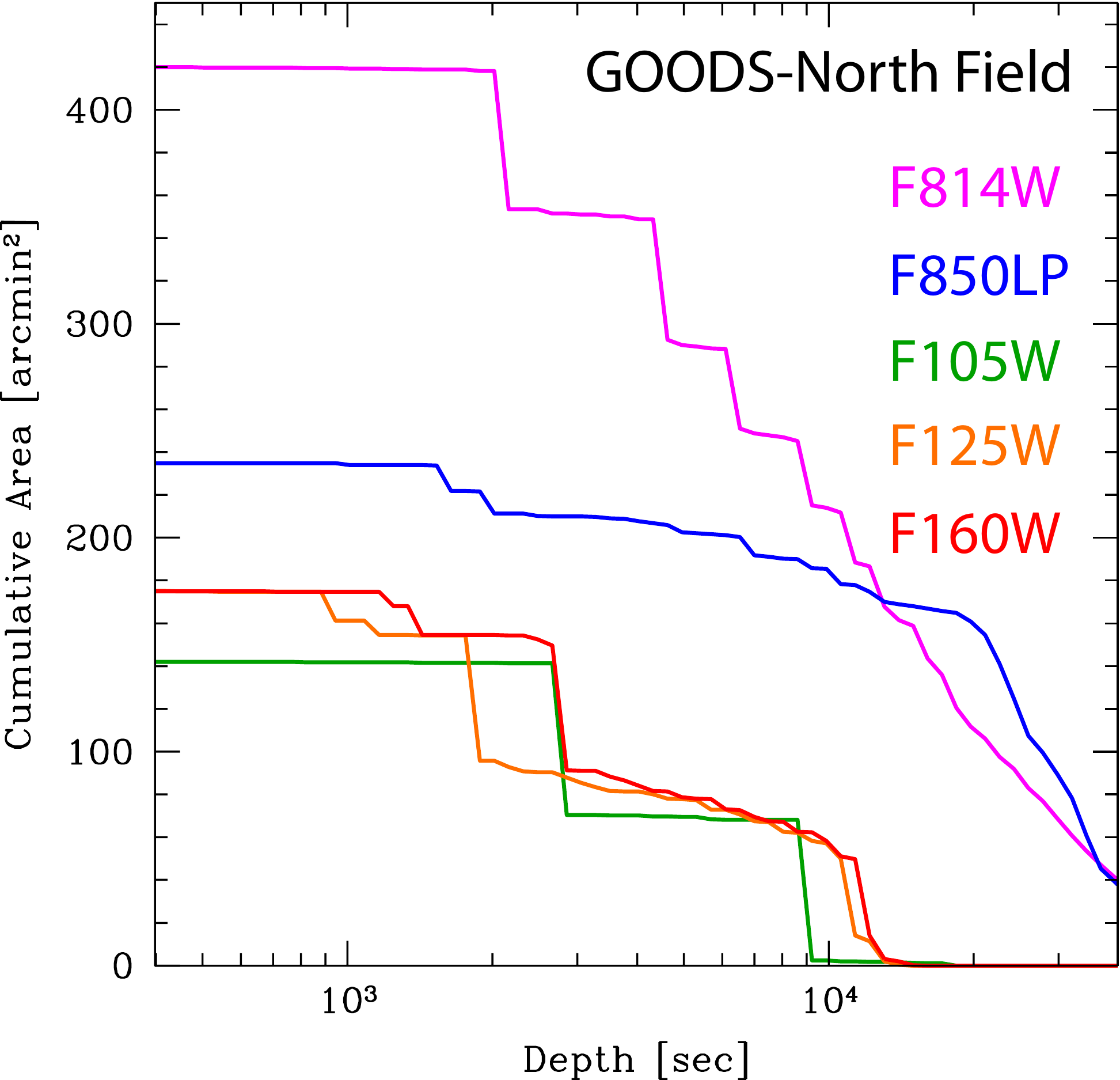}
\caption{\figgdnad}
\end{figure*}
\clearpage
\insertfigend{MasterSchedule.eps}{\figepochs}
\clearpage
\fi

\end{document}